\documentclass[fleqn,usenatbib]{mnras}

\usepackage{newtxtext,newtxmath}
\usepackage{xcolor}
\usepackage{soul}
\usepackage[flushleft]{threeparttable}
 
\usepackage[T1]{fontenc}

\DeclareRobustCommand{\VAN}[3]{#2}
\let\VANthebibliography\thebibliography
\def\thebibliography{\DeclareRobustCommand{\VAN}[3]{##3}\VANthebibliography}

\usepackage{graphicx}	
\usepackage{amsmath}	
\usepackage{stackengine}
\usepackage{threeparttable}

\def\SPSB#1#2{\rlap{\textsuperscript{\textcolor{black}{#1}}}\SB{#2}}

\def\SB#1{\textsubscript{\textcolor{black}{#1}}}
\usepackage[normalem]{ulem}

\newcommand{\ergs}{\mbox{ erg s}^{-1}}

\newcommand{\kms}{\mbox{ km s}^{-1}}

\title[Jets in magnetized media]{A polarization study of jets interacting with turbulent magnetic fields}

\author[Meenakshi et. al]{Moun Meenakshi$^{1}$,\thanks{E-mail:mounmeenakshi@iucaa.in}
Dipanjan Mukherjee$^{1}$,\thanks{E-mail:dipanjan@iucaa.in} Gianluigi Bodo$^{2}$ and Paola Rossi$^{2}$\\
$^{1}$Inter-University Centre for Astronomy and Astrophysics, Pune- 411007, India\\
$^{2}$INAF, Osservatorio Astroﬁsico di Torino, Strada Osservatorio 20, I-10025 Pino Torinese, Italy
}

\date{Accepted XXX. Received YYY; in original form ZZZ}

\pubyear{2023}

\begin{document}
\label{firstpage}
\pagerange{\pageref{firstpage}--\pageref{lastpage}}
\maketitle

\begin{abstract}

We investigate the effect of the jet's immediate surroundings on the non-thermal synchrotron emission and its polarization properties. The ambient medium is equipped with a turbulent magnetic field, which is compressed and amplified by the jets as they progress. This leads to high polarization at the forward shock surface. The randomness in the magnetic polarities of the external fields in the shocked ambient medium (SAM) results in vector cancellation of the polarized components from the jet, thereby causing depolarization of the radiation from the cocoon. We find that due to the slow decay of the fields in the SAM, such depolarization by the fields with large correlation lengths is more prominent when compared to the small-scale fields. Also, the low-power jets, which have magnetic fields comparable in strength to those in the SAM, are more severely affected by the SAM's depolarizing effect, than the high-power ones. The turbulent backflows in the cocoon, as well as the shearing of fields near the contact discontinuity, strengthen the poloidal component in the jet. This causes internal depolarization due to the cancellation of the orthogonally polarized components along the Line of Sight as the field transitions from ordered toroidal to poloidal. The synchrotron maps display high-emission filaments in the cocoon with magnetic fields aligned along them. The kink instability leads to the wiggling motion of the jet's spine, resulting in hotspot complexes in low-power sources.

\end{abstract}




\section{Introduction}

Observations of radio jets emitted from  supermassive black holes at the center of galaxies have been made in systems at various redshifts \citep[see][for a review]{Saikia_2022}. These jets are thought to originate from either magneto-hydrodynamic outflows from the accretion disk or the extraction of rotational energy from the black hole \citep{blandford_1977, blandford_1982}. Non-thermal synchrotron emission from  electrons is considered to be the primary emission mechanism in these sources \citep{burbidge_1956,bless_1962}. Therefore, understanding the polarization configuration of this radiation is useful for comprehending the field distribution inside the jets. If produced from the regions of ordered magnetic fields, the radiation can be expected to be highly polarized, or vice versa. However, the observed net polarization fraction may be reduced due to internal processes, such as cancellation of orthogonal polarization \citep{swain_1998}, or internal Faraday rotation due to mixing with thermal plasma \citep{Sullivan_2013,knuettel_2019,anderson_2018}. Moreover, the polarized radiation from the jets can be influenced by the magneto-ionized medium surrounding it when passing through these regions before reaching the observer \citep{burn_1966}.

Numerical studies \citep{bovino_2013,federrath_2016,seta_2020} suggest that the seed magnetic fields in the interstellar medium (ISM) can be amplified by the turbulent small-scale dynamo in galaxies, which is driven by supernovae or other stellar processes. Similarly, sub-cluster mergers can increase the strength of magnetic fields in the intracluster medium (ICM) \citep{kandu_2018}. Magnetic fields in and around galaxies and clusters are shown to follow a Gaussian distribution \citep{Guidetti_2008,guidetti_2010}, or a combination of a Gaussian and a non-Gaussian component \citep{govoni_2006,seta_2018}. Comparatively, the magnetic fields in the ISM are found to be stronger than the ICM \citep{roy_2003, Guidetti_2008,chyzy_2011}, and have smaller correlation lengths i.e. $\sim~$300~pc \citep{livingston_2022}, than those in the ICM \citep[up to 60~kpc, as shown in][]{guidetti_2010}.

Observations have shown that the linear polarization increases towards the outer edges of the radio lobes, while low values are observed inside the lobes \citep{saikia_1988,bridle_1994,pushkarev_2017}. This is attributed to the backflows from the jet's head, which cause the alignment of magnetic fields along the edges of the lobes. The compression of the external randomly oriented magnetic field by the bow shock of the jet is also expected to produce high-polarization values \citep{laing_1980}, with the magnetic fields aligned along the surface of the shock.

 Resolved radio observations have enabled to investigate the spatial effect of external medium on the depolarization of synchrotron radiation from jets at high redshifts \citep{goodlet_2004,kharb_2008,hammond_2012,lerato_2020,cantwell_2020}. The `Laing-Garrington effect' is a well-known observed phenomenon \citep{morganti_1997, morganti_1997a}, where polarization asymmetries can be seen in the advancing and receding lobes of jets viewed at an inclination \citep{laing_1988,garrington_1988}, although it is still debatable whether these asymmetries arise due to the external Faraday screen or internal lobe depolarization \citep{bell_2013}. Corresponding numerical 3D magneto-hydrodynamics studies with jets launched in a random magnetic field, which closely resembles the typical field distributions in the ambient medium \citep[see][]{tribble91,murgia_2004}, have also been performed \citep{laing_2008,guidetti_2011,huarte_2011b}. These studies show that the presence of random magnetic fields in the foreground can enhance the rotation measures (RM), resulting in Faraday rotations, and depolarization of the synchrotron radiation at the observer's image plane. \citet{huarte_2011b} found that the external magnetized medium is compressed and amplified at the bow shock by the jets, which has a prominent effect on enhancing the RM.

Almost all of the previous studies focus on how the spatial distribution of the magnetic fields at large scales can affect the RM and deduce its effect on the depolarization \citep[e.g.][]{govoni_2006,huarte_2011b,guidetti_2011,hardcastle_2014}. In this study, we focus on how the contribution of the immediate surrounding medium is affecting the integrated non-thermal emission and net polarization from the jet on the observer's image plane. To address this issue, we perform a set of 3D relativistic magneto-hydrodynamics (RMHD) simulations of jets interacting with a random magnetic field. The scope of our simulations is confined to jets of linear dimension 4~kpc. Thus the simulated jets represent young evolving jets of different powers, similar to Gigahertz Peak Spectrum (GPS) or Compact Steep Spectrum (CSS) sources, which are reported to be of sizes below 20~kpc \citep{odea_2021}. In this study, however, we do not focus on the spectral evolution (or absorption processes) of these sources. Instead, our primary aim is to understand how the dynamics of these young jets affect the morphology of emission, and how the presence of turbulent external magnetic fields affects the estimates of polarization.

An important motivation for this study has been the recent observations of AGN-related origin (jets/winds) of the radio emission in several compact and extended sources \citep{jarvis_2019,silpa_2020,jarvis_2021}. However, the nature of emission is still unclear in numerous sources. It is plausible to expect that studying the polarized radio emission from such sources could aid in deciphering the diverse outflow mechanisms at play. In our study, we primarily focus on analyzing the dynamics, synchrotron emission, and polarization originating from the jets. This will assist in understanding the magnetic field distribution and evolution in the jets, along with the effect of the external magnetized medium on the observable polarization characteristics from the radio lobes. We will extend the current investigation to explore the nature of synchrotron emission from AGN-driven winds in a subsequent study (Meenakshi et al., in prep). The winds are wide-angled outflows that are expected to be weakly magnetized. On the other hand, jets are collimated and can carry strong magnetic fields along them. Thus, the interaction of these outflows with the external medium can affect the emission and polarization features in different ways, which we intend to explore in the next phase of the study.

The paper is structured as follows. We describe the numerical setup for the simulations in Sec.~\ref{setup} and discuss in brief the post-processing analysis to estimate the synchrotron emission and polarization fraction in Sec.~\ref{theory_synch}. Since the observable features can vary with the inclination angle between the jet and the observer \citep[see e.g.][]{meenakshi_2022}, we also perform analysis for different viewing angles and present our results in Sec.~\ref{results}. Finally, we discuss the main findings from our study and their implications in Sec.~\ref{discussion}.

\section{Simulation Setup}
\label{setup}
We use the RMHD module in \textsc{pluto} \citep{mignone_2007}. The jets are launched with toroidal fields, and the ambient medium is equipped with turbulent magnetic fields at different correlation lengths. The setup for the various components in the simulations is discussed in detail in the following sections.

\subsection{Ambient atmosphere}

\label{ambient_atm}

\begin{figure*}
 \centerline{
\def\arraystretch{1.0}
\setlength{\tabcolsep}{0.0pt}
\begin{tabular}{lcr}
      \includegraphics[width=0.33\linewidth,height=0.28\linewidth]{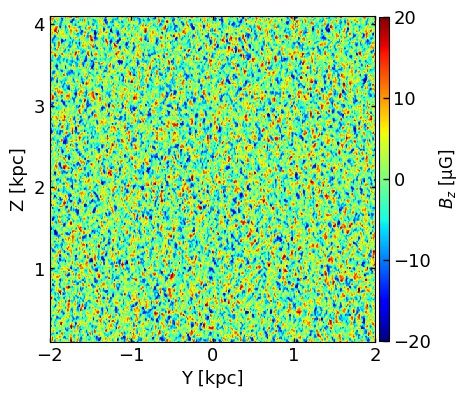} &
     \includegraphics[width=0.33\linewidth,height=0.28\linewidth]{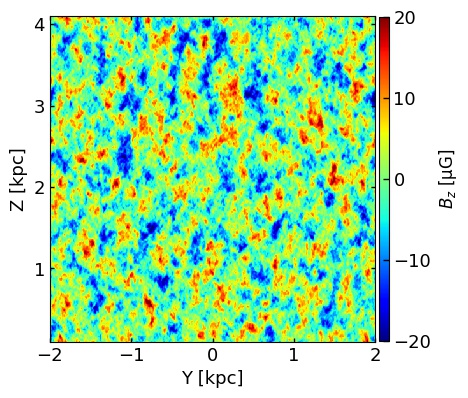} & 
    \includegraphics[width=0.33\linewidth,height=0.28\linewidth]{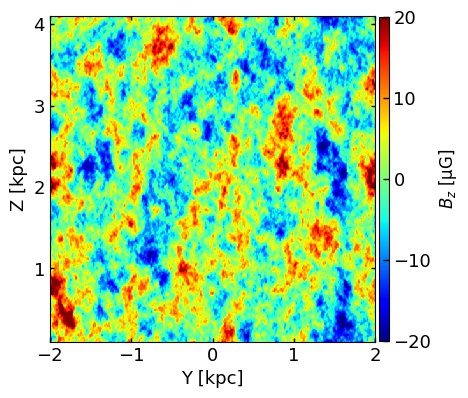}
     \end{tabular}}
          \centerline{
\def\arraystretch{1.0}
\setlength{\tabcolsep}{0.0pt}
\begin{tabular}{lcr}
      \includegraphics[width=0.33\linewidth,height=0.28\linewidth]{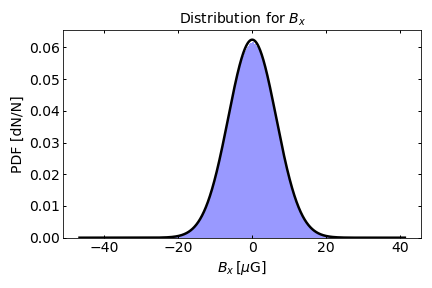} &
     \includegraphics[width=0.33\linewidth,height=0.28\linewidth]{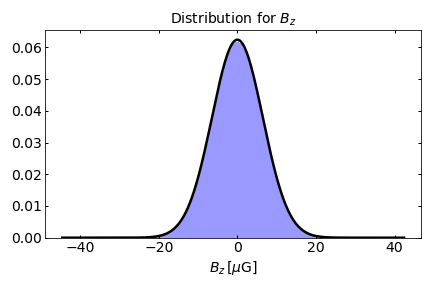} & 
    \includegraphics[width=0.33\linewidth,height=0.28\linewidth]{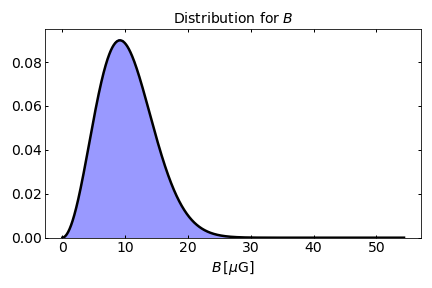}
     \end{tabular}}
\caption{\textbf{Top:} $B_z$ component of the magnetic field in the $Y-Z$ Plane at $t=0$ for the correlation length $\mathrm{L=100}, 500 \,\mathrm{and}\, 1000$~pc. \textbf{Bottom}: Distribution of the $B_x$ and $B_z$ components and magnitude of the total magnetic field for $\mathrm{L=500}$~pc case only,  the other two cases are expected to follow similar distributions. The black curves show the fits to the distribution which are Gaussian for the first two panels and Maxwell-Boltzmann for the last panel.}
 \label{fig:initial}
\end{figure*}

\begin{table*}
	\centering
	\caption{Parameters for simulations performed in this study.}
	\label{tab:sim_table}
	\begin{threeparttable}
	
	\begin{tabular}{|c|l|l|l|l|l|l|l|l|} 
		\hline
Simulation & Domain $(X\times Y \times Z)$ & Grid point &     $\eta_j$ & $\gamma_j$ & $P_j [\mathrm{\ergs}]$ &  $B_0 [\mathrm{mG}]$ & $M_j$ \\
		\hline
		$\mathrm{J42\_L500}$ & 5~kpc $\times$ 5~kpc $\times$ 4~kpc & $640 \times 640 \times 512$ & $4 \times 10^{-5}$ &  1.17 & $1.04 \times 10^{42}$ & 0.034 & 2.43 \\
		\hline
			$\mathrm{J43\_(L100,L500,L1000)}$ & 4~kpc $\times$ 4~kpc $\times$ 4~kpc & $512 \times 512 \times 512$ & $4 \times 10^{-5}$ &  2 & $1.25 \times 10^{43}$ & 0.09 & 6.9 \\
	\hline
	$\mathrm{J44\_(L100,L500,L1000)}$ & 4~kpc $\times$ 4~kpc $\times$ 4~kpc & $512 \times 512 \times 512$ & $4 \times 10^{-4}$ &  2 & $1.07 \times 10^{44}$ & 0.26 & 21.6 \\
		\hline
			$\mathrm{J45\_(L100,L500,L1000)}$ & 4~kpc $\times$ 4~kpc $\times$ 4~kpc & $512 \times 512 \times 512$ & $4 \times 10^{-4}$ &  5 & $1.2 \times 10^{45}$ & 0.83 & 61 \\
			\hline
		$\mathrm{J44\_L500a}$ & 5~kpc $\times$ 5~kpc $\times$ 16~kpc & $500 \times 500 \times 1600$ & $4 \times 10^{-4}$ &  2 & $1.07 \times 10^{44}$ & 0.26 & 21.6 \\
		\hline
	\end{tabular}
	\begin{tablenotes}
\item The simulation label shows the jet power and the correlation length of the ambient magnetic field. For example, $\mathrm{J42\_L500}$ denotes the simulation with a jet of power $\sim10^{42}\,\ergs$ launched in an ambient medium with a mean magnetic field of approx 10~$\mu$G and a correlation length of 500~pc. We also perform `no-jet' or control simulations for each correlation length and name them $\mathrm{L(100,5000,1000))\_nj}$ in the text. \\
$\eta_j$: Ratio of jet density to the ambient gas density at the radius of injection. \\
 $\gamma_j$: Bulk Lorentz factor of the jet.\\
 $P_j$: Mechanical power of the jet.\\
 $B_0 $: Maximum strength of the toroidal magnetic field of the jet at the injection zone. Note that the magnetization of the jet ($\sigma$) as defined in Eq.~\ref{eq:sigma} is taken as 0.1 for all the simulations.\\
$M_j$ : Mach number of the jet.
	\end{tablenotes}
	\end{threeparttable}
\end{table*}

\subsubsection{Hydrodynamics}
The setup of the ambient medium is very similar to that used in \citet{dipanjan_2020}, where the halo is in pressure equilibrium in the presence of an external static gravitational potential. For the stellar (baryonic) contribution we use the Hernquist potential \citep{Hernquist_1990}, given as:
\begin{equation}
    \phi_\mathrm{B} = -\frac{G M_B}{r+a_H}
\end{equation}
where $G$ is the gravitational constant, $M_B = 10^{11} M_{\odot}$ is the stellar (baryonic) mass of the galaxy, $a_H=2$~kpc is the scale radius, and $r$ is the spherical radius.

The contribution of the dark-matter component in the external potential is set using the Navarro-Frank-White (NFW) profile \citep{NFW_1996}:
\begin{align}
   \phi_\mathrm{DM} = \frac{-G M_{200}}{[\ln{(1+\tilde{c})}+\tilde{c}/(1+\tilde{c})]}\left( \frac{1}{r+d}\right) \ln{\left( 1+\frac{r}{r_s}\right)} \notag \\
\end{align}

Here, $M_{200}= 200 \rho_\mathrm{cr}\frac{4\pi}{3}\tilde{c}^3 r\SPSB{3}{s}$,  $r_s = r_{200}/\tilde{c}$, $\rho_\mathrm{cr} = 3H^2/(8\pi G) \approx 8.506\times 10^{-30}\, \mathrm{g\,cm^{-3}}$ is the critical density of the universe at $z=0$ with a Hubble constant $H = 67\, \mathrm{kms^{-1} Mpc^{-1}}$ \citep{planck_2015}, $\tilde{c}$ is the concentration parameter, and $r_{200}$ is the radius where the mean density of the dark matter halo is 200 times the critical density of the universe. The value of $r_{200}$ is chosen as 1~Mpc and $\tilde{c}=10$, and the virial mass of the system is $M_{200} = 10^{14} M_\odot (r_{200}/\mathrm{Mpc})^3$. Thus, these parameters represent a giant elliptical galaxy at the center of the clusters \citep{Croston_2008,bohringer_2022}. The value of $d$ is set to $10^{-3}$ to avoid the singularity at $r=0$. 
The halo is kept at a constant temperature $T_h = 10^7$~K. 
The density and pressure values for the halo are then estimated by assuming hydrostatic equilibrium in the external gravitational field as:

\begin{equation}
\frac{d p_a}{d r} = \rho_a (r)\frac{d \phi (r)}{dr}; \qquad p_a(r) = \frac{\rho_a (r)}{\mu m_a} k_B T_h     
\end{equation}
from which we derive
\begin{equation}\label{eq:halo}
p_a(r) = (n_0 k_B T_h) \exp{\left[- \int_0^r \left( \frac{\mu m_a}{k_B T_h}\right)\frac{d\phi (r)}{dr} dr \right]}    
\end{equation}
 where $p_a$ and $\rho_a = \mu m_a n_h$ are respectively the pressure and density of the halo gas, $\mu=0.6$ is the mean molecular weight of the fully ionized gas, $m_a$ is the atomic weight, and $n_0$ is the number density at $r=0$. The total gravitational potential ($\phi$) is the sum of the baryonic ($\phi_\mathrm{B}$) and dark-matter potential ($\phi_\mathrm{DM}$). We solve Eq.~\ref{eq:halo} numerically and generate a table with values of density and pressure as a function of radius, which is then interpolated onto the \textsc{pluto} grid.
 
\subsubsection{Magnetic field}
 In our study, we use a turbulent magnetic field in the ambient atmosphere. For this, we follow the approach given in \citet{tribble91} and \citet{murgia_2004}. The Fourier components of the vector-potential $\tilde{\textbf{A}}\textbf{(k)}$ i.e. the amplitude ($A$) and phase ($\phi$) in polar coordinates, are randomly drawn from the distribution given below,
\begin{align}
    P(A,\phi)dA d\phi=\frac{A}{2\pi|A_k|^2}\exp\left[\frac{-A^2}{2|A_k|^2}\right]dA d\phi
\end{align}
i.e. $A$ is estimated from a Rayleigh distribution and $\phi$ is uniformly distributed between 0 and 2$\pi$. The power-spectra of the vector potential ($A$) is chosen as $|A_k|^2\propto k^{-\zeta}$.
Then, the magnetic field components can be estimated as,
\begin{equation*}
    \tilde{\textbf{B}}\textbf{(k)}=i\textbf{k}\times \tilde{\textbf{A}}\textbf{(k)}
\end{equation*}
 where the components of the field follow Gaussian distribution with zero mean value, and the magnitude of the total field has a Maxwell-Boltzmann distribution. The power spectrum for a turbulent system with free energy cascade to lower spatial scales is expected as $|B_k|^2\propto k^{-(\zeta-2)}$. We chose the exponent for the power spectrum of the vector field to be $\zeta=17/3\approx 5.667$, such that for the magnetic field it is $\zeta - 2=11/3 \approx 3.667$, which corresponds to turbulent Kolmogorov power spectrum. In Fourier space, the different correlation lengths of the magnetic field are set by taking the minimum value of the wave number $k$ i.e. $k_\mathrm{min}$ as 40, 8, and 4 in the grid units to represent the lengths of $\mathrm{L=}$100~pc, 500~pc, and 1000~pc, respectively in a simulation cube of dimensions $512^3$ and length 4~kpc. The maximum value $k_\mathrm{max}$=256 is limited by the Nyquist sampling.
 
 We show slices of the initial ($t=0$) magnetic field in the $Y-Z$ plane in the top panel of Fig.~\ref{fig:initial} for different correlation lengths.
 The vector potential spectral distribution is given as an input to compute the corresponding spatial distribution of the  magnetic field in the initial configuration. In the bottom panels, we show the distribution of $B_x$, $B_z$, and the total magnetic field obtained from the $\textsc{pluto}$ output, where the first two show  a Gaussian distribution and the latter shows a Maxwell-Boltzmann distribution. From the fits (black curves) we can see that in real space the magnetic field components and their intensity follow the expected behaviors. The mean magnitude of the ambient magnetic field at $t=0$ is $\sim10\,\mu$G, which is similar to the typical small-scale magnetic field values in spirals, irregulars, and ellipticals \citep{kepley_2010,chyzy_2011,hilay_2021}. This corresponds to a mean plasma $\beta$, i.e. the ratio of the thermal energy to the magnetic energy, of 12 in our simulations.

\subsection{\textsc{pluto} Setup}
\label{pluto_setup}
We employ the RMHD module in \textsc{pluto} \citep{mignone_2007} for performing simulations in 3D Cartesian geometry for this study. 
The different parameters for the simulations performed for this study are listed in Table~\ref{tab:sim_table}. For comparison, we also perform `no-jet' or control simulations, as mentioned in the table notes. The resolution element for the standard simulations is 7.8~pc, and 10~pc for a large-box run with a jet of power $10^{44}\,\ergs$ (i.e. $\mathrm{J44\_L500a}$).
The RMHD module solves the following set of conservation equations:

\begin{equation}
\centering
\frac{\partial}{\partial t} 
\begin{pmatrix} D \\ \pmb{m} \\ E_t \\ \pmb{B} \end{pmatrix}\,
+\,
\nabla \cdot
  \begin{pmatrix}
  D \pmb{v}\\ w_t\gamma^2 \pmb{v v} - \pmb{b b} + I p_t \\ \pmb{m} \\ \pmb{vB-Bv} 
  \end{pmatrix}^T  \,  = \, 
  \begin{pmatrix}
  0 \\ \pmb{f}_g \\ \pmb{v \cdot f}_g \\ 0 
  \end{pmatrix},
\end{equation}
where $D$ is the laboratory density, $\pmb{m}$ is the momentum density, $E_t$ is the total energy, $\pmb{B}$ is the magnetic field and \pmb{f} is an acceleration term.
The different variables in terms of fluid density $\rho$, velocity $\pmb{v}$, pressure $p$ and Lorentz factor $\gamma$ are given as:
\begin{equation} \notag
\begin{matrix}
      D = \gamma \rho \\
    \pmb{m} = w_t \gamma^2 \pmb{v}- b^0 \pmb{b}\,\,, \quad  \\
    E_t = w_t \gamma^2 - b^0 b^0 - p_t 
\end{matrix}
\begin{cases}
\begin{matrix}
        b^0 = \gamma \pmb{v \cdot B} \\
      \pmb{b} = \pmb{B}/\gamma + \gamma (\pmb{v \cdot B})\pmb{v}\\
      w_t = \rho h + \pmb{B^2}/\gamma^2 + (\pmb{v \cdot B})^2 \\
      p_t = p + \frac{\pmb{B^2}/\gamma^2 + (\pmb{v \cdot B})^2}{2}
\end{matrix}
\end{cases}
\end{equation}

We use a piece-wise Parabolic scheme for reconstruction \citep{mignone_2005}, third-order Runge-Kutta method for time integration, and five-wave HLLD Riemann solver \citep{mignone_2009}. We utilize a more diffusive Riemann solver (HLL) and MIN-MOD limiter for cells that are recognized as shocked in regions with $Z\leq0.5$~kpc to ensure the numerical stability of the solutions. A computational cell is identified as shocked if the pressure jumps $\delta p$ with respect to neighbouring cells satisfies the condition $\delta p/p_\mathrm{min}>4$, where $p_\mathrm{min}$ is the minimum pressure among all the surrounding cells. We use the outflow boundary condition at all the faces except at lower $Z$, where the outflow is injected and the boundary, outside the injection region, is set as reflective.

At initialization, the ambient medium is set up with the density and pressure profiles as discussed in Sec.~\ref{ambient_atm}. The halo is also equipped with a divergence-free random magnetic field. This is achieved by specifying the vector potential at cell edges. We use the Constrained Transport approach \citep{balsara_1999,mignone_2021} to preserve the divergence-free magnetic field during the simulation. The electromotive forces are reconstructed at the edges using the CT-CONTACT scheme in \textsc{pluto} \citep{gardiner_2005}. 

The jet is injected as a conical outflow from the lower $Z-$axis. We assume the jets to be in pressure equilibrium with the ambient medium at the injection zone. For a specific power of the outflow, one needs to specify different injection parameters which are \citep[also see Sec.~2.3 in][]{dipanjan_2020}: 

\begin{itemize}
    \item $\eta_j$: The ratio between the gas density of the jet and the ambient medium at the region of injection. We use values between $4\times 10^{-5}-4\times 10^{-4}$ for the simulations in this paper, similar to the values used in previous simulations \citep{rossi_2008,perucho_2022}.
    \item $\gamma_j$: The bulk Lorentz factor of the jet. The jet is injected with a half-cone opening angle of $7^\circ$ centered at the $Z$-axis.
    \item $\sigma$: Jet-magnetization, defined as the ratio of the Poynting flux ($S_j$) to the jet enthalpy flux ($F_j$), and is given as:
    \begin{equation}\label{eq:sigma}
        \sigma\, =\, \frac{|\textbf{S}_j \cdot \hat{z}|}{|\textbf{F}_j \cdot \hat{z}|} \,=\,\frac{(|{\textbf{B}_j}\times({\textbf{v}_j}\times {\textbf{B}_j})) \cdot \hat{z}|}{4\pi({\gamma}^2 \rho_j h_j -\gamma \rho_j c^2)(|\textbf{v}_j \cdot \hat{z}|)},
    \end{equation}
    where $\textbf{B}_j$ and $\textbf{v}_j$ are the magnetic field and the velocity vector of the jet. The density and pressure of the jet are denoted by $\rho_j$ and $p_j$, respectively. The relativistic enthalpy of the jet per unit volume is $\rho_j h_j$, which for an ideal fluid is given as
    \begin{equation}\label{eq:enthalpy}
        \rho_j h_j = \rho_j c^2 + \frac{p_j \Gamma}{\Gamma - 1},
    \end{equation}
    where a constant adiabatic index ($\Gamma=5/3$) is used for the ideal gas. We use a fixed value of 0.1 for the magnetization ($\sigma$) of the jets in our simulations. This value is comparable to the magnetization of a jet launched with an initial value of 10 \citep{chatterjee_2019} by a black hole with  a mass of $3\times 10^{8}~M_{\odot}$ \citep{laor_2000}. For the jets injected with the toroidal field, the peak value of the magnetic field is determined using a simplified form of the Eq.~\ref{eq:sigma} i.e. $B_0 = \sqrt{(4\pi \sigma F_j)/v_j}$, and the values are also mentioned in Table~\ref{tab:sim_table}. The $X$ and $Y$ components of the jet's magnetic field are calculated as:
\begin{equation}\label{eq:Bmag}
B_{(x,y)} = \begin{cases}
\begin{matrix}
     B_{(x0,y0)} (r/R_j)  \quad \mathrm{for}\,\, r<=R_j\\
    B_{(x0,y0)} (r/R_j) \cosh{(r/R_j)}^6 \quad \mathrm{for}\,\, r>R_j
\end{matrix}
\end{cases}
\end{equation}
where $B_{(x0,y0)}$ are the magnetic field components estimated using the peak strength value $B_0$.

    \item $R_j:$ The radius of the jet. This defines the extent of the region where the outflow is injected in the simulation domain. We use a value of 50~pc for the jet's radius. For a resolution of 7.8~pc in our standard simulations, the diameter of the jet is resolved with 12 cells.
    
\end{itemize}
The total mechanical power of the jet $P_j$ is estimated by integrating the total energy (after subtracting the rest mass energy) flux over the injection region of the jet. The energy flux per unit area along the $Z-$ direction is given as \citep{mignone_2009}:
\begin{equation}\label{eq:jet-power}
   F_z = ({\gamma}^2\rho_j h_j - \gamma\rho_j c^2)v_z + \frac{B^2}{4\pi}v_z - (\textbf{v}\cdot \textbf{B})\frac{B_z}{4\pi}
\end{equation}

Since the magnetization value for the jets is 0.1 in our study, we use only the first term in the expression above to estimate the mechanical power of the jets, which are given in Table~\ref{tab:sim_table}.
The density and pressure of the jet are also smoothed with the cosh function in Eq.~\ref{eq:Bmag} for $R>R_j$ to avoid any spurious instabilities at the edges. For velocity, the injection is sharply cut at the jet's radius ($R_j$) to avoid any extra influx from the outer regions so that the injected power of the jet is limited to the values estimated using Eq.~\ref{eq:jet-power}.

\section{Post-processing: Synchrotron emission and polarization}
\label{theory_synch}

In the following section, we provide a concise overview of the mathematical representation of synchrotron emissivity. For a more detailed and comprehensive understanding, we recommend the readers refer, for example, to \citet{vaidya_2018}.

The synchrotron emissivity by an ensemble of relativistic electrons with an energy range between $E_i$ and $E_f$ is estimated as \citep{Ginzburg_1965}
\begin{equation}\label{eq:synch2}
    {J'}_{syn} (\nu',n',B') = \frac{\sqrt{3}e^3}{ 4\pi m_e c^2} |\vec{B}'\times \hat{n}'| \int_{E_i}^{E_f} \mathcal{N}'(E') F(x)dE',
\end{equation}
where all the prime quantities are measured in a local comoving frame with respect to the observer.
The mass and charge of an electron are denoted by $m_e$ and $e$, the speed of light is $c$, and $\nu'$, $\Vec{B'}$, and $\Vec{n'}$ are the frequency of emitted radiation, magnetic field, and the LOS vector.
The spectrum of energy of the electrons is postulated to adhere to a power-law distribution with an index $-\alpha$ and is represented by $\mathcal{N'}$. We also assume isotropic condition, i.e. $\mathcal{N'} (E') = 4\pi N'(E',\hat{n'})$, where $N'(E',\hat{n}')$ represents the count of particles with energy $E'$ and velocities directed around $\hat{n}'$. 

The function $F(x)$ is the Bessel function integral, given by
\begin{equation}
    F(x) = x \int_{x}^{\infty} K_{2/3}(z) dz,
\end{equation}
where $x=\frac{\nu'}{{\nu}^{'}_{cr}} = \frac{4\pi m_e^3 c^5 \nu'}{3e {E'}^2 |\vec{B}'\times \hat{n}'|}$
where ${\nu}^{'}_{cr}$  is the critical frequency where the function $F(x)$ peaks.  In our calculations, we simplify the integral in Eq.~\ref{eq:synch2} by using the Bessel integrals given in \citet{rybicki_1979}. Thus, the expression for the synchrotron emissivity in the observer's frame is given as:

 \begin{align}\label{eq:jsyn}
\mathcal{J}_{syn} =  \mathcal{D}^2 N_0\frac{3^{\alpha/2} e^2 \nu'^{-(\alpha-1)/2} |\Vec{B'} \times \hat{n'}|^{(\alpha+1)/2}}{2c(\alpha+1)} \notag \\ \Gamma\left(\frac{\alpha}{4}+\frac{19}{12}\right)\Gamma\left(\frac{\alpha}{4}-\frac{1}{12}\right)
\left(\frac{e}{2\pi mc}\right)^{(\alpha+1)/2}
 \end{align}

Here

\begin{equation}\label{eq:N_0}
    N_0 = \frac{\epsilon E_n}{m_ec^2\int_{\gamma_i}^{\gamma_f} {\gamma_e}^{(-\alpha+1)}d\gamma},
\end{equation}
and $\epsilon$ shows the fraction of the internal energy density of the fluid ($E_n = p/(\Gamma -1)$) carried by the electrons, which is taken as 0.1 for our study. The Doppler factor is denoted by $\mathcal{D} = 1/(\gamma (1-\beta \cdot \Vec{n}))$, where $\gamma$ is the Lorentz factor of the fluid. $\gamma_i$ and $\gamma_f$ are the minimum and maximum limits of the Lorentz factor  ($\gamma_e$) for the accelerated electrons which are taken as $100$ and $10^6$ for our calculations.
The emissivity obtained from Eq.~\ref{eq:jsyn} is integrated along the LOS to estimate the total synchrotron flux ($I_\nu$) on the image plane.

Similarly, the linear polarized emissivity in the local comoving frame is given as
\begin{equation}\label{eq:polar}
     {J'}_{pol} (\nu',n',B') = \frac{\sqrt{3}e^3}{4\pi m_e c^2} |\vec{B}'\times \hat{n}'| \int_{E_i}^{E_f} \mathcal{N}'(E') G(x)dE',
\end{equation}
where $G(x)=xK_{2/3}$ is the Bessel function \citep{vaidya_2018}. After simplifying the Bessel integral, the expression for the emissivity of the polarized radiation in the observer's frame is
\begin{align}\label{eq:jpol}
\mathcal{J}_{pol} = \mathcal{D}^2 N_0\frac{3^{\alpha/2} e^2 \nu'^{-(\alpha-1)/2} |\Vec{B'} \times \hat{n}'|^{(\alpha+1)/2}}{8 c} \notag \\ \Gamma\left(\frac{\alpha}{4}+\frac{7}{12}\right)\Gamma\left(\frac{\alpha}{4}-\frac{1}{12}\right)
\left(\frac{e}{2 \pi m c}\right)^{(\alpha+1)/2}
\end{align}
Using Eq.~\ref{eq:jpol} the Stokes parameters for linear polarization $Q_\nu$ and $U_\nu$ can be computed by integrating along the line of sight ($\mathcal{Z}$) as \citep{lyutikov_2003,Zanna_2006} 
\begin{align} \label{eq:QU}
     Q_\nu = \int_{-\infty}^{\infty} \mathcal{J}_{pol} \cos{2\chi} d\mathcal{Z}, \quad \mathrm{where}\, \cos{2\chi} = \frac{q\SPSB{2}{$\mathcal{X}$}-q\SPSB{2}{$\mathcal{Y}$}}{q\SPSB{2}{$\mathcal{X}$}+q\SPSB{2}{$\mathcal{Y}$}} \notag \\
   U_\nu = \int_{-\infty}^{\infty} \mathcal{J}_{pol} \sin{2\chi} d\mathcal{Z},\, \quad \mathrm{where}\, \sin{2\chi} = -\frac{2q_{\mathcal{X}} q_{\mathcal{Y}}}
  {q\SPSB{2}{$\mathcal{X}$}+q\SPSB{2}{$\mathcal{Y}$}}
\end{align}
Here $q_\mathcal{X} = (1-\beta_\mathcal{Z})B_\mathcal{X} + \beta_\mathcal{X} B_\mathcal{Z}$ and $q_\mathcal{Y} = (1-\beta_\mathcal{Z})B_\mathcal{Y} + \beta_\mathcal{Y} B_\mathcal{Z}$. 
The $\mathcal{X}$ and $\mathcal{Y}$ axes are taken in the plane of the sky in the observer's frame with $\mathcal{Y}$ facing north and $\chi$ is the local polarization angle, measured clockwise from the $\mathcal{Y}$-axis. $\beta_\mathcal{X,Y,Z}$
are the velocity components along the $\mathcal{X,Y}$ and $\mathcal{Z}$ axes.
The total polarization fraction along the LOS can be estimated as:
\begin{equation}
    \Pi = \frac{\sqrt{{Q_\nu}^2+{U_\nu}^2}}{I_{\nu}}
\end{equation}

The output from $\textsc{pluto}$ is in the lab frame, and the primed measures in the co-moving frame in Eq.~\ref{eq:jsyn} and~\ref{eq:jpol} are estimated using the transformation equations given in \citet{vaidya_2018}. Since the simulations in this study do not include the evolution of any non-thermal particles, we assume a distribution of electron population with energy spectrum following a power-law index 2.2 (i.e. $\alpha$ in Eq.~\ref{eq:N_0}). This corresponds to a steep synchrotron spectrum with an index of 0.6, typically observed in the radio lobes of the jets \citep{jarvis_2019,silpa_2022}. The observed frequency for the emitted radiation in the lab frame is taken as 1~GHz for all the results presented in the following sections.

\begin{figure}
 \centerline{
\def\arraystretch{1.0}
\setlength{\tabcolsep}{0.0pt}
\begin{tabular}{lcr}
    \includegraphics[width=0.99\linewidth]{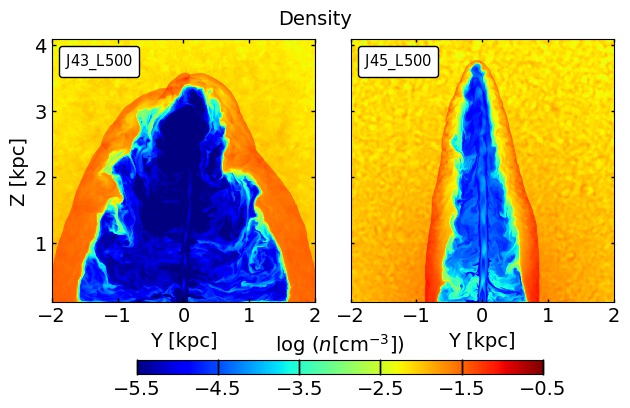} \\
    \includegraphics[width=0.99\linewidth]{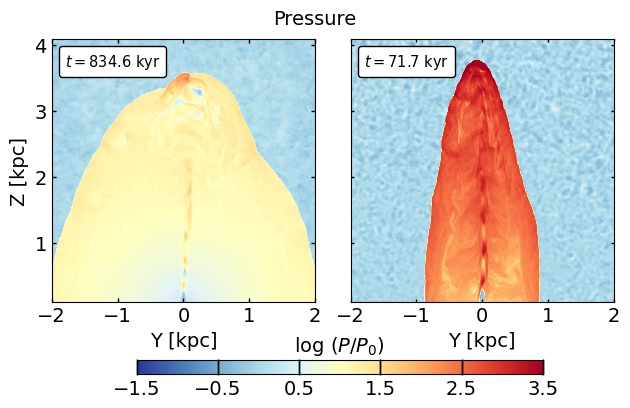} 
     \end{tabular}}    
     \caption{\textbf{Left to Right:} Logarithmic density (top) and pressure (bottom) in $\mathrm{J43\_L500}$ and $\mathrm{J45\_L500}$ at $t =$ 834.6~kyr and 71.7~kyr, respectively.}\label{fig:rho_prs}
\end{figure}

\begin{figure}
 \centerline{
\def\arraystretch{1.0}
\setlength{\tabcolsep}{0.0pt}
\begin{tabular}{lcr}
    \includegraphics[width=0.7\linewidth]{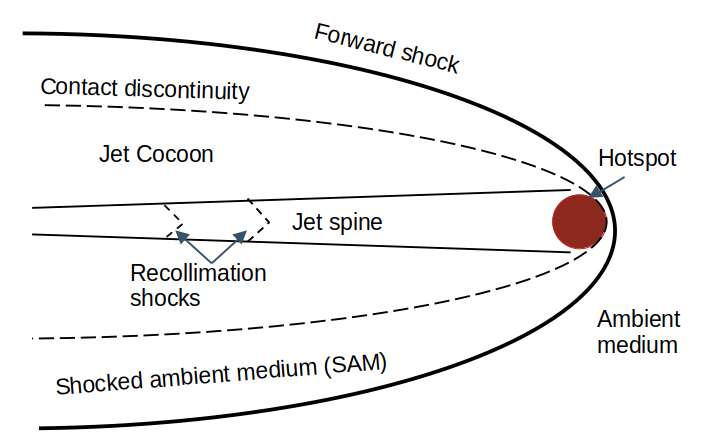}
     \end{tabular}}    
     \caption{A schematic diagram showing different elements of the radio jet, cocoon and shocked ambient medium (SAM).}\label{fig:cartoon}
\end{figure}

\begin{figure*}
 \centerline{
\def\arraystretch{1.0}
\setlength{\tabcolsep}{0.0pt}
\begin{tabular}{lcr}
      \includegraphics[width=0.33\linewidth,height=0.28\linewidth]{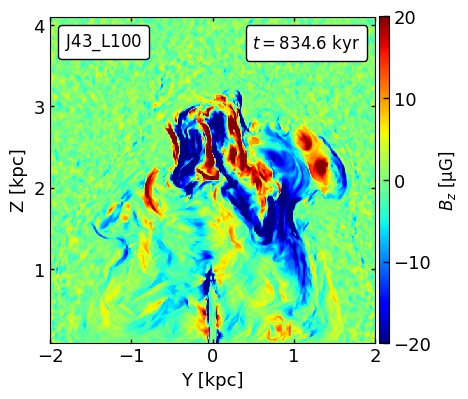} &
     \includegraphics[width=0.33\linewidth,height=0.28\linewidth]{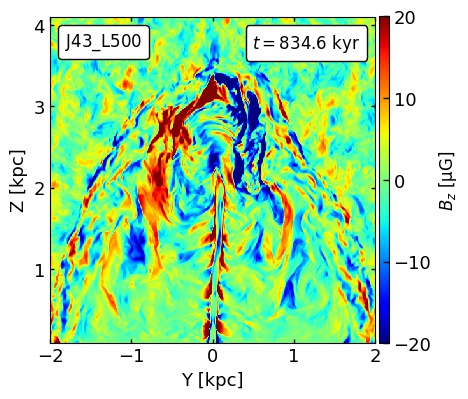} & 
    \includegraphics[width=0.33\linewidth,height=0.28\linewidth]{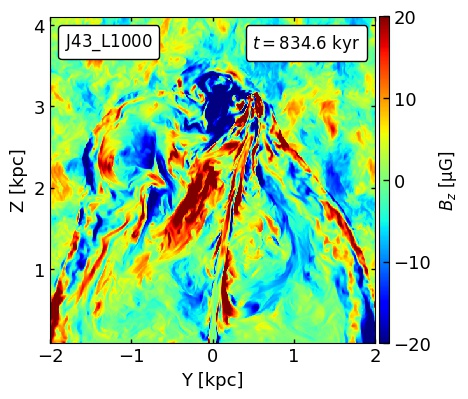}
     \end{tabular}}
     \centerline{
\def\arraystretch{1.0}
\setlength{\tabcolsep}{0.0pt}
\begin{tabular}{lcr}
    \includegraphics[width=0.33\linewidth,height=0.28\linewidth]{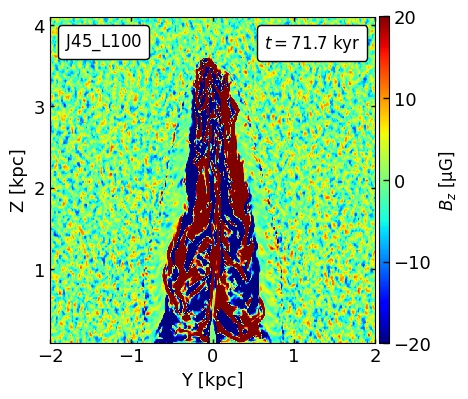} &
     \includegraphics[width=0.33\linewidth,height=0.28\linewidth]{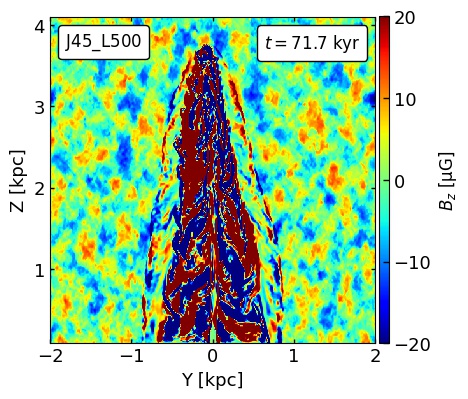} & 
    \includegraphics[width=0.33\linewidth,height=0.28\linewidth]{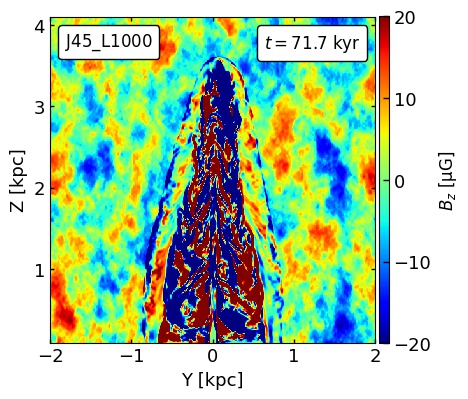}
     \end{tabular}}
     \caption{\textbf{Top:} $Z$ component of the magnetic field ($B_z$) for $\mathrm{J43\_L(100,500,1000)}$ at 834.6~kyr. \textbf{Bottom:} Same as above for $\mathrm{J45\_L(100,500,1000)}$ at 71.7~kyr.}
\label{fig:Bmag}
\end{figure*}

\begin{figure*}
 \centerline{
\def\arraystretch{1.0}
\setlength{\tabcolsep}{0.0pt}
\begin{tabular}{lcr}
 \includegraphics[width=0.25\linewidth]{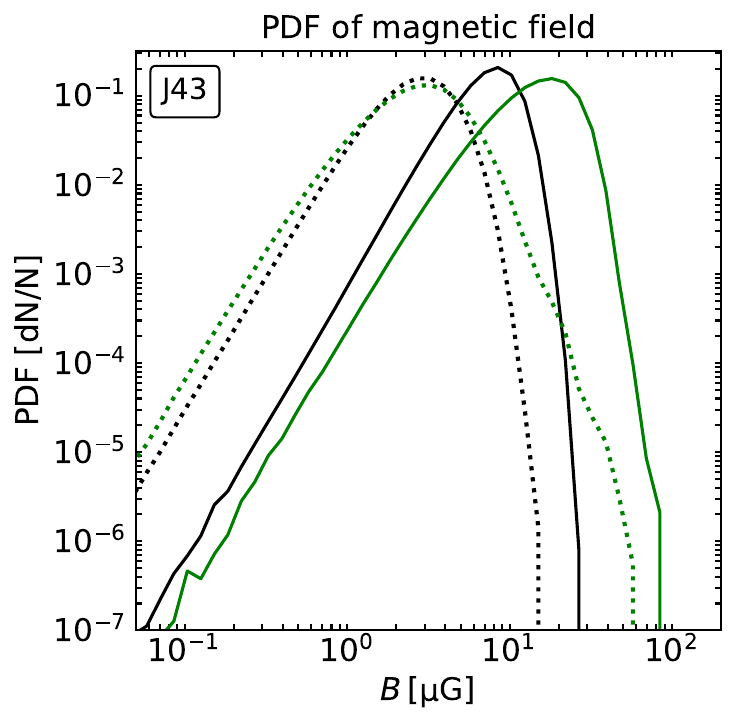}
      \hspace{-0.23cm}
    \includegraphics[width=0.25\linewidth]{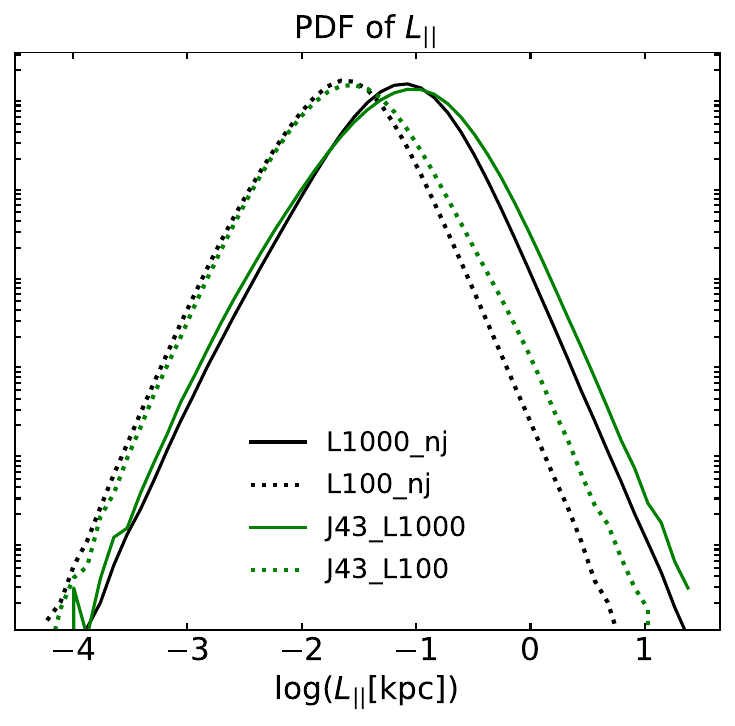} \hspace{0.15cm}
   \includegraphics[width=0.25\linewidth]{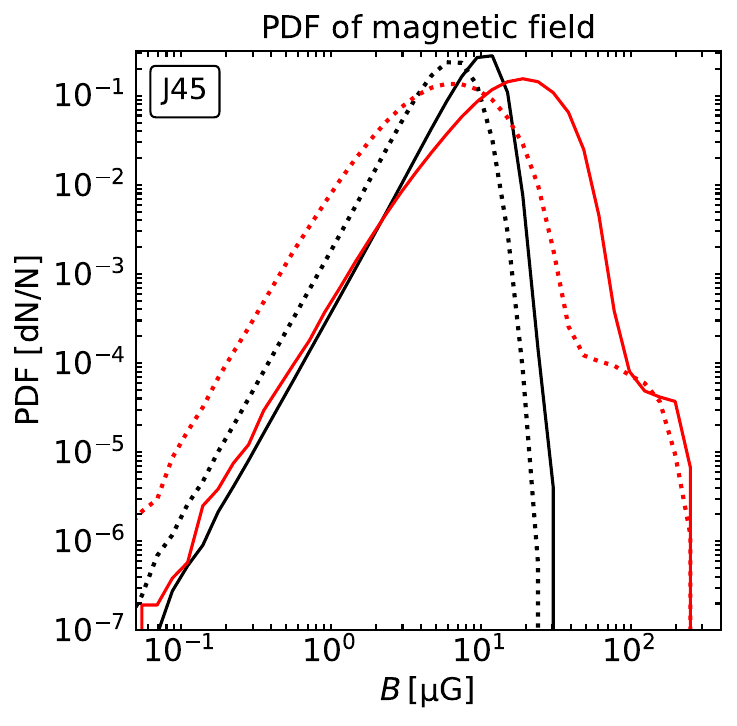}
   \hspace{-0.15cm}
    \includegraphics[width=0.25\linewidth]{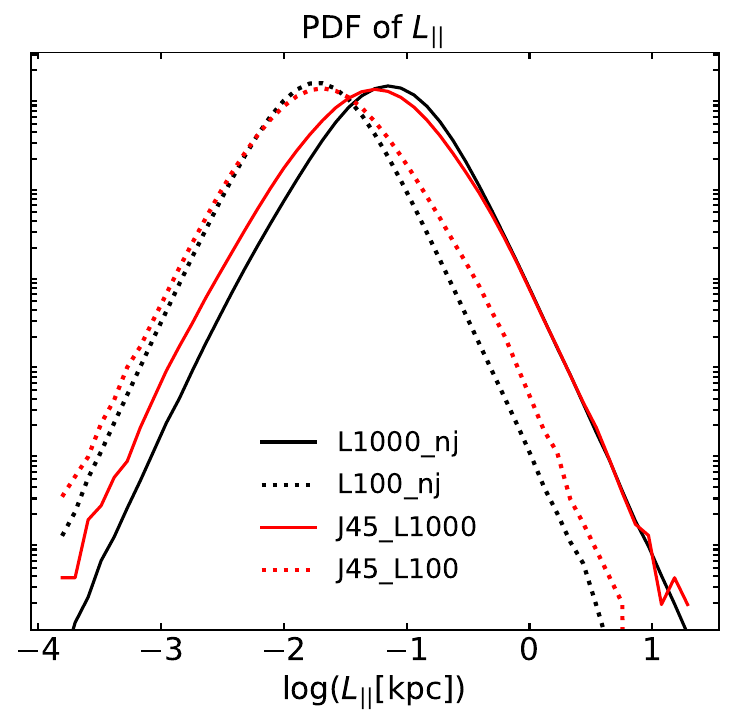}
     \end{tabular}}  
    \caption{\textbf{Left:} Probability distribution of the magnetic field and $L_{||}$ in the shocked ambient medium (SAM) for $\mathrm{J43\_L(100,1000)}$ (green) and `no-jet' $L(100,1000)\_\mathrm{nj}$ (black) at 834.6~kyr. \textbf{Right:} Same as the left panels for $\mathrm{J45\_L(100,1000)}$ (red) and `no-jet' (black) at 71.7~kyr. The forward shock of the jet amplifies as well as elongates the magnetic fields. This amplification can go up to $200~\mathrm{\mu G}$ and is enhanced with the correlation length of the magnetic field.}
     \label{fig:dist_43_45}
\end{figure*}

\begin{figure*}
     \centerline{
\def\arraystretch{1.0}
\setlength{\tabcolsep}{0.0pt}
\begin{tabular}{lcr}
    \includegraphics[width=0.3\linewidth,height=0.28\linewidth]{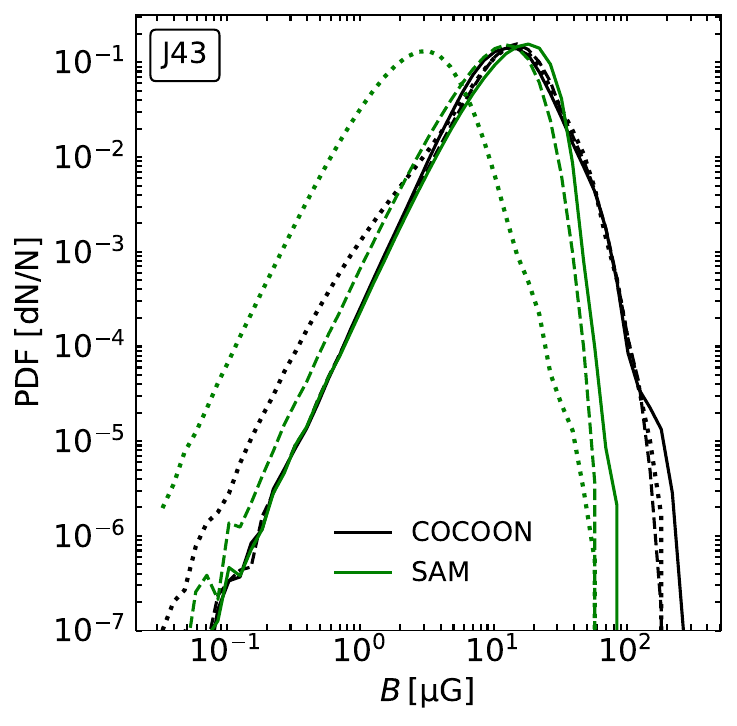} 
     \includegraphics[width=0.3\linewidth,height=0.3\linewidth]{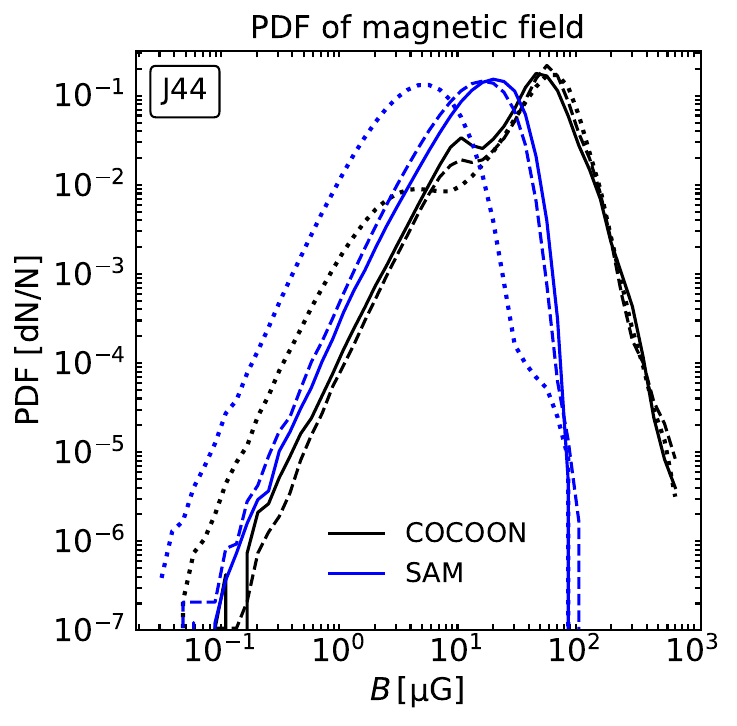}
     \includegraphics[width=0.3\linewidth,height=0.28\linewidth]{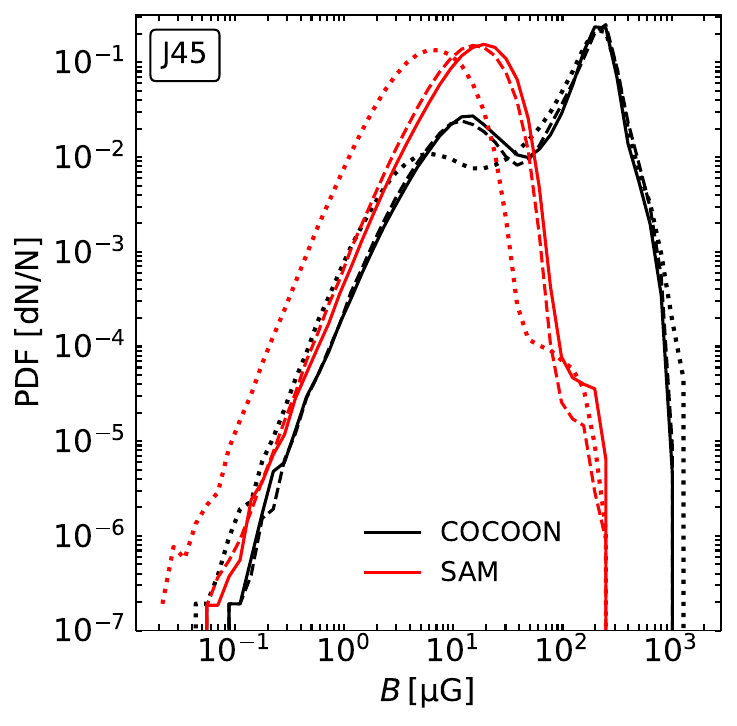}
     \end{tabular}}
\caption{\textbf{Left to right:} Distribution of magnetic field in the cocoon (tr1$>10^{-7}$) and SAM (see Eq.~\ref{eq:forward}) for $\mathrm{J43}$, $\mathrm{J44}$ and $\mathrm{J45}$ at times when the jet in each simulation is about to escape from the top of the simulation box. The different line styles denote different correlation lengths: $\mathrm{L=1000}~$pc with solid lines, $\mathrm{L=500}~$pc in dashed and $\mathrm{L=100}~$pc in dotted lines. The black curves represent corresponding results for the `no-jet' simulations, to serve as a comparison. The peaks in the PDFs of the cocoon (jet) and SAM for $\mathrm{J44}$ and $\mathrm{J45}$ are clearly separated while overlapping peaks can be seen for $\mathrm{J43}$. The fast decay of the fields in $\mathrm{L=100}$~pc cases with expansion (see Appendix~\ref{no_jet_test}) has shifted the distribution in the SAM towards the low values in all the jetted simulations.}
\label{fig:dist_Bj}
\end{figure*}

\begin{figure}
 \centerline{
\def\arraystretch{1.0}
\setlength{\tabcolsep}{0.0pt}
\begin{tabular}{lcr}
    \includegraphics[width=0.5\linewidth]{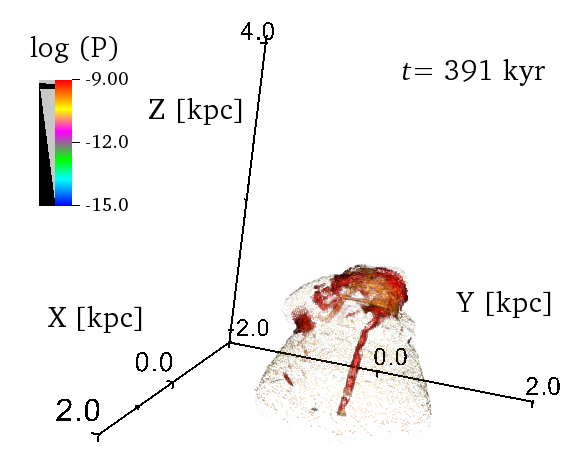} 
     \includegraphics[width=0.5\linewidth]{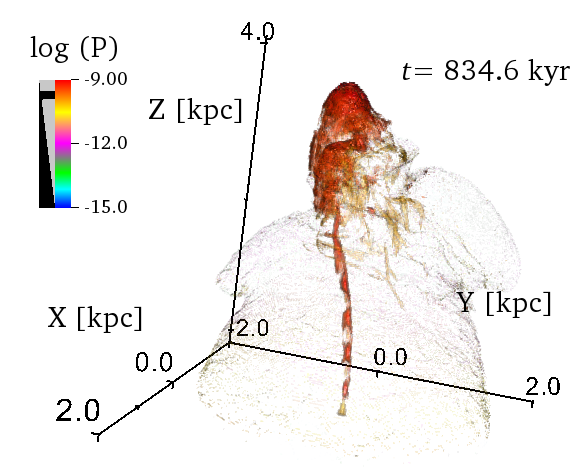} 
     \end{tabular}}    
     \caption{3D volume rendering of logarithmic pressure [$\mathrm{dyne ~cm^{-2}}$] in the cocoon for $\mathrm{J43\_L100}$ at different times. The color bar shows the logarithmic pressure, accompanied by the gray bar indicating the corresponding opacity. The gray bar is made specifically to emphasize the pressure along the spine of the jet.}\label{fig:prs_3d_43}
\end{figure}

\begin{figure*}
\centerline{
\def\arraystretch{1.0}
\setlength{\tabcolsep}{0.0pt}
\begin{tabular}{lcr}
    \includegraphics[scale=0.6,keepaspectratio]{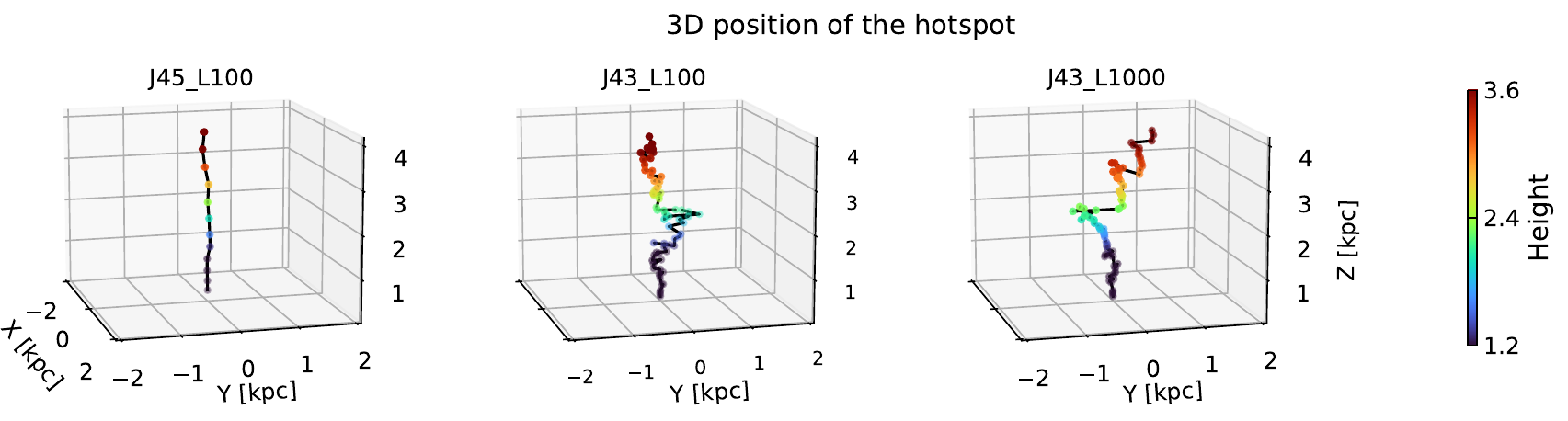}
    \end{tabular}}
     \caption{Time evolution of the 3D positions of the hotspots for the jets in different simulations. The color bar represents the $Z$ component of the hotspot's position.}
\label{fig:kinks}
\end{figure*}

\section{Results}
\label{results}
We perform a suite of simulations with jets launched in an ambient medium with a turbulent magnetic field. The aim of this work is to study the evolution of magnetized jets with varying mechanical powers as they interact with the fields in their immediate surroundings and predict the effect on the observed non-thermal synchrotron emission and polarization. In Fig.~\ref{fig:rho_prs}, the top panel shows  slices of the gas density distributions in the $Y-Z$ plane for $\mathrm{J43\_L500}$ and $\mathrm{J45\_L500}$, and the bottom panel shows the corresponding pressure distributions. The jet structure shown in the maps exhibits similarity to that from previous studies \citep[see e.g.][]{kaiser_1997,marti_1997,komissarov_1998,perucho_2014,perucho_2019}, as can also be seen from the labelled schematic diagram in Fig.~\ref{fig:cartoon}. The diagram depicts recollimation shocks along the spine of the jet, the contact discontinuity separating the jet's cocoon and SAM, and forward shock sweeping the ambient medium. Jets of varying powers can interact differently with their immediate surroundings, thus affecting the observable features, which are discussed in the sections below.

\subsection{Dynamics of the jet in a turbulent magnetic field}
\label{sec:bmag}
In this section, we discuss how jets of different mechanical powers interact with the magnetized ambient medium which is equipped with turbulent magnetic fields of different correlation lengths. Our main findings are outlined below:

\begin{itemize}
    \item \textbf{Compression of magnetic fields:} We show the $Z-$component of the magnetic field ($B_z$) in the $Y-Z$ plane for $\mathrm{J43\_L(100,500,1000)}$ and $\mathrm{J45\_L(100,500,1000)}$ in Fig.~\ref{fig:Bmag}. One can see that, as the jet progresses, the jet-driven bubble compresses the external magnetic field at the forward shock leading to the enhancement of the field in these regions. This amplification is more pronounced for large-scale fields than the small-scale ones. The magnetic field in the SAM decays as it undergoes stretching and expansion with the jet's evolution. 
    
    
    \item \textbf{PDF of fields in the SAM:} To understand how the jet-ISM interaction affects the distribution of the magnetic field in the SAM, we evaluate the probability distribution (PDF) for the magnetic field and their length scales parallel to the magnetic field $(L_{||}$) in the SAM. The SAM is identified using the expressions given in Eq.~\ref{eq:forward}. The $L_{||}$ values are calculated as \citep{schekochinin_2004, Bodo_2011,dipanjan_2020}:
\begin{equation}
    L_{||} = \left( \frac{|B|^4}{|(B\cdot\nabla)B|^2}\right)^{1/2}
\end{equation}
In Fig.~\ref{fig:dist_43_45}, we show the distribution of magnetic fields and $L_{||}$ for J43 (in green) and J45 (in red) cases. The `no-jet' simulation is represented using black curves, to serve as a control for comparison. We show results only for  $\mathrm{L=100~pc}$ and 1000~pc cases and confirm that the distribution for L~=~500~pc cases shows a similar trend as L~=~1000~pc. 

It can be seen that the compression of the magnetic fields in the SAM shifts the PDF of the fields towards the right when compared to the `no-jet' cases. The elongation of $L_{||}$ seems more prominent for the low-power jets (J43) than the high-power ones, likely because of the long time-scale interaction between the bow shock and the external magnetic fields. Due to small coherence lengths, the fields with $\mathrm{L=100~pc}$ are more stretched with the expansion of the SAM when compared to the large-scale ones.


\item \textbf{Comparison of fields in the SAM and cocoon:} In Fig.~\ref{fig:dist_Bj}, the distribution of the magnetic field in the cocoon and the SAM is shown for different simulations. The cocoon regions are selected using a lower cutoff of $10^{-7}$ for the jet tracer (tr1). We use a small value of jet-tracer to identify the contact discontinuity so that no magnetic fields from the jet are included in the SAM (see Appendix~\ref{appendix:forward_shock}), which may affect our result for depolarization, which we discuss later in Sec.~\ref{sec:SAM _depolarization}. It can be noticed in Fig.~\ref{fig:dist_Bj} that the cocoon in $\mathrm{J44}$ and $\mathrm{J45}$ carry much higher magnetic fields than the SAM, as the peaks in the former are at around $100~\mathrm{\mu G}$ and latter is at $\sim 20~\mathrm{\mu G}$. In contrast, the distributions for $\mathrm{J43}$ are closer with overlapping peaks. Such variation in the relative strengths of the magnetic fields in the jet and the SAM can have a different impact on the observed emission and polarization characteristics in different simulations, which we discuss in detail in Sec.~\ref{sec_synch_edge}.

\item \textbf{Kink instabilities in low-power jets:} Simulations have found kink modes to be dominating in jets carrying strong toroidal fields \citep[e.g.][]{mignone_2010,bodo_2013,bodo_2021,chen_2023}. In our study, kinks are observed for the low-power jets for the time scale of the simulations. This can be seen from the evolution of 3D volume rendering of logarithmic pressure for the $\mathrm{J43\_L100}$, prepared using \textsc{visit}\footnote{\url{https://wci.llnl.gov/simulation/computer-codes/visit}} visualization software, as is shown in Fig.~\ref{fig:prs_3d_43}. The wiggles and bends along the jet spine can lead to different emission features with time \citep[see][]{massaglia_2022}, and can affect the evolution of the hotspot's position with time, which we show in  Fig.~\ref{fig:kinks}. Here, the time evolution of the location of the hotspots for the high-power ($\mathrm{J45}$) and low-power jet ($\mathrm{J43}$) is presented. The hotspots are identified by averaging over the positions of the regions where pressure exceeds 0.8 times the maximum pressure in the jet cocoon ($\mathrm{tr1>10^{-7}}$). Additionally, a constraint is put to select regions where the $Z$ coordinate is greater than the height of the previous hotspot minus 0.5~kpc. This accounts for any lowering in the hotspot position due to the wiggling of the spine, which can create hotspot complexes with high pressures at regions below the maximum height of the jet beam. One can clearly see that the jet in $\mathrm{J45}$ is stable and propagates almost linearly along the $Z-$axis with only minor wiggles. Contrarily, the kinks in the $\mathrm{J43}$ jets are much more prominent (see Fig.~\ref{fig:prs_3d_43}), resulting in frequent changes in the direction of the hotspots. 

\end{itemize}

\begin{figure*}
     \centerline{
\def\arraystretch{1.0}
\setlength{\tabcolsep}{0.0pt}
\begin{tabular}{lcr}\hspace{0.29cm}
    \includegraphics[height=4.1cm,keepaspectratio]
    {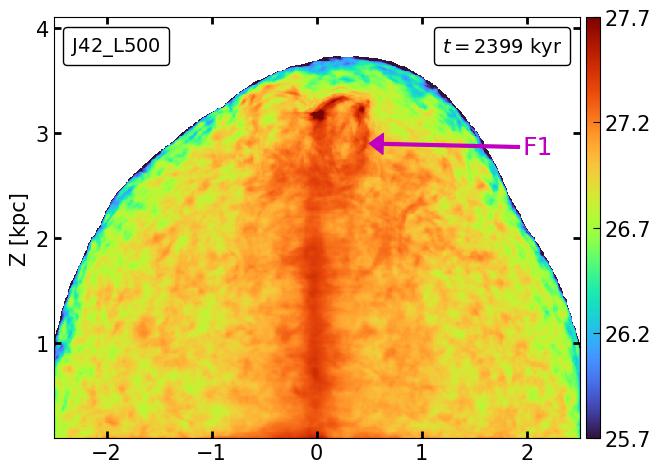}\hspace{-0.17cm}
    \includegraphics[height=4.1cm,keepaspectratio]{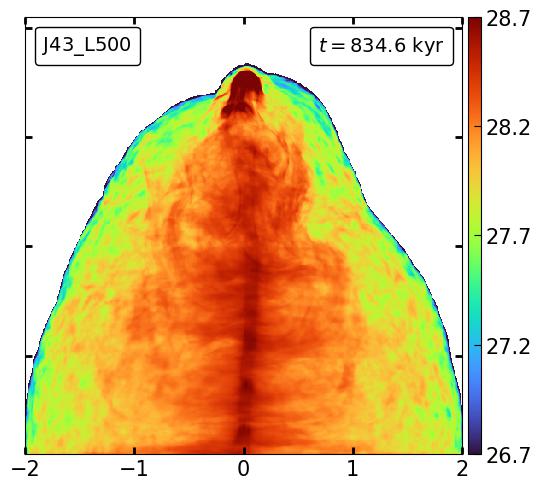} \hspace{-0.23cm}
 \includegraphics[height=4.1cm,keepaspectratio]{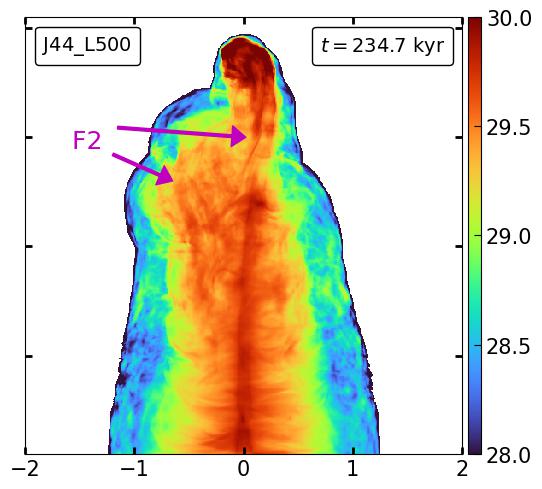} \hspace{-0.24cm}
 \includegraphics[height=4.1cm,keepaspectratio]{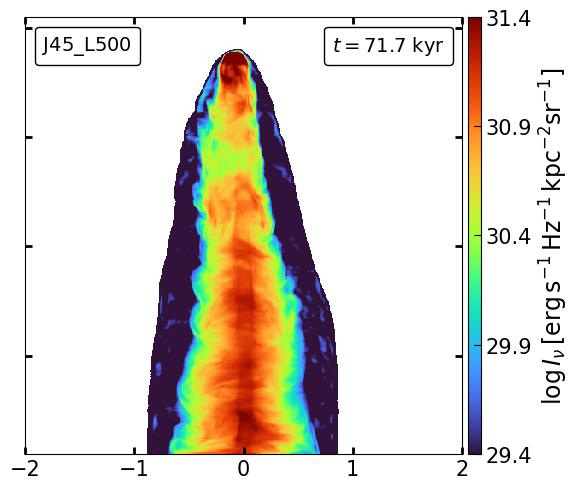} 
     \end{tabular}}
     \centerline{
\def\arraystretch{1.0}
\setlength{\tabcolsep}{0.0pt}
\begin{tabular}{lcr}
\hspace{0.22cm}
\includegraphics[height=4.1cm,keepaspectratio]{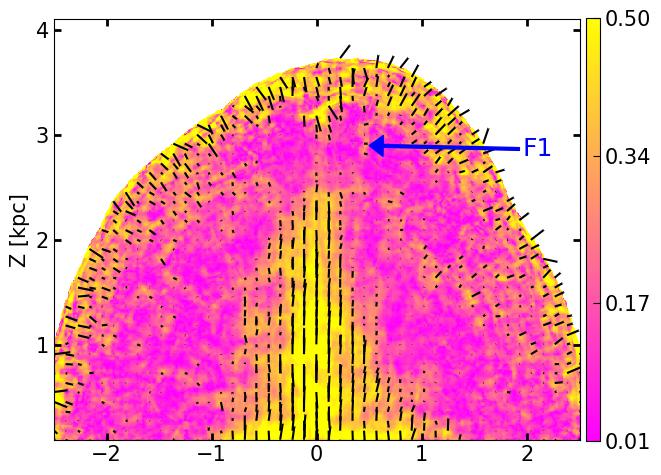} \hspace{-0.16cm}\includegraphics[height=4.1cm,keepaspectratio]{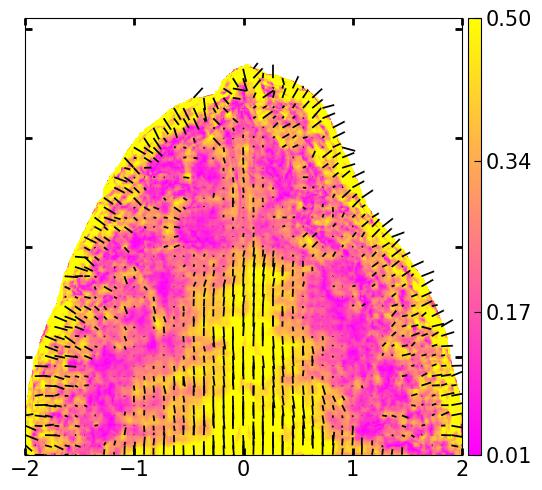} \hspace{-0.2cm}
 \includegraphics[height=4.1cm,keepaspectratio]{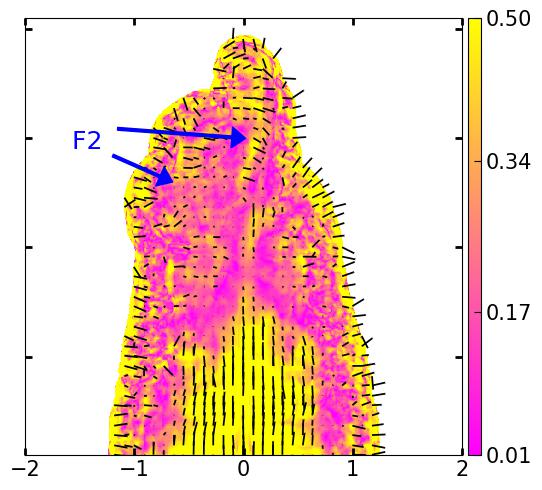}  \hspace{-0.2cm}
 \includegraphics[height=4.1cm,keepaspectratio]{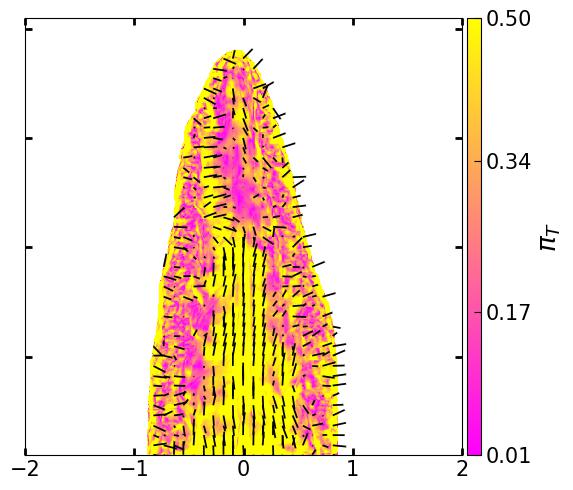} 
     \end{tabular}}
      \centerline{
\def\arraystretch{1.0}
\setlength{\tabcolsep}{0.0pt}
\begin{tabular}{lcr}
\hspace{0.22cm}
\includegraphics[height=4.1cm,keepaspectratio]{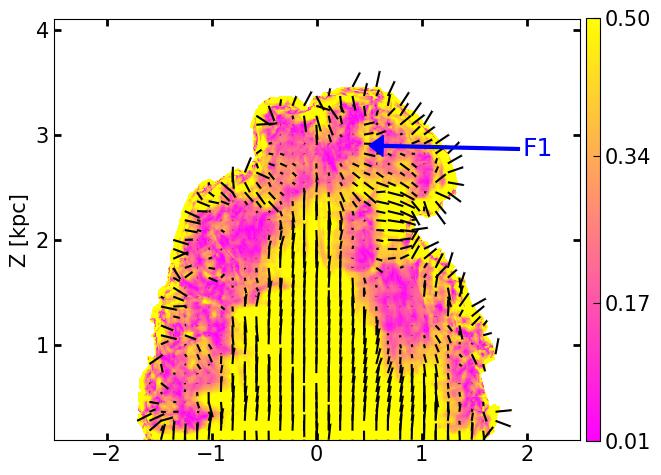} \hspace{-0.16cm}\includegraphics[height=4.1cm,keepaspectratio]{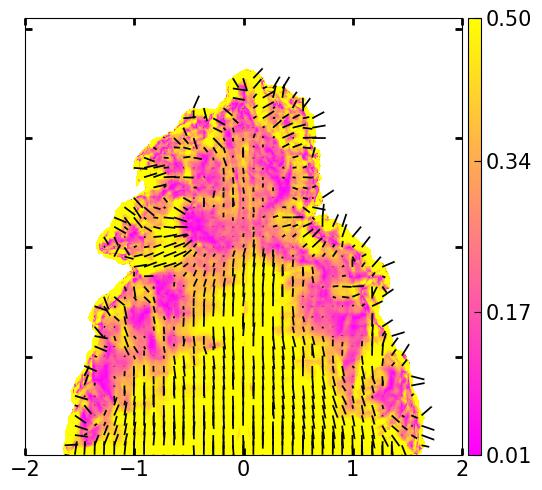} \hspace{-0.2cm}
 \includegraphics[height=4.1cm,keepaspectratio]{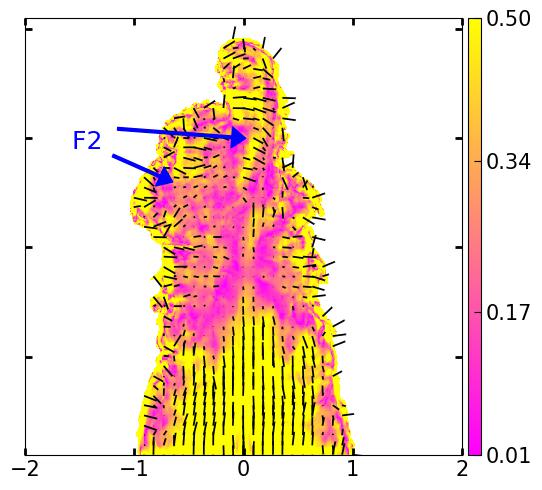}  \hspace{-0.2cm}
 \includegraphics[height=4.1cm,keepaspectratio]{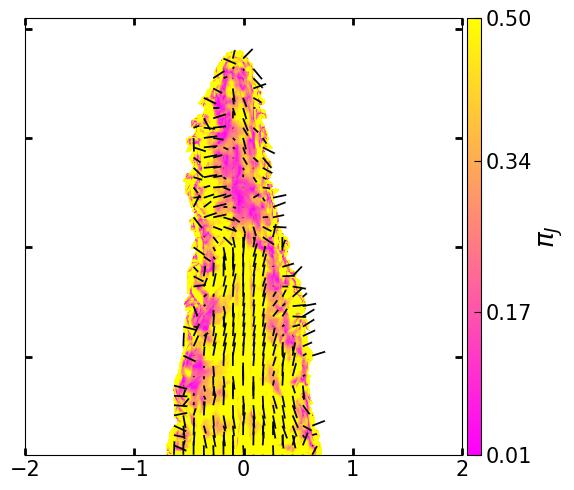} 
     \end{tabular}}
     \centerline{
\def\arraystretch{1.0}
\setlength{\tabcolsep}{0.0pt}
\begin{tabular}{lcr}
\hspace{0.37cm}
\includegraphics[height=4.2cm,keepaspectratio]{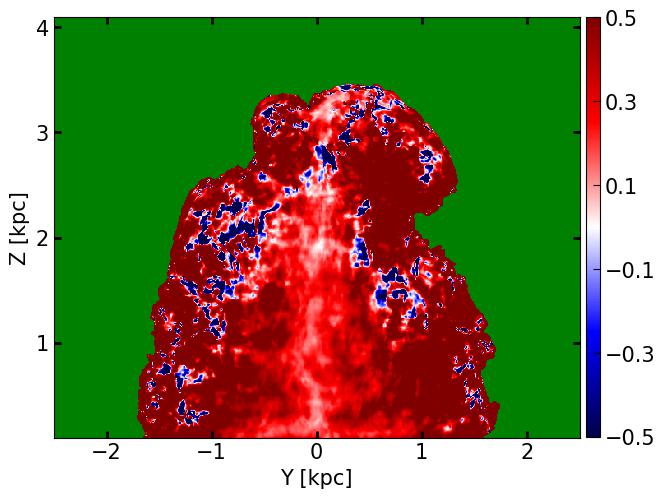}\hspace{0.01cm}
     \includegraphics[height=4.2cm,keepaspectratio]{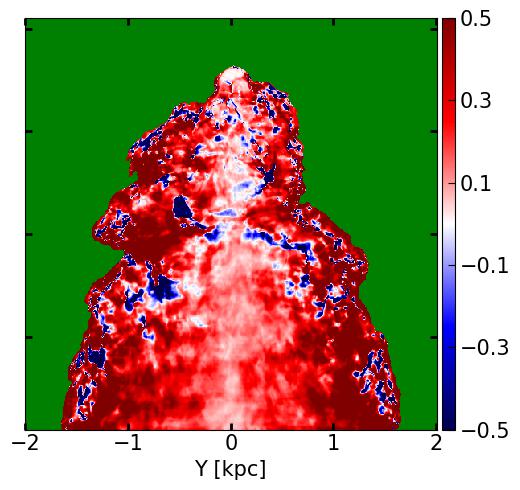} \hspace{-0.09cm}
    \includegraphics[height=4.2cm,keepaspectratio]{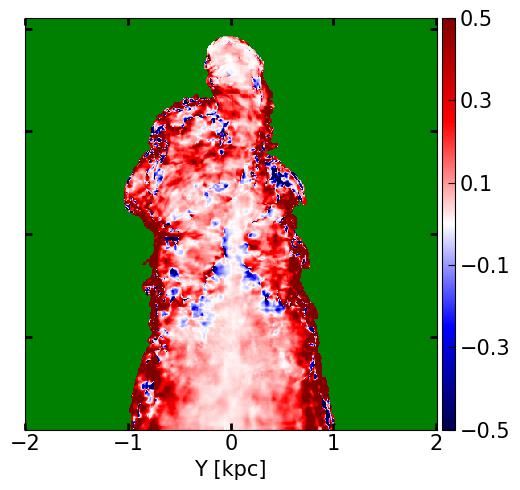} \hspace{-0.15cm}
      \includegraphics[height=4.2cm,keepaspectratio]{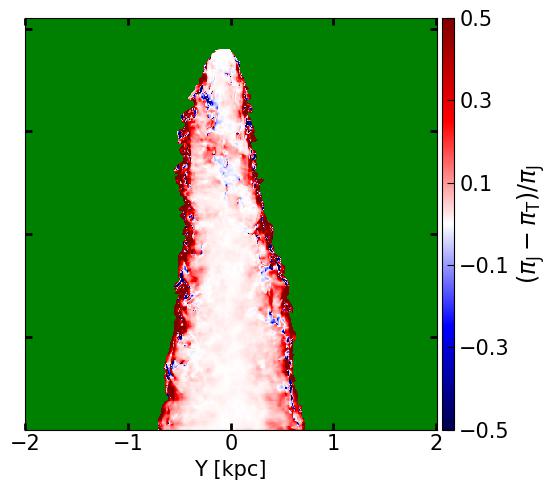}
     \end{tabular}}
     \caption{\textbf{Top:} Maps of logarithmic total synchrotron emission. \textbf{Second and third row:} Polarization fraction maps ($\pi_T$ and $\pi_J$) overlaid with polarization vectors and \textbf{Bottom:} Fractional change in polarization ($\Delta\pi$) at $\theta_I=90^\circ$ image plane in different simulations. The plots in the bottom row are shown only for regions with $\pi_J>0$ on the image plane. \textbf{F1} and \textbf{F2} depict filaments in the cocoon of $\mathrm{J42}$ and $\mathrm{J44}$, respectively.}
\label{fig:inc_42}
\end{figure*}

\begin{figure}
 \centerline{
\def\arraystretch{1.0}
\setlength{\tabcolsep}{0.0pt}
\begin{tabular}{lcr}
    \includegraphics[width=0.7\linewidth,keepaspectratio]{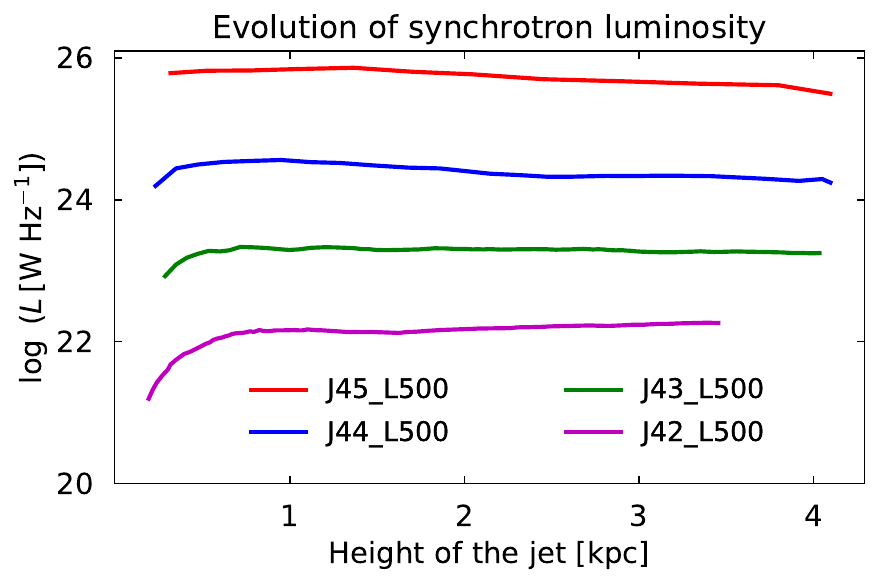}
     \end{tabular}}    
     \caption{Evolution of total synchrotron luminosity ($\mathrm{Watts~Hz^{-1}}$) as a function of the jet's height for different simulations. The values are estimated for the $\theta_I=90^\circ$ image plane in Fig.~\ref{fig:inc_42}.}
     \label{synch_lum}
\end{figure}

\begin{figure*}
     \centerline{
\def\arraystretch{1.0}
\setlength{\tabcolsep}{0.0pt}
\begin{tabular}{lcr}
    \includegraphics[height=5.7cm,keepaspectratio]{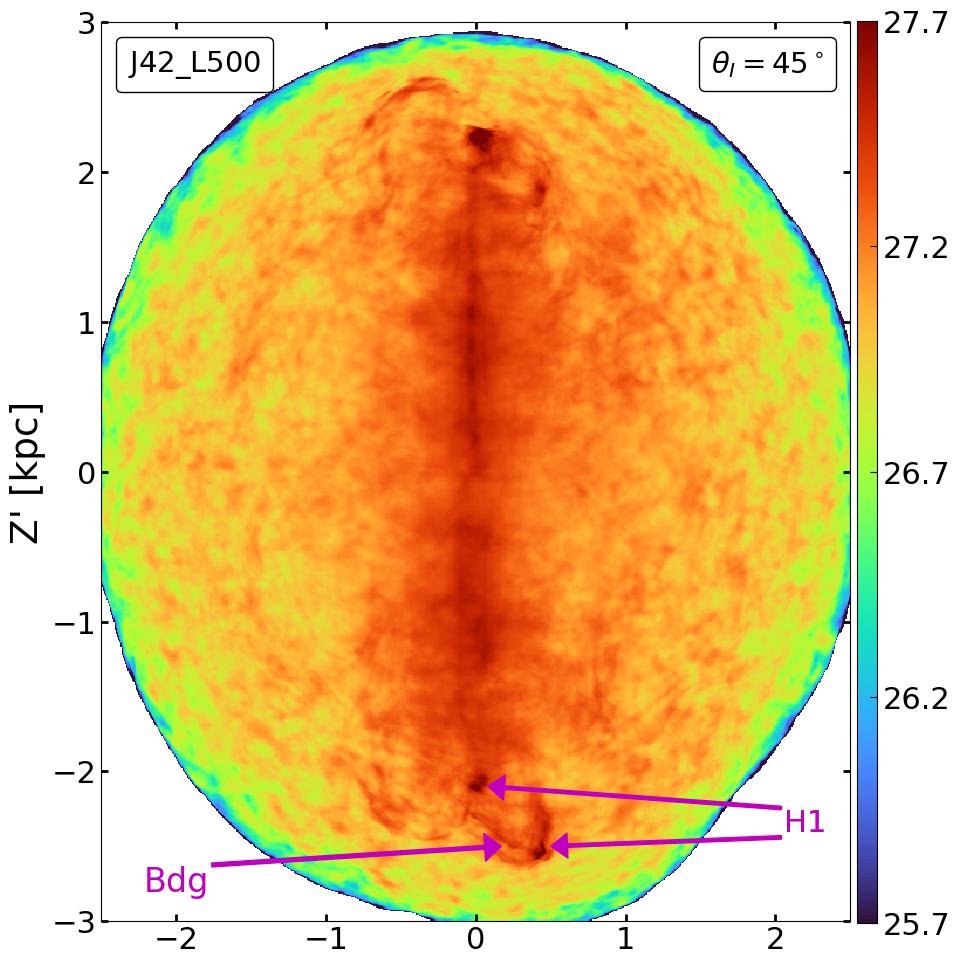} \hspace{-0.15cm}
     \includegraphics[height=5.7cm,keepaspectratio]{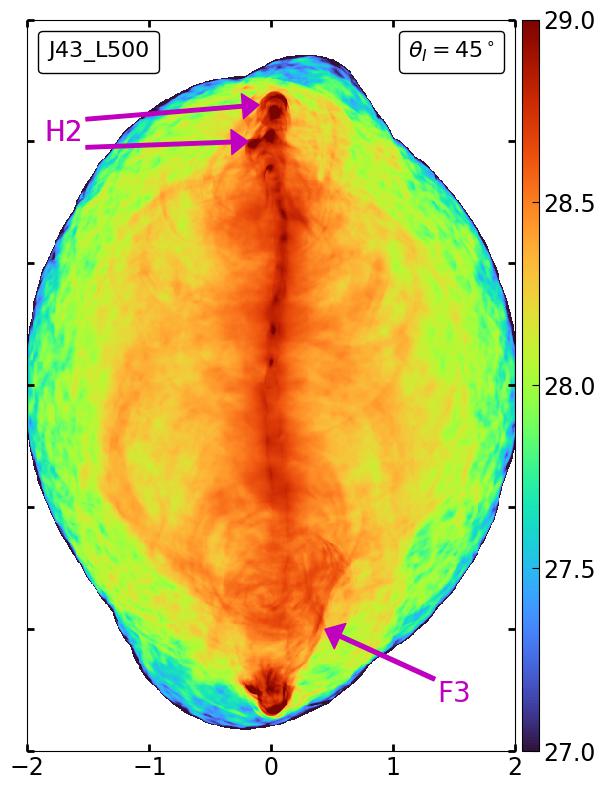}\hspace{-0.15cm}
     \includegraphics[height=5.7cm,keepaspectratio]{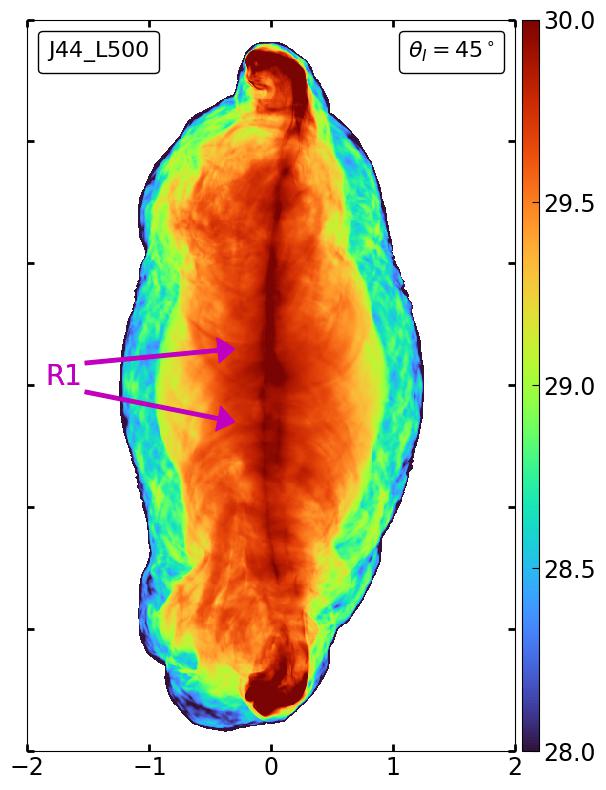} \hspace{-0.2cm}
     \includegraphics[height=5.7cm,keepaspectratio]{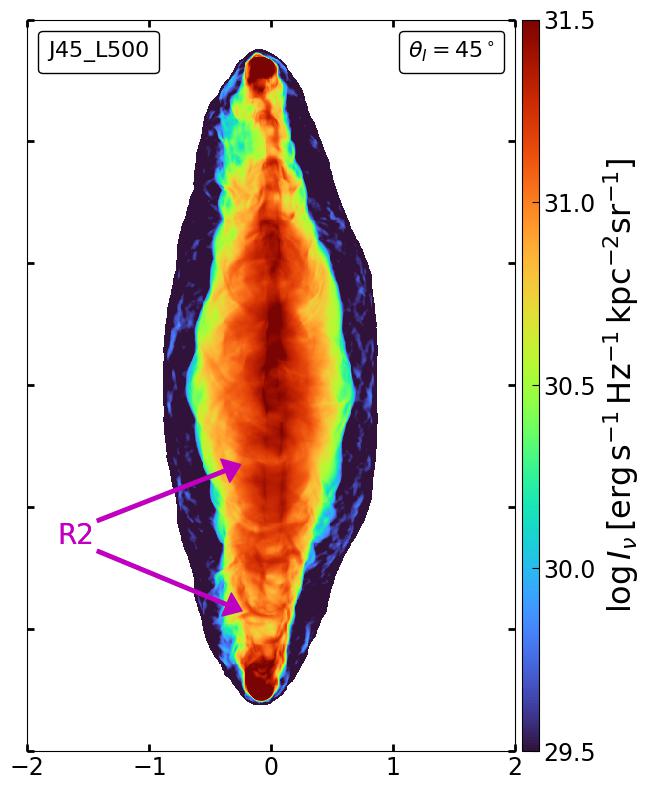}
    
     \end{tabular}}
        \centerline{
\def\arraystretch{1.0}
\setlength{\tabcolsep}{0.0pt}
\begin{tabular}{lcr}
     \hspace{0.15cm}
 \includegraphics[height=5.7cm,keepaspectratio]{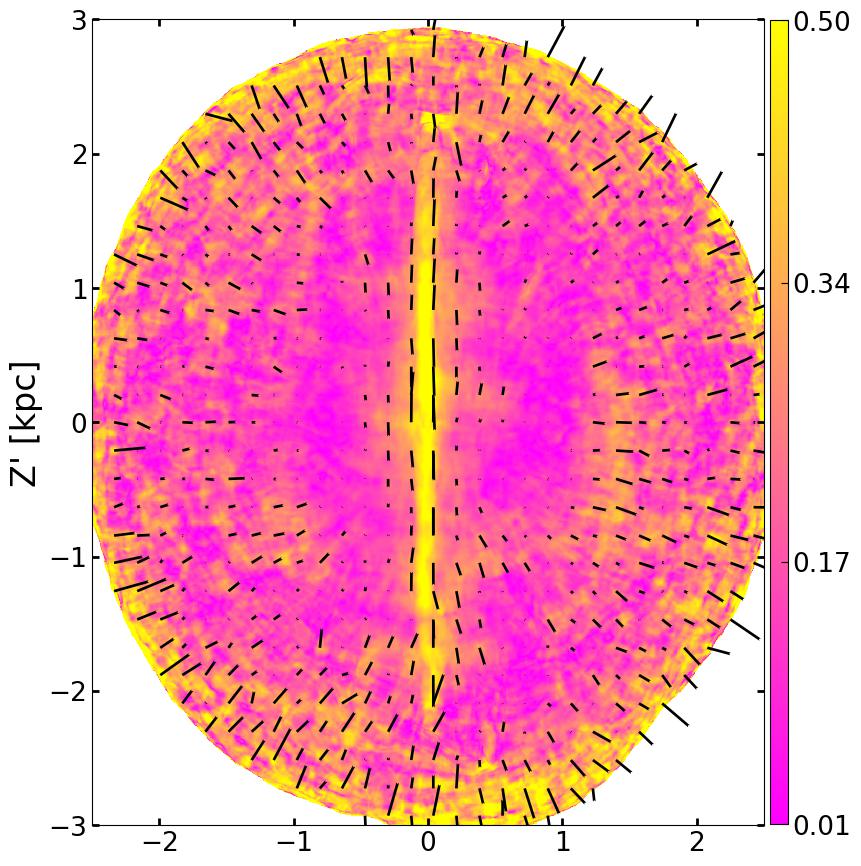}  
  \includegraphics[height=5.7cm,keepaspectratio]{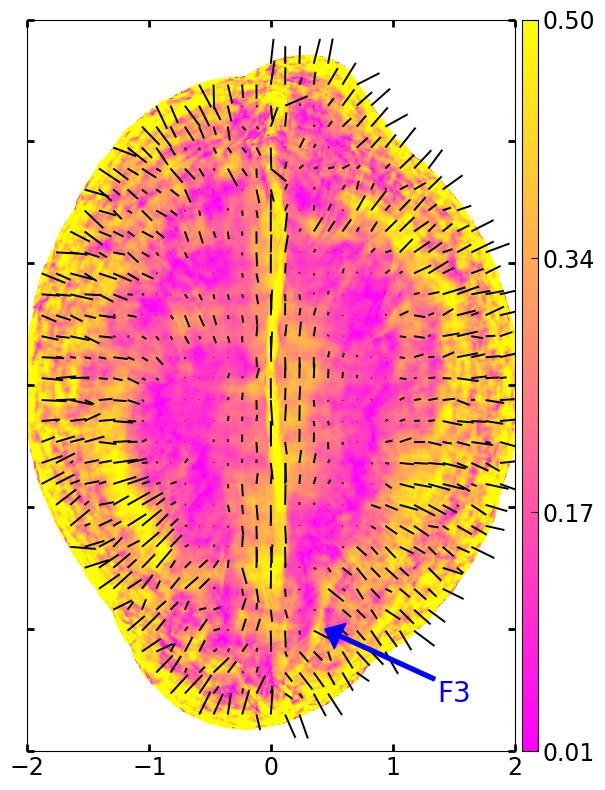} \hspace{-0.1cm}
 \includegraphics[height=5.7cm,keepaspectratio]{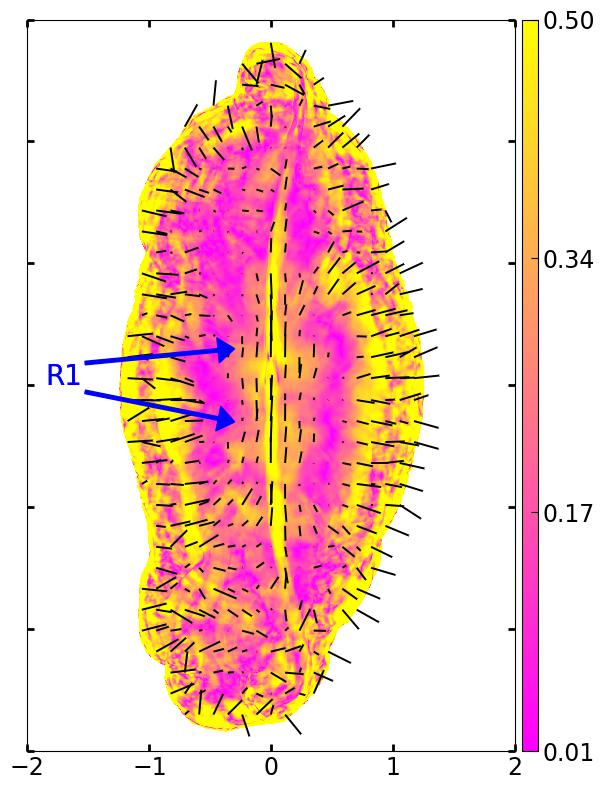} \hspace{-0.1cm}
 \includegraphics[height=5.7cm,keepaspectratio]{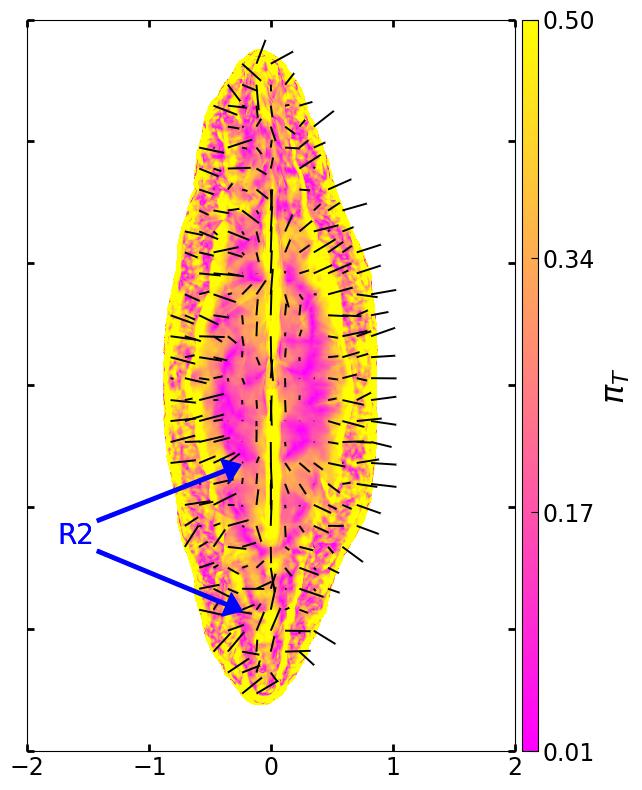} 
     \end{tabular}}
    
 \centerline{
\def\arraystretch{1.0}
\setlength{\tabcolsep}{0.0pt}
\begin{tabular}{lcr}
\hspace{0.16cm}
    \includegraphics[height=6cm,keepaspectratio]{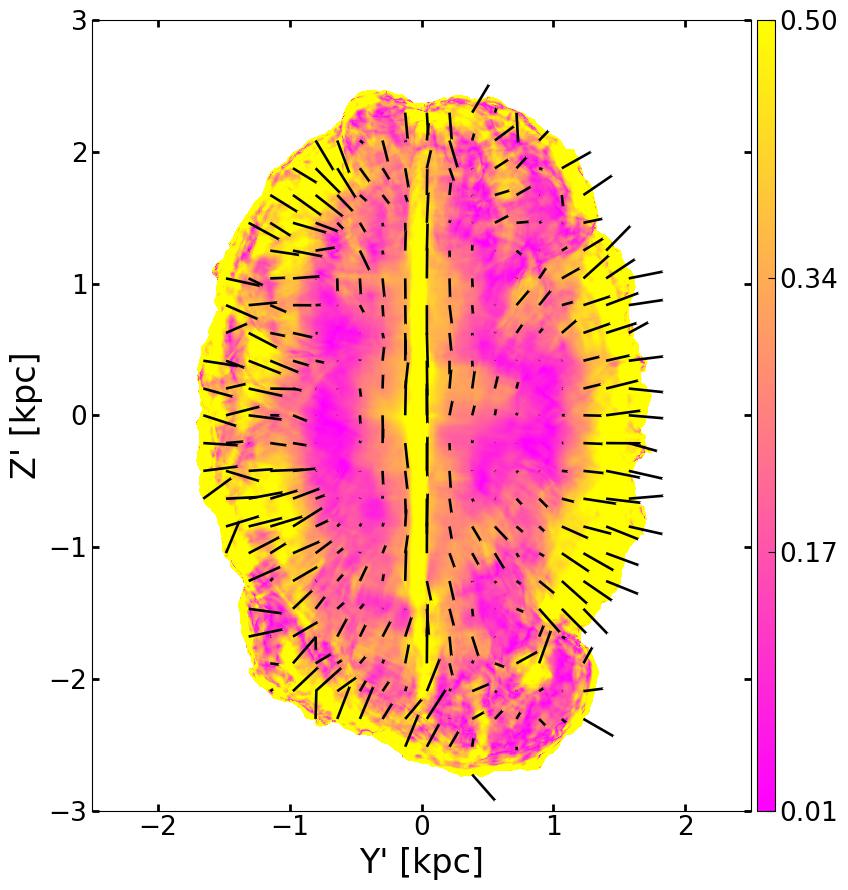}
    
   \includegraphics[height=6cm,keepaspectratio]{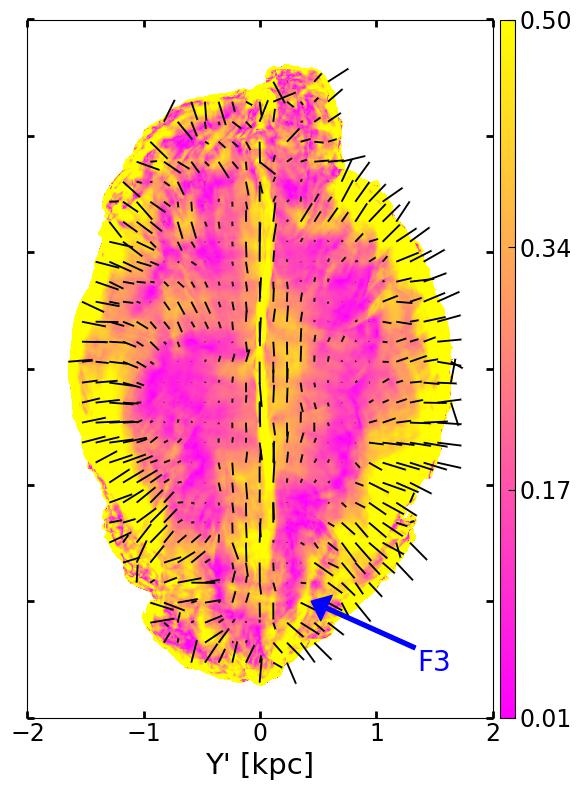}
   \hspace{-0.1cm}
    \includegraphics[height=6cm,keepaspectratio]{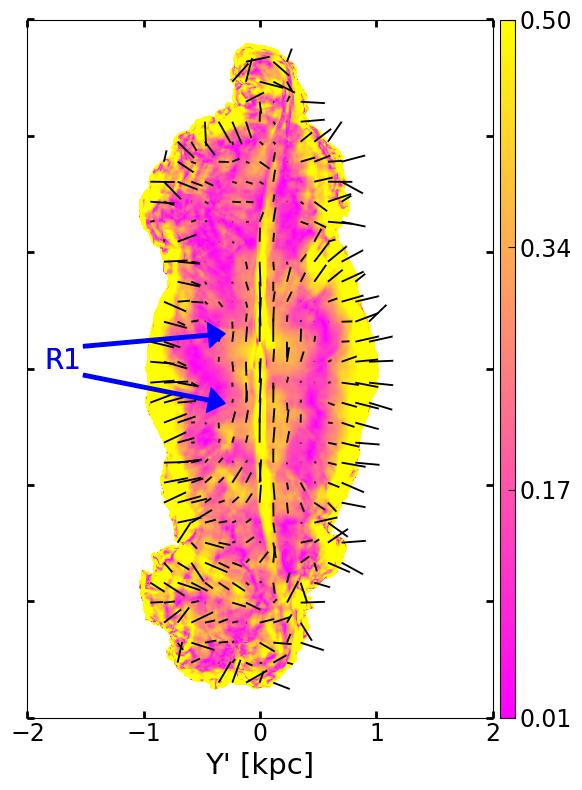} \hspace{-0.1cm}
    \includegraphics[height=6cm,keepaspectratio]{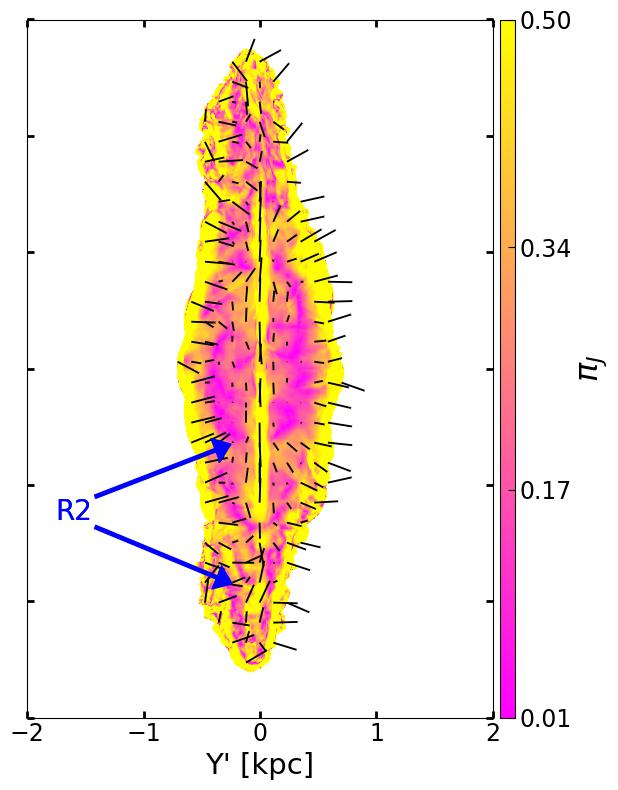} 
     \end{tabular}}
     \caption{\textbf{Top:} Logarithmic total synchrotron emission \textbf{Middle:} Polarization fraction ($\pi_T$) and \textbf{Bottom:} Polarization fraction ($\pi_J$) for different simulations at $\theta_I=45^\circ$ at the similar times as shown in Fig.~\ref{fig:inc_42}. \textbf{F3} indicates a filamentary structure inside the jet cocoon for $\mathrm{J43}$, and \textbf{H1} and \textbf{H2} show the complex hotspots in $\mathrm{J42}$ and $\mathrm{J43}$, respectively. The two hotspots in H1 are connected by a bridge-like structure (labelled as \textbf{Bdg}). \textbf{R1} and \textbf{R2} shows some ``ring-like'' features in J44 and J45, respectively.}
\label{fig:inc_43}
\end{figure*}

\begin{figure*}
\centerline{
\def\arraystretch{1.0}
\setlength{\tabcolsep}{0.0pt}
\begin{tabular}{lcr}
   \includegraphics[height=5.7cm,keepaspectratio]{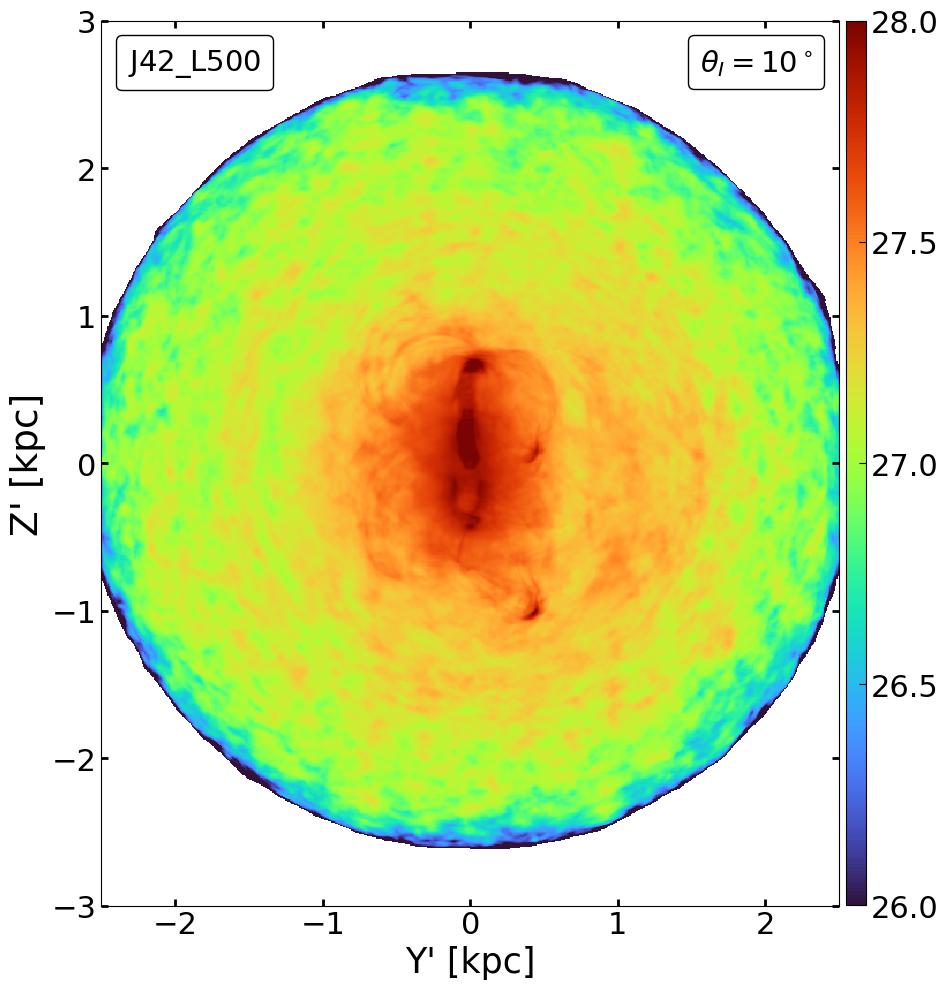} 
    \includegraphics[height=5.7cm,keepaspectratio]{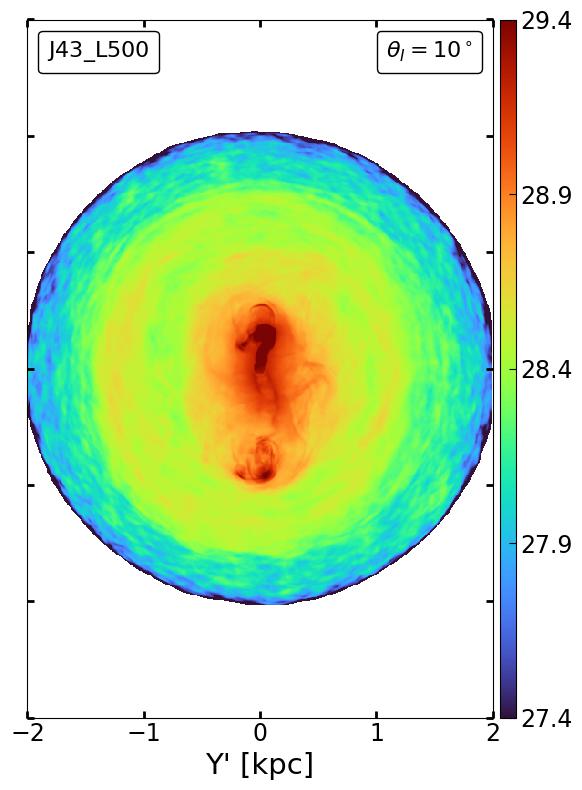}
     \includegraphics[height=5.7cm,keepaspectratio]{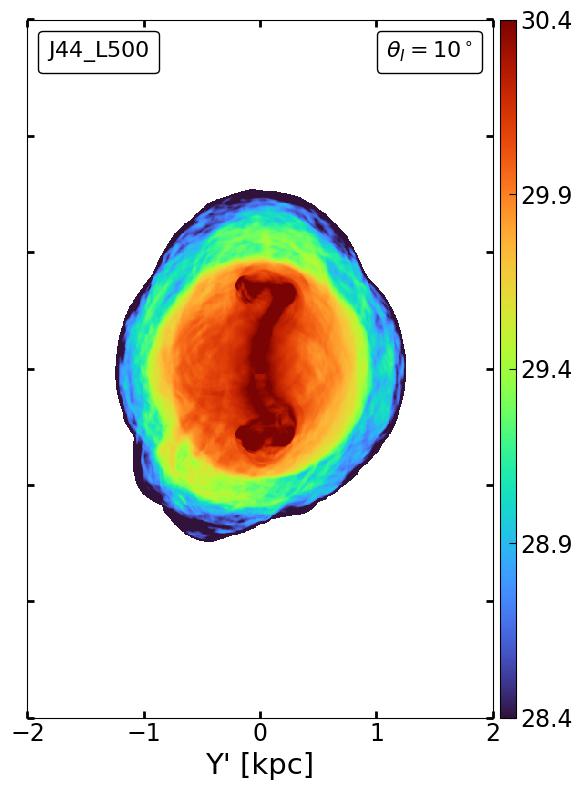} 
     \includegraphics[height=5.7cm,keepaspectratio]{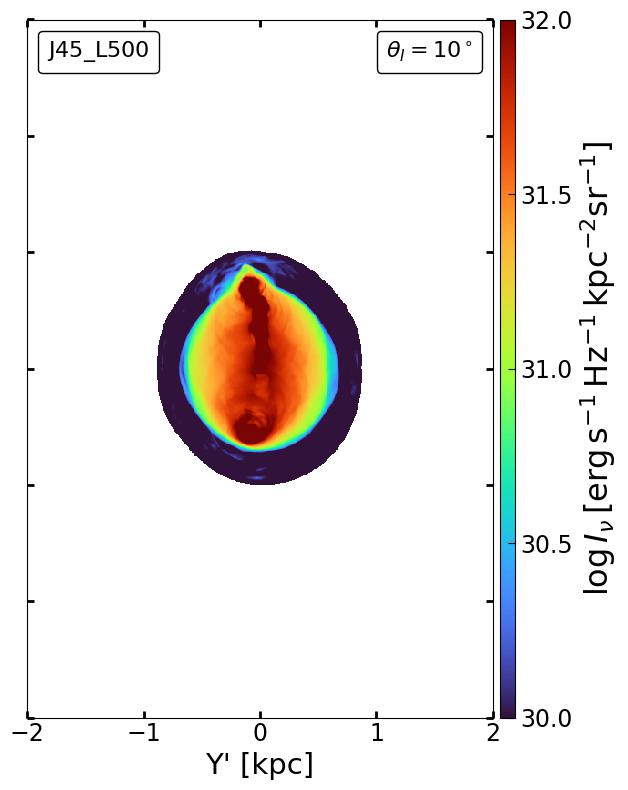}
     \end{tabular}}
\caption{\textbf{Top:} Logarithmic total synchrotron emission for different simulations at $\theta_I=10^\circ$ at the similar times as shown in Fig.~\ref{fig:inc_42}.}
\label{fig:inc_44}
\end{figure*}

\subsection{Synchrotron emission and polarization from jets}
\label{sec:synch_results}
In this section, we discuss several diagnostics to understand how the immediate surrounding magnetized medium can affect the synchrotron emission and polarization from the jet. As seen previously, the swept-up SAM carries amplified magnetic fields (see Fig.~\ref{fig:Bmag}). These regions can contribute to the total observed synchrotron emission when the electrons get accelerated at these sites. Moreover, such compressed fields from the external medium around the radio lobes can cause external depolarization \citep{burn_1966}. The different observable features can vary depending on the viewing angle of the observer with respect to the jet, and thus we also perform an analysis by changing the orientation of the observer with respect to the jet. The LOS of an observer is defined using the angles $\theta_I$ and $\phi_I$, which are calculated counter-clockwise from the positive $Z$-axis and $X-Z$ plane, respectively. We fix the azimuthal angle $\phi_I$ at $0^\circ$, such that the observer is facing towards the positive $X$-axis and is viewing the $Y-Z$ plane in the conventions of the simulation's grid. The angle $\theta_I$ is varied to define the orientation of different image planes.

In Fig.~\ref{fig:inc_42} we show the images for the synchrotron flux, and polarization fraction maps for $\mathrm{J42},~\mathrm{J43}$, $\mathrm{J44}$ and $\mathrm{J45}$ for L~=~500~pc cases at $\theta_I = 90^\circ$ image plane. All these snapshots correspond to the times when the jet's head reaches close to the top of the simulation box. The first row shows the total synchrotron flux emitted from the regions inside the forward shock. The total synchrotron luminosity values in our simulations span a range of $10^{22} -10^{26}~\mathrm{Watts~Hz^{-1}}$, as can be seen from Fig.~\ref{synch_lum}. These values are consistent with observed radio luminosity estimates in numerous compact sources \citep[see e.g.][and references there in]{odea_2021}. Furthermore, using the luminosity-jet power relations listed in \citet{kondapally_2023} , we confirm that the total synchrotron luminosity values are in agreement with the injected jet power within a span of one order of magnitude. The polarization maps are shown from the cocoon only and cocoon + SAM regions in the second and third row, which we call $\pi_J$ and $\pi_T$, respectively. The former includes the contribution from the cocoon or regions inside the contact discontinuity ($\mathrm{tr1} > 10^{-7}$) and the latter includes all the regions inside the forward shock (see Appendix.~\ref{appendix:forward_shock}). The fourth row shows the fractional change in polarization ($\Delta\pi$), which is defined as $\Delta\pi = (\pi_J-\pi_T)/\pi_J$. A positive value of $\Delta\pi$ indicates that the cocoon leads to high-polarization values ($\pi_J$), which is reduced after including the contribution from the SAM ($\pi_T < \pi_J$). Moreover, $\Delta\pi=1$ implies complete cancellation of polarization from the jet's cocoon by the SAM (i.e. $\pi_T=0$), while $\Delta\pi =0$ indicates that the contribution from the SAM is negligible ($\pi_T = \pi_J$).

Fig.~\ref{fig:inc_43},~\ref{fig:inc_44},~\ref{fig:diff_len_43},~\ref{fig:P43_L1000} in the paper, and Fig.~\ref{fig:diff_len_44} and \ref{fig:diff_len_45} in Appendix~\ref{app:figures} display emission and polarization maps for different magnetic field correlation lengths and orientations of the observer. In Fig.~\ref{fig:inc_43} and~\ref{fig:inc_44}, we present maps for distinct jet runs at image planes of $\theta_I=45^\circ$ and $10^\circ$, respectively. Fig.~\ref{fig:diff_len_43} exhibits synchrotron flux, polarization maps, and fractional polarization change for $\mathrm{J43\_L100}$ and $\mathrm{J43\_L1000}$ at $\theta_I=90^\circ$ and at $\theta_I=45^\circ$ in Fig.~\ref{fig:P43_L1000}. The emission and polarization maps for $\mathrm{J44\_L(100,1000)}$ and $\mathrm{J45\_L(100,1000)}$ at $\theta_I=90^\circ$ are shown in Fig.~\ref{fig:diff_len_44} and \ref{fig:diff_len_45} (refer to Appendix~\ref{app:figures}). For $\theta_I=90^\circ$, we show the maps for the upper half plane only, whereas for the low-inclinations we reflect the quantities in the lower-half plane to map the final observed values on the image plane.
Additionally, in Fig.~\ref{synch_lum}, we illustrate the evolution of the total synchrotron luminosity as a function of the jet's height for different simulations. Below, we elucidate the inferences drawn from these figures.

\subsubsection{Morphology of the emission}
\label{sec_synch_edge}

\begin{figure*}
         \centerline{
\def\arraystretch{1.0}
\setlength{\tabcolsep}{0.0pt}
\begin{tabular}{lcr}
    \includegraphics[height=4.5cm,keepaspectratio]{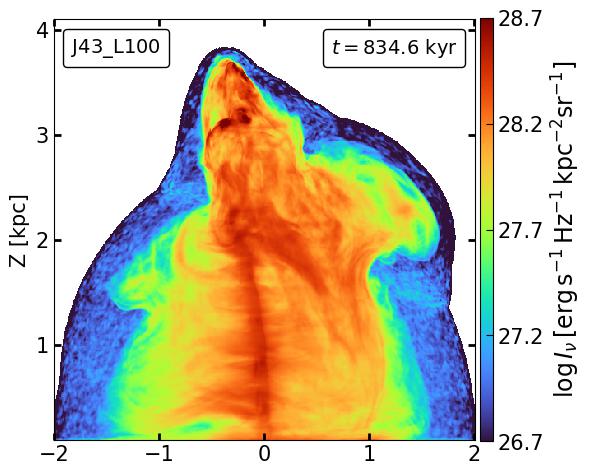} 
    \includegraphics[height=4.5cm,keepaspectratio]{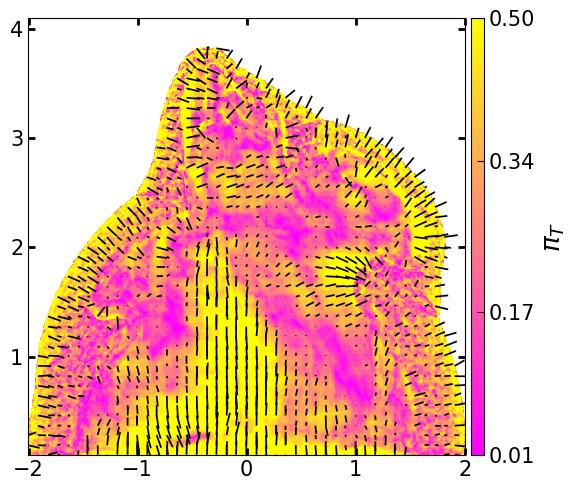} 
    \includegraphics[height=4.5cm,keepaspectratio]{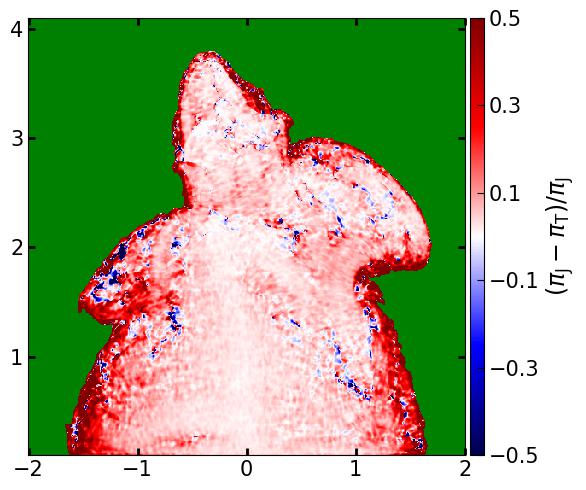} 
     \end{tabular}}
    \centerline{
\def\arraystretch{1.0}
\setlength{\tabcolsep}{0.0pt}
\begin{tabular}{lcr}
      \includegraphics[height=4.65cm,keepaspectratio]{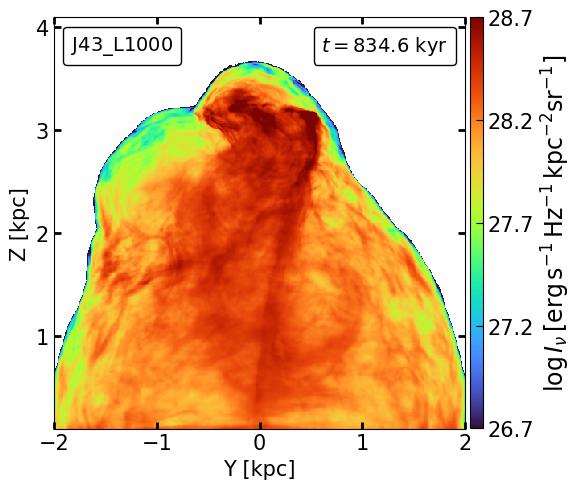} 
    \includegraphics[height=4.65cm,keepaspectratio]{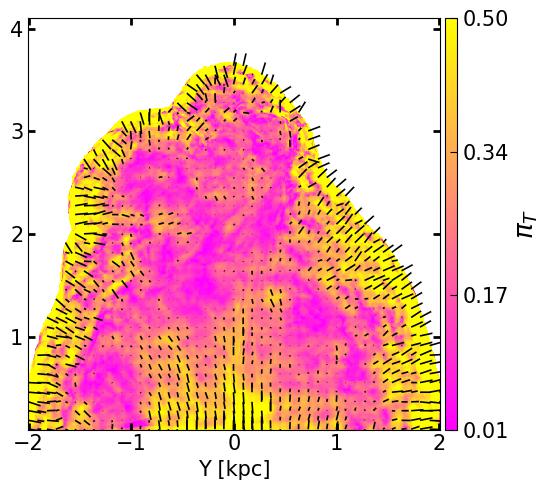} 
   \includegraphics[height=4.65cm,keepaspectratio]{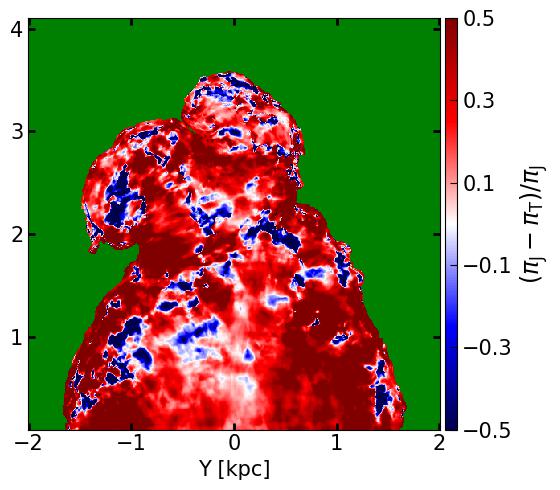} 
     \end{tabular}}
      \caption{\textbf{Top:} Logarithmic total synchrotron emission, polarization fraction ($\pi_T$) and fractional change in polarization ($\Delta\pi$) for $\mathrm{J43\_L100}$ at $\theta_I=90^\circ$ image plane. \textbf{Bottom:} Same as above for $\mathrm{J43\_L1000}$.}
\label{fig:diff_len_43}
\end{figure*}

\begin{figure*}
 \centerline{
\def\arraystretch{1.0}
\setlength{\tabcolsep}{0.0pt}
\begin{tabular}{lcr}
      \includegraphics[height=5.7cm,keepaspectratio]{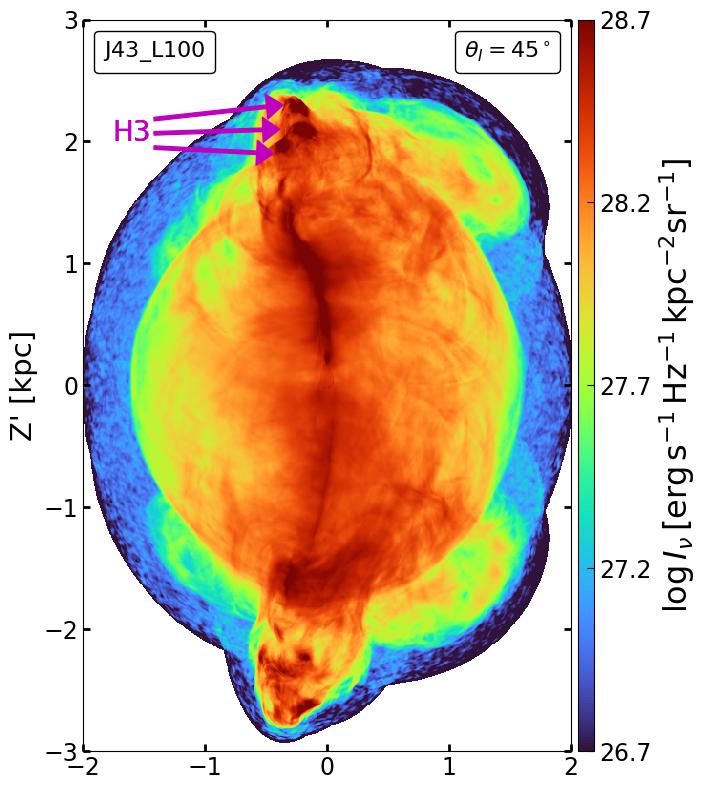} 
    \includegraphics[height=5.7cm,keepaspectratio]{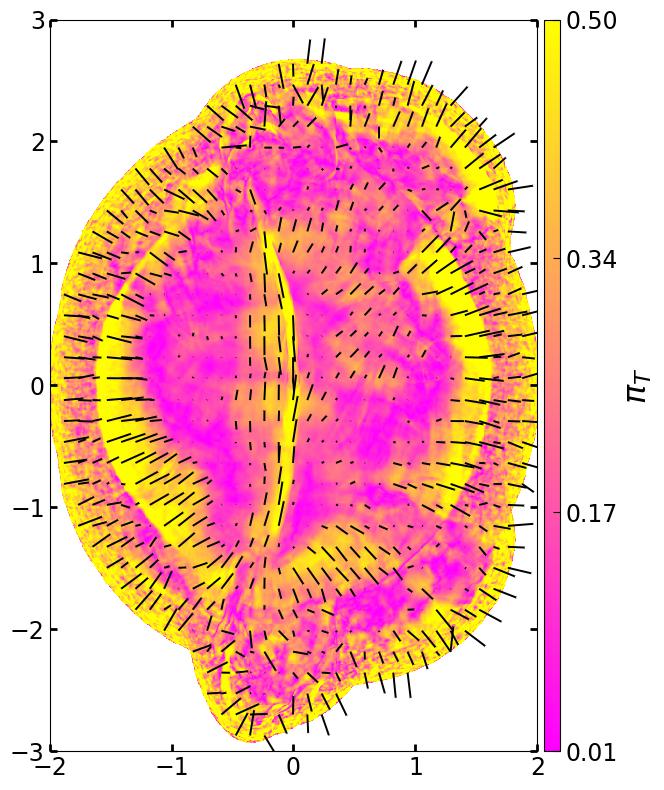} 
    \includegraphics[height=5.7cm,keepaspectratio]{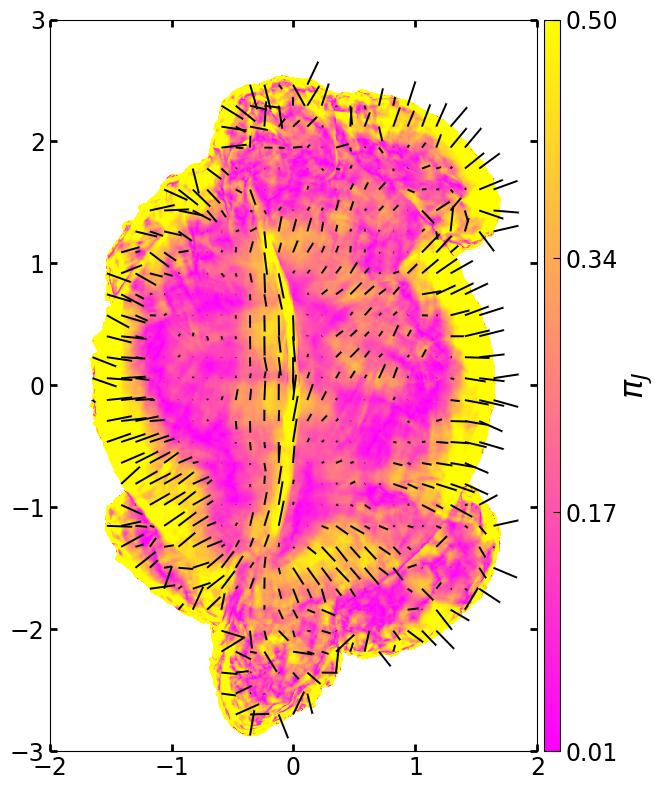} 
     \end{tabular}}
      \centerline{
\def\arraystretch{1.0}
\setlength{\tabcolsep}{0.0pt}
\begin{tabular}{lcr}
      \includegraphics[height=5.7cm,keepaspectratio]{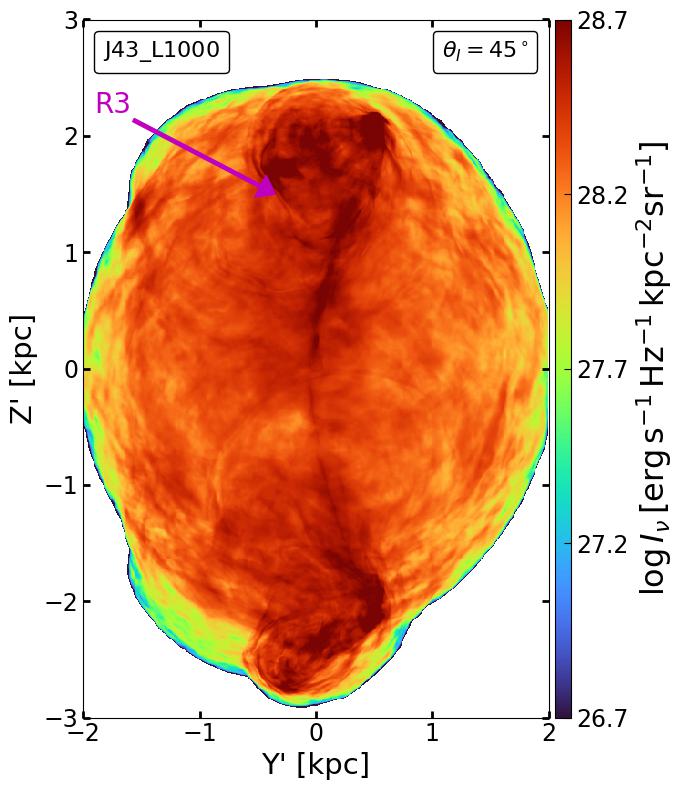} 
    \includegraphics[height=5.7cm,keepaspectratio]{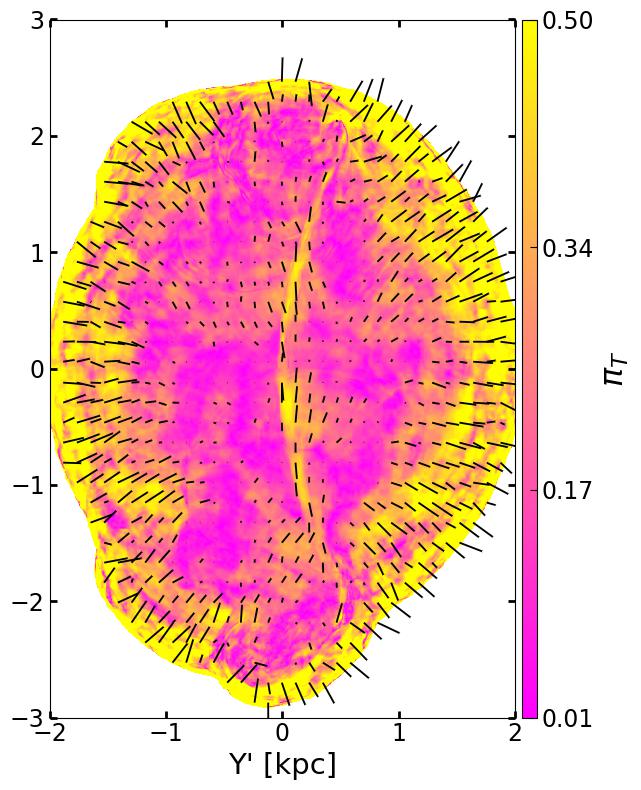} 
    \includegraphics[height=5.7cm,keepaspectratio]{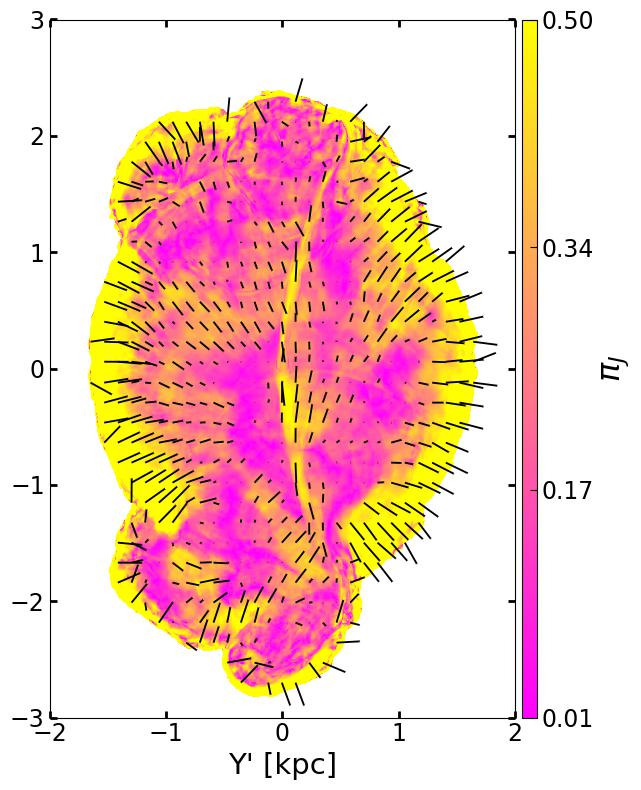} 
     \end{tabular}}
    \caption{\textbf{Top:} Logarithmic total synchrotron emission, polarization fraction ($\pi_T$ and $\pi_J$) for $\mathrm{J43\_L100}$ at $\theta_I=45^\circ$ at similar times as Fig.~\ref{fig:diff_len_43}. \textbf{Bottom:} Same as above for $\mathrm{J43\_L1000}$. \textbf{H3} shows hotspot complexes in $\mathrm{J43}$ with three hotspots, and \textbf{R3} shows arc-like features.}
\label{fig:P43_L1000}
\end{figure*}

\begin{itemize}
    \item \textbf{Dependence on the viewing angle:} At $\theta_I=45^\circ$ and~$10^\circ$ in Fig.~\ref{fig:inc_43} and~\ref{fig:inc_44}, respectively, the morphology of emission is different than at $\theta_I=90^\circ$ in Fig.~\ref{fig:inc_42}. Higher flux is observed from the approaching jet than the receding one due to Doppler boosting; although the highest emission is obtained from the hotspots at both of the jets' heads. Sources with jets inclined at small angles ($\lesssim 20^\circ$) to the observer are also not so uncommon \citep{hovatta_2009}. At a small angle of $\theta_I=10^\circ$ in Fig.~\ref{fig:inc_44}, the emission from the cocoon is much brighter than the higher inclination angles. This gives rise to an almost circular emission morphology in all cases. For high-power jets, the narrow cocoon appears as a compact circular emitting region surrounded by a ring-shaped glowing SAM. 
    
 
  \item \textbf{Dependence on the correlation length of the external magnetic field:} For different correlation lengths in Fig.~\ref{fig:diff_len_43} and~\ref{fig:P43_L1000}, one can see that the emission is lower in the SAM in $\mathrm{J43\_L100}$ than $\mathrm{J43\_L1000}$. Such lowering in the flux can also be seen in the emission maps in Fig.~\ref{fig:diff_len_44} and~\ref{fig:diff_len_45} in Appendix~\ref{app:figures} for $\mathrm{J44\_L100}$ and $\mathrm{J45\_L100}$, respectively. This happens as the compressed external magnetic fields with small coherence lengths decay fast in the SAM (see Appendix.~\ref{no_jet_test}). Contrarily for $\mathrm{J43\_L1000}$, the compressed regions in the SAM attain high-magnetic field values due to strong compression and slow decay, giving rise to comparatively high synchrotron emission. Such enhanced emission from the SAM also leads to a prominent effect on the depolarization of the radiation from the jet's cocoon, as discussed later in Sec.~\ref{sec:SAM _depolarization}. For $\mathrm{J44\_L1000}$ and $\mathrm{J45\_L1000}$, the synchrotron flux in the SAM is lower than that from the jet cocoon because these high-power jets, at the early stages of evolution, carry stronger magnetic fields when compared to the SAM, as inferred from PDFs in Fig.~\ref{fig:dist_Bj}.
    
    \item \textbf{Kinks and multiple hotspots:} It is observed that the emission morphology for the low-power jets is broader when compared to the high-power jet in $\mathrm{J45}$ (see Fig.~\ref{fig:inc_42} and~\ref{fig:inc_43}). The jet plasma in $\mathrm{J42}$ and $\mathrm{J43}$ (see also Fig.~\ref{fig:diff_len_43} and~\ref{fig:P43_L1000}) spreads widely, shaping a broader turbulent cocoon. This happens because these jets suffer from kink instabilities \citep{mignone_2010, bromberg_2016}, leading to slow propagation of the jet head and a large expansion of the cocoon radius. Furthermore, the emission maps indicate kinks and wiggling motions along the spine of the jet, as inferred from Fig.~\ref{fig:diff_len_43} and~\ref{fig:P43_L1000}. Such features can also be seen in the emission maps from previous studies of kink modes in jets \citep{massaglia_2022}. On the contrary, the kink modes have minimal or insignificant influence on the high-power jets \citep{dipanjan_2020}. The slow growth rate of kink modes in highly relativistic jets keeps them stable and collimated for a long duration \citep{bodo_2013}. Therefore, the J45 jet in Fig.~\ref{fig:inc_42} exhibits a hotspot at its head, resembling an FRII-type source (3C~223 in \citet{massaglia_2022}). On the other hand, the other jets show ``bends'' or ``hook-like'' features at the head, similar to what is found for 3C~200, 3C~098, and 3C~334 in their study.
     
    Multiple hotspots for the low-power jets ($\mathrm{J42}$ and $\mathrm{J43}$) can also be seen in Fig.~\ref{fig:inc_43}, which are caused by wiggling motions along the jet's spine, resulting in visible `hotspot complexes' (see H1 and H2). The bridge (see region labelled `Bdg') connecting the hotspots in $\mathrm{J42}$'s receding jet shows that the secondary hotspot is fueled by the primary one. Similar multiple hotspots can also be seen for $\mathrm{J43\_L100}$ in Fig.~\ref{fig:P43_L1000} (see H3).

    \item \textbf{Emission structures inside the cocoon:} Some high emission filaments can also be seen in the cocoon, particularly near the hotspot; for example, see F1, F2, and F3 in Fig.~\ref{fig:inc_42} and~\ref{fig:inc_43}, respectively. Filamentary structures have been observed in the radio lobes of several FRII sources, such as Cygnus A \citep{perley_1984}, 3C~34 and 3C~320 \citep{mahatma_2023}. Also, one can notice distinct ``ring or arc-like'' features in the emission maps; e.g. see R1 and R2 in J44 and J45 in Fig.~\ref{fig:inc_43}. These rings are likely a result of annular shocks caused by the backflows in the cocoon \citep{saxton_2002}. We also notice these rings in the emission maps from the cocoon in low-power jets (see R3 in Fig.~\ref{fig:P43_L1000} for $\mathrm{J43\_L1000}$); however, their prominence is diminished due to the contribution from the SAM.
\end{itemize}

\begin{figure}
 \centerline{
\def\arraystretch{1.0}
\setlength{\tabcolsep}{0.0pt}
\begin{tabular}{lcr}
    \includegraphics[width=0.99\linewidth]{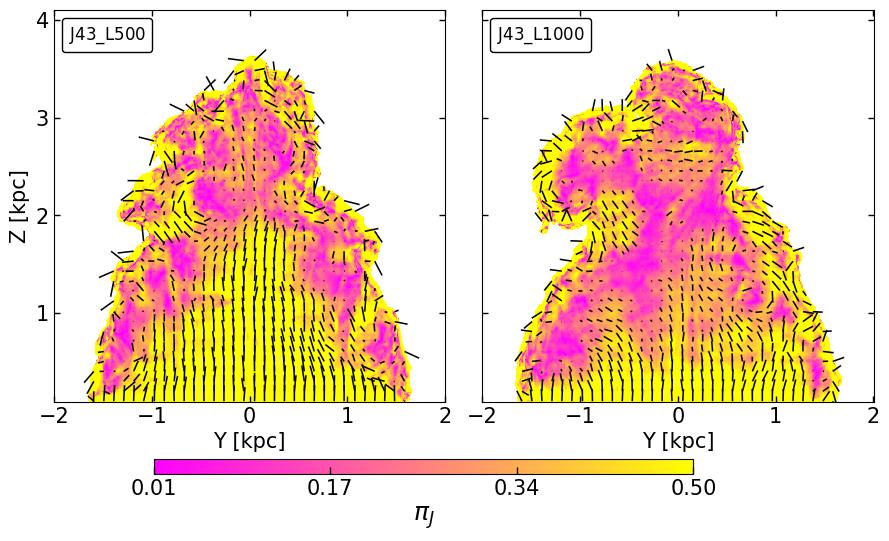}
     \end{tabular}}    
     \caption{Polarization maps ($\pi_J$) for $\mathrm{J43\_L500}$ (Left) and $\mathrm{J43\_L1000}$ (Right) at $\theta_I=90^\circ$ image plane and $t=~$834.6~kyr.}
     \label{pi_J_43}
\end{figure}

\subsubsection{Distribution of polarization characteristics}
\label{sec_pol_edge}
In our calculations, the spectral index for the population of emitting electrons is assumed to be $\alpha=2.2$, which implies a local polarization fraction for an ordered magnetic field to be \citep{burn_1966}:
\begin{align}
    \pi (\alpha) = \frac{3\alpha + 3}{3\alpha + 7} \approx 0.706
\end{align}
However, the polarization can vary depending on the orientation of the local magnetic field and electrons/fluid motion with respect to the LOS. These spatial variations cause changes in the local Stokes parameters in Eq.~\ref{eq:QU}, thus affecting the integrated polarized emission on the image plane which we discuss below. In this section, we limit our focus on the distribution of polarization characteristics in the cocoon and cocoon+SAM regions, and the depolarization effect from the SAM is discussed in the next section (Sec.~\ref{sec:SAM _depolarization}).

\begin{itemize}
  \item \textbf{Morphology of polarization:} 
 One can clearly see that in Fig.~\ref{fig:inc_42} and~\ref{fig:inc_43}, the polarization plot $\pi_T$ appears more extended, especially for the low-power jets when compared to $\pi_J$. This is due to wider SAM for the low-power jets when compared to the high-power ones.
The alignment of the polarization vectors gives inference about the distribution of the magnetic field inside the cocoon and the SAM. These vectors show that in the central regions of the cocoon, most of the magnetic field is aligned perpendicular to the jet (i.e. injected toroidal field), which gives rise to vertically polarized radiation (more clearly visible in $\pi_J$ maps at $\theta_I=90^\circ$ in Fig.~\ref{fig:inc_42}). 

In $\pi_T$ maps, the polarization vectors in the SAM are perpendicular to the forward shock surface, which implies that the compressed magnetic fields are parallel to the forward shock \citep{laing_1980,kollgaard_1990,guidetti_2011,fernandez_2021}. High polarization values with a range of 60-70\% can be seen in these regions. Similar vectors  normal to the outer edges of the contact discontinuity surface can also be seen in the $\pi_J$ maps, indicating fields parallel to the contact surface. Such alignments are commonly observed in regions towards the edges of the radio lobes \citep{bridle1984,leahy_1986}, and are attributed to the backflows in regions close to the contact discontinuity \citep[see][]{huarte_2011a}. In our study, this is caused by the velocity shears due to the turbulent motions in the backflowing plasma as well as shearing of the fields in regions near the contact discontinuity, as shown in Appendix.~\ref{appendix:forward_shock}. At $\theta_I=45^\circ$, the injected toroidal field leads to high polarization along the thin jet's spine. Low polarization in a wide region here is because, at a larger inclination, a LOS passes through a larger volume of the turbulent cocoon, as can be seen for $\pi_J$ maps in Fig.~\ref{fig:inc_43}. The cause of this internal depolarization in the cocoon is discussed below in detail. Some ``arc-like'' features (see R1, R2 in Fig.~\ref{fig:inc_43} and R3 in Fig.~\ref{fig:P43_L1000}) showing polarization values of around 35\~\%, with vectors indicating magnetic fields aligned along their circumferential surfaces, can also be clearly noticed. The polarization map ($\pi_T$) for J44 and J45 in Fig.~\ref{fig:inc_43} shows two bands of high polarization. The inner band is the high polarization region due to fields aligned along the contact surface by virtue of the shearing of fields in these regions (see Fig.~\ref{fig:marks3}). The outer band is due to the compression of fields at the forward shock. Such two distinct bands are not visible for J42 and J43 as depolarization from the SAM lowers the prominence of inner bands (see bottom row).

\textbf{Low-polarization in the cocoon:} Turbulent motions including the backflows with velocities ranging from $3,000~\kms$ to $30,000~\kms$ in J43 and J45, respectively are observed particularly in the upper regions of the cocoon (see Fig.~\ref{fig:marks3}). Several of these regions show high values of poloidal fields, which are comparable to the toroidal component. However, positive velocities (up to 300~$\kms$) can be seen in regions near the contact discontinuity (in the lower parts of the cocoon), which also coincides with elongated poloidal fields, indicating shearing of fields. This all leads to the generation of strong poloidal fields in the cocoon. Such a change in orientation of the magnetic field direction in the cocoon (from toroidal to poloidal) causes depolarization due to the vector cancellation of the orthogonally polarized components \citep[also observed in][]{hardcastle_2014}. This happens due to the sign flip in the cosine and sine components of the Stokes emissivities (see Eq.~\ref{eq:QU}), while integrating along the LOS. 

The vector cancellation due to poloidal transitions of the fields in the upper regions of the cocoon results in low-polarization values, i.e. less than $10\%$ in these regions, as can be seen in the $\pi_J$ maps of Fig.~\ref{fig:inc_42}. The turbulent backflows can also lead to the formation of high-emission filaments in the cocoon extending from or near the hotspots towards the lobes (see F1, F2 in Fig.~\ref{fig:inc_42} and F3 in Fig.~\ref{fig:inc_43}). The polarization vectors indicate that the fields are aligned along the ridges of these filaments. High polarization can be seen along these filaments in $\pi_J$ maps, which is later reduced after the contribution from the SAM is included (see Sec.~\ref{sec:SAM _depolarization} for a detailed discussion). In the lower part of the cocoon, i.e. close to the jet's spine, the cancellation of the polarization from the poloidal fields is not strong due to the large contribution from the toroidal component. However, this cancellation becomes prominent radially outwards, as the integration length through the cocoon decreases. Near the edges of the cocoon, the LOS passes through regions dominated by poloidal fields, leading to high polarization.

In contrast to the findings of \citet{huarte_2011a}, we do not observe backflows at the edges of the cocoon. However, this disparity can be attributed to the fact that the cocoon boundary in their study does not encompass the outer parts of the mixing layer near the contact discontinuity where the backflows are not as prominent as it is for the inner regions (see Appendix~\ref{appendix:forward_shock}). This can be seen by comparing the density map in Fig.~1 and the contour in Fig.~2 of their paper. Nonetheless, similar to their study, we do identify the presence of poloidal fields at the boundaries of the cocoon.


\item \textbf{Dependence on the correlation length of external magnetic field:}
At $\theta_I=90^\circ$ image plane in Fig.~\ref{fig:diff_len_43} (see also Fig.~\ref{fig:diff_len_44} and~\ref{fig:diff_len_45} in Appendix~\ref{app:figures}), the spatial configuration of the polarization vectors from the cocoon ($\pi_J$) for L~=~100 and 1000~pc cases appears similar to Fig.~\ref{fig:inc_42} for L~=~500~pc cases. However, the presence of turbulent backflows and the mixing with the ambient gas can result in distinct local internal depolarization effects for jets possessing similar magnetic fields but launched into an external medium with varying magnetic configurations. This happens as a result of mixing caused by Kelvin-Helmholtz instabilities at the contact surface, which introduces seeds of poloidal fields in the cocoon from the SAM. These seeds are then amplified by the turbulent motions, consequently augmenting the overall poloidal field strength within the cocoon. As an example, the cocoons in $\mathrm{J43\_L500}$ and $\mathrm{J43\_L1000}$ exhibit different polarization features (see Fig.~\ref{pi_J_43}). In $\pi_T$ maps, low polarization values along with some patches of high values can be seen in the SAM for L~=100~ and 1000~pc cases in Fig.~\ref{fig:diff_len_43} and~\ref{fig:P43_L1000}. The low polarization in the SAM is more prominent for the L~=~100~pc case due to the small-scale coherence when compared to the large-scale fields.

\end{itemize}

\begin{figure*}
 \centerline{
\def\arraystretch{1.0}
\setlength{\tabcolsep}{0.0pt}
\begin{tabular}{lcr}
    \includegraphics[scale=0.4]{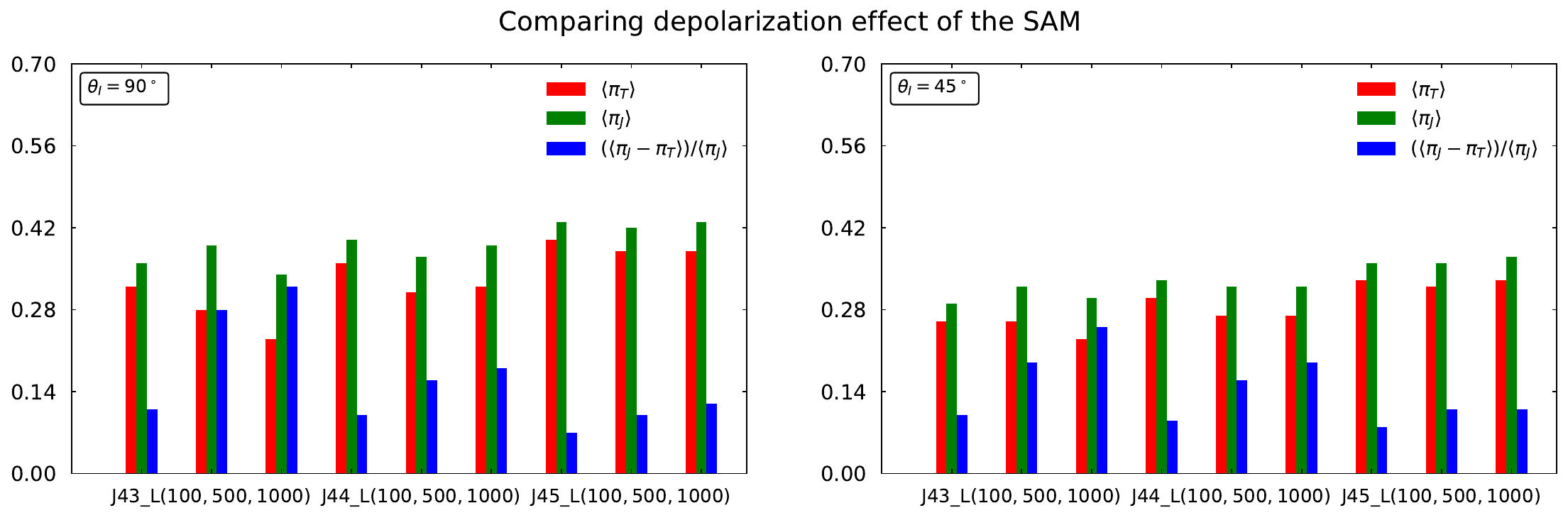}
     \end{tabular}}    
     \caption{Mean polarization fractions at $\theta_I=90^\circ$ (left) and $45^\circ$ (right) image planes for jet simulations with different correlation lengths at similar times as Figs.~\ref{fig:inc_42} and ~\ref{fig:diff_len_43} in the paper, and~\ref{fig:diff_len_44} and~\ref{fig:diff_len_45} in Appendix~\ref{app:figures}. One set of three bars shows the mean of the polarization fractions i.e. $\pi_T$ (red) and $\pi_J$ (green), and the fractional change in mean polarization (blue). The three consecutive sets show the same quantities for a fixed power of the jet interacting with an external magnetic field of correlation lengths 100~pc, 500~pc, and 1000~pc, respectively.}
     \label{fig:hist_pol}
\end{figure*}

\begin{figure*}
\centerline{
\def\arraystretch{1.0}
\setlength{\tabcolsep}{0.0pt}
\begin{tabular}{lcr}
\hspace{0.01cm}
    \includegraphics[height=7cm,keepaspectratio]{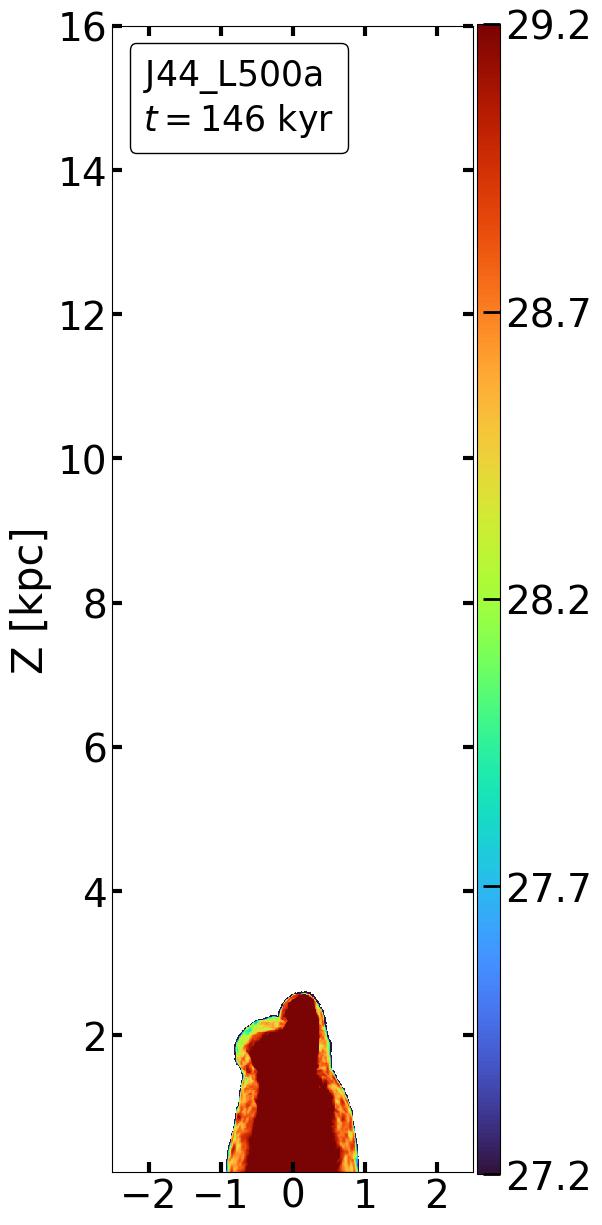}
    \includegraphics[height=7cm,keepaspectratio]{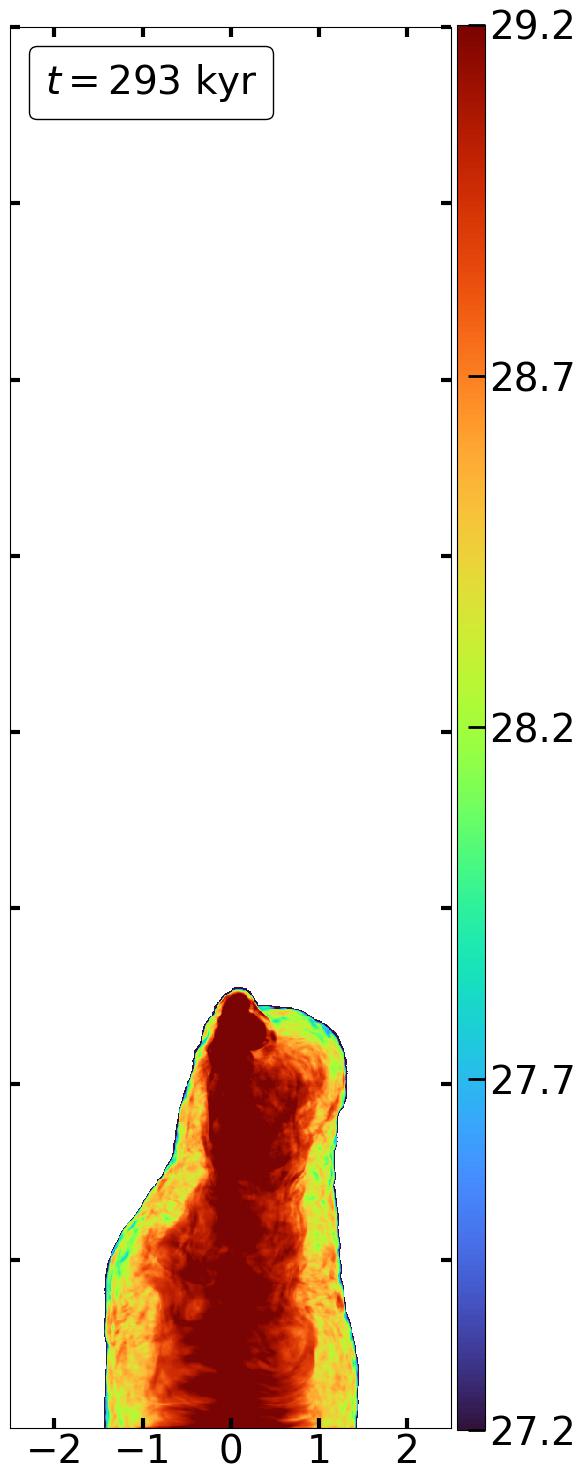}
    \includegraphics[height=7cm,keepaspectratio]{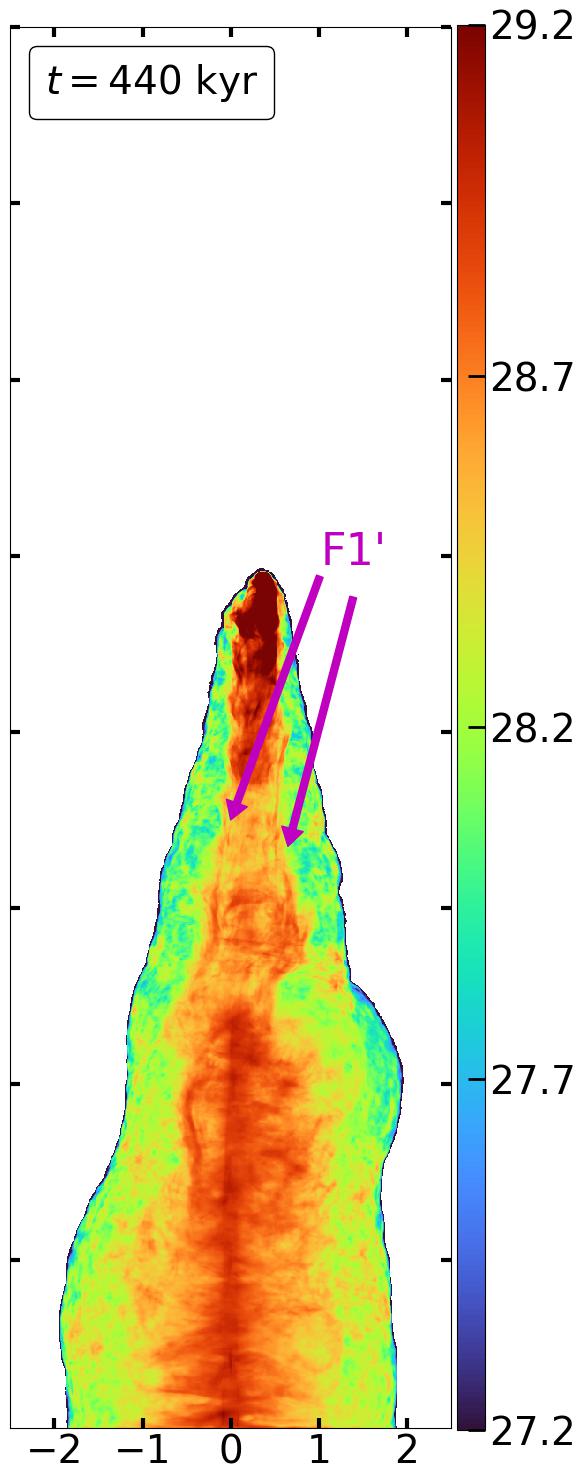}
    \includegraphics[height=7cm,keepaspectratio]{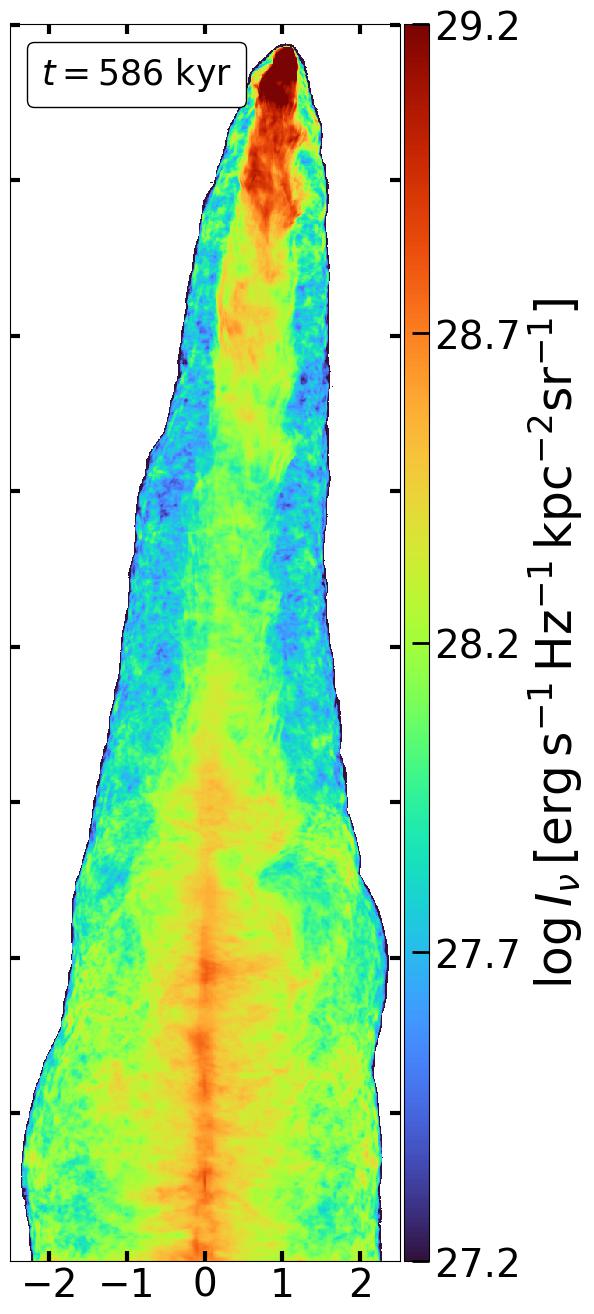}
    \end{tabular}}

    \centerline{
\def\arraystretch{1.0}
\setlength{\tabcolsep}{0.0pt}
\begin{tabular}{lcr}
    \includegraphics[height=7cm,keepaspectratio]{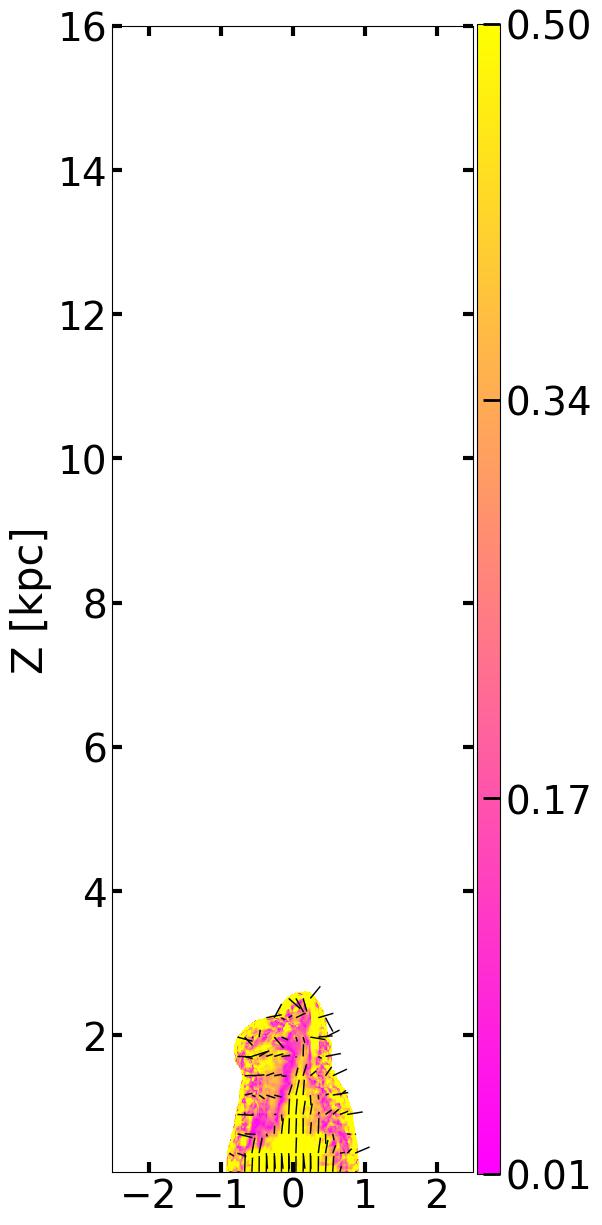}
    \includegraphics[height=7cm,keepaspectratio]{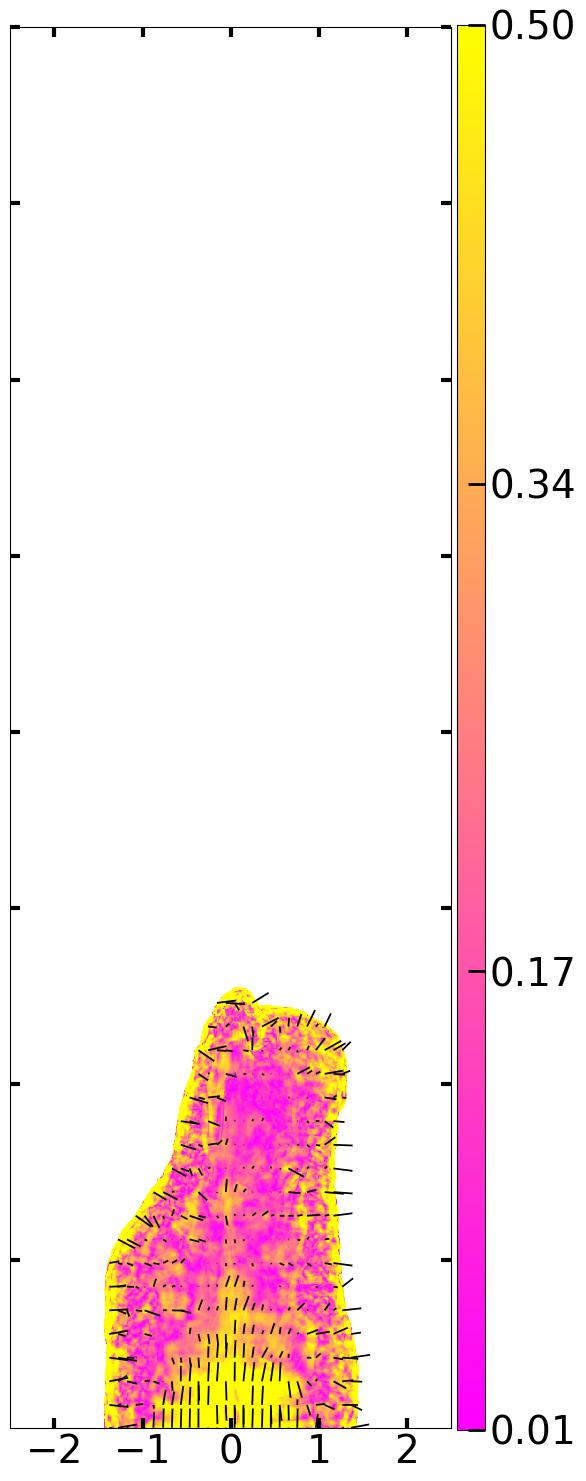}
    \includegraphics[height=7cm,keepaspectratio]{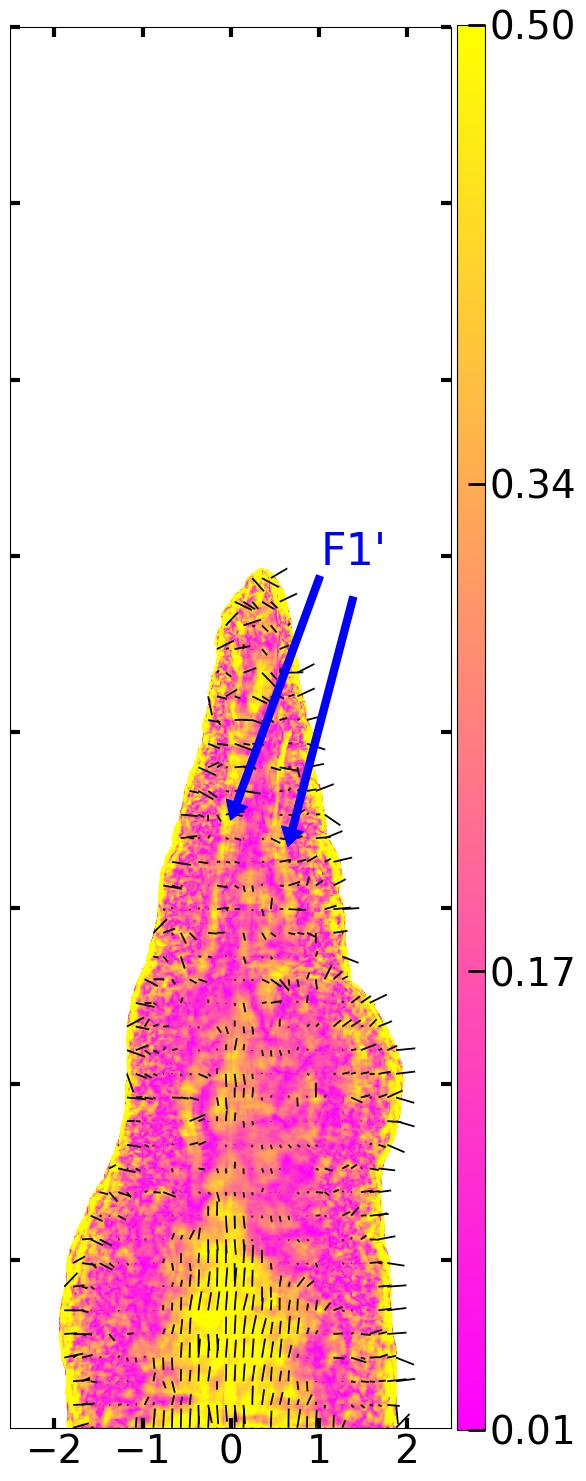}
    \includegraphics[height=7cm,keepaspectratio]{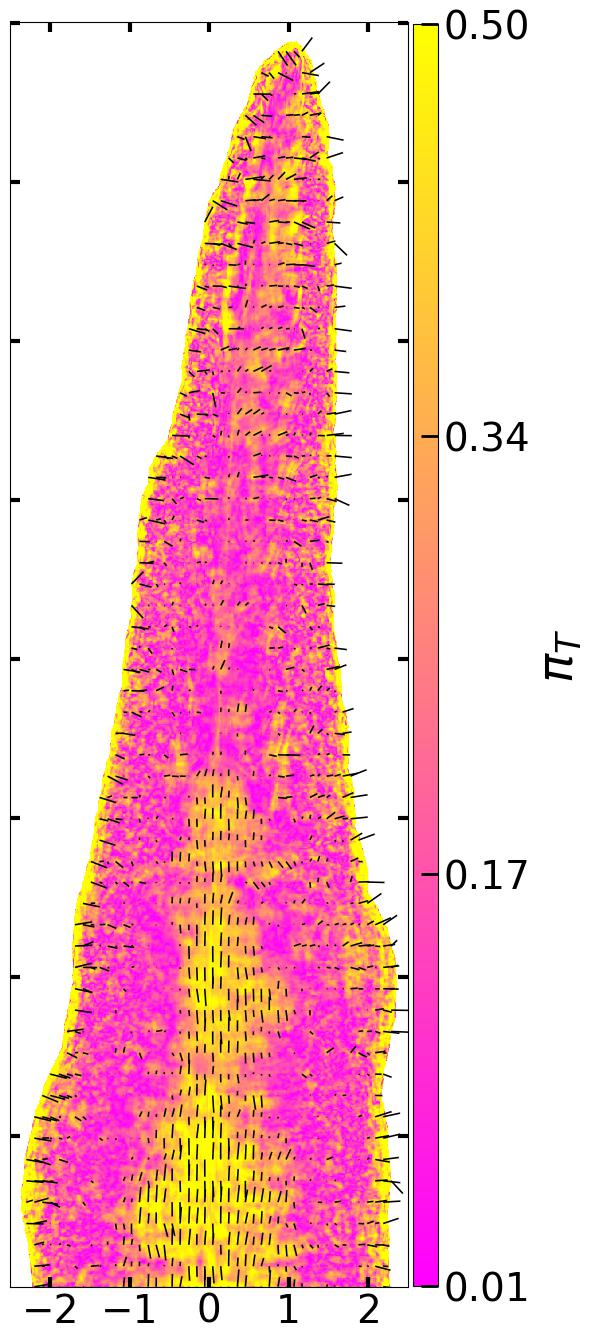}
    \end{tabular}}
     \centerline{
\def\arraystretch{1.0}
\setlength{\tabcolsep}{0.0pt}
\begin{tabular}{lcr}\hspace{0.1cm}
    \includegraphics[height=7.2cm,keepaspectratio]{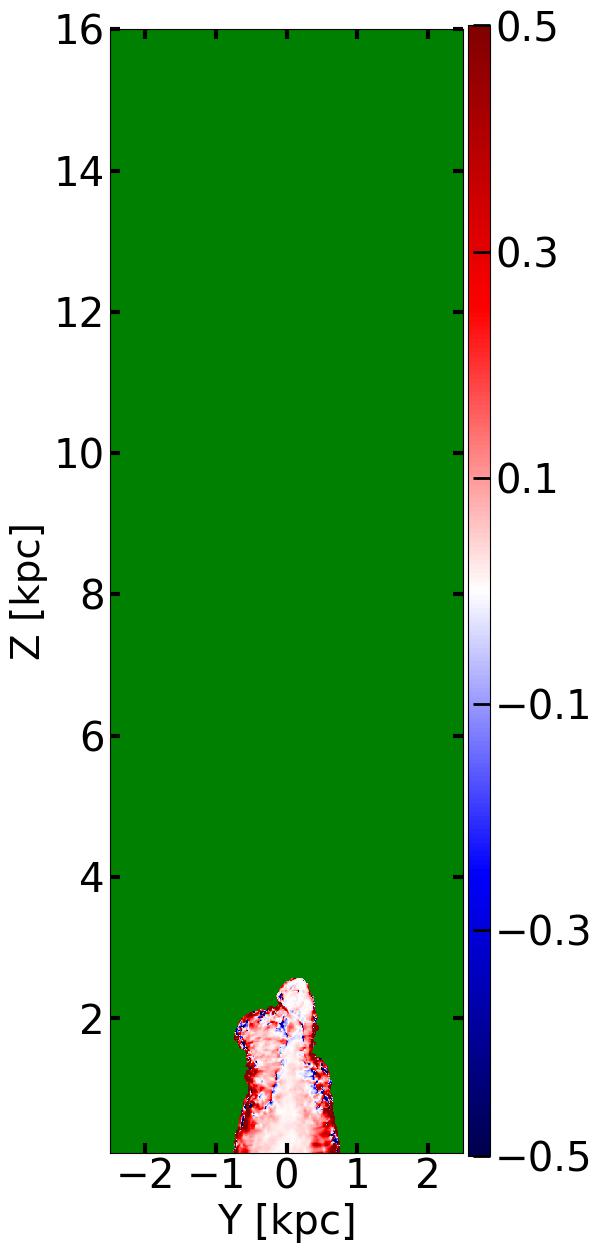}
    \includegraphics[height=7.2cm,keepaspectratio]{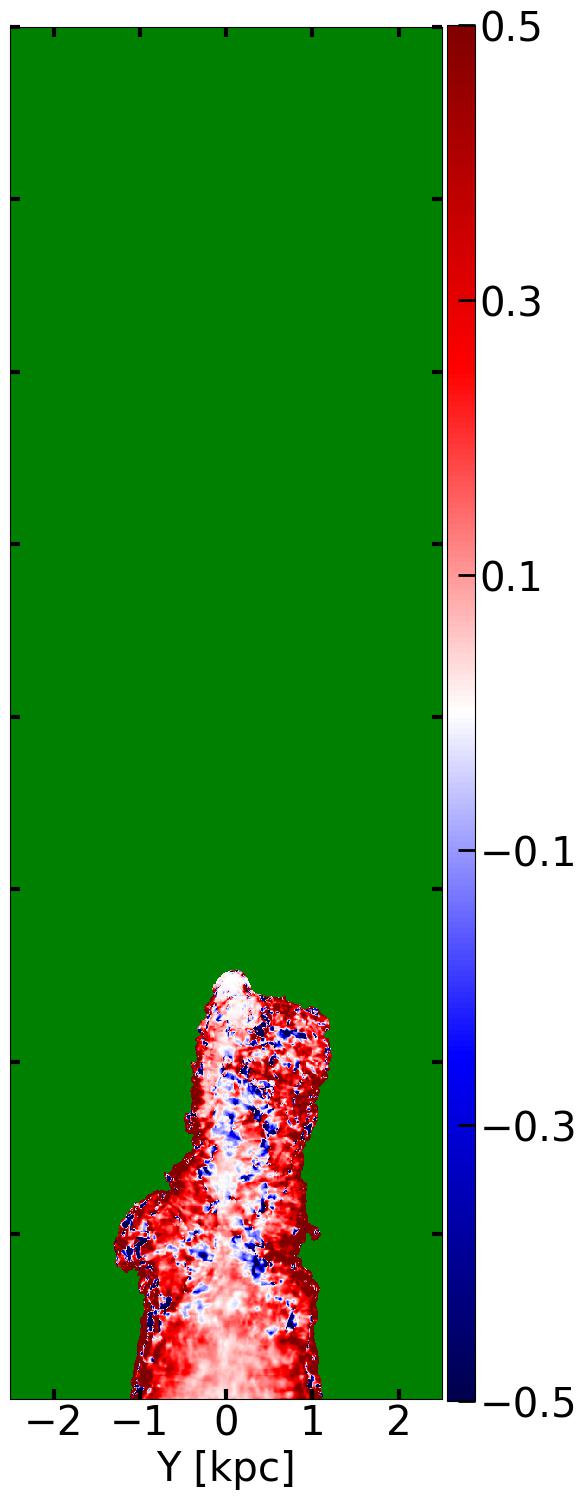}
    \includegraphics[height=7.2cm,keepaspectratio]{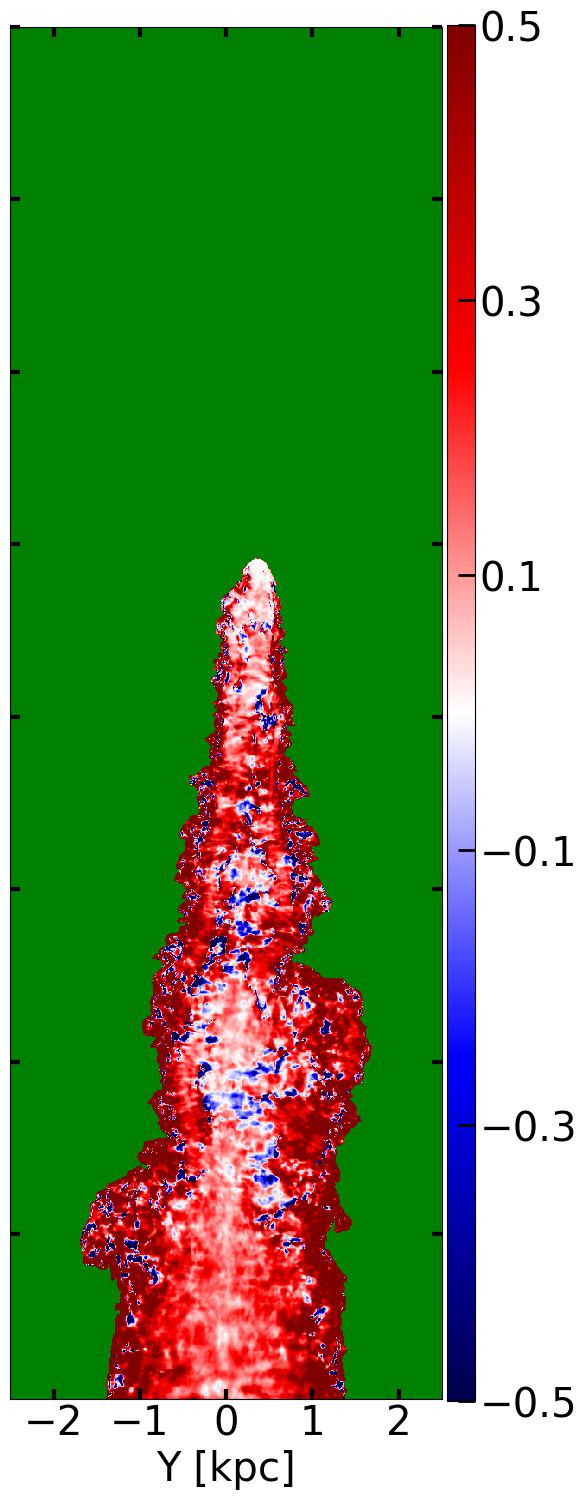}
    \includegraphics[height=7.2cm,keepaspectratio]{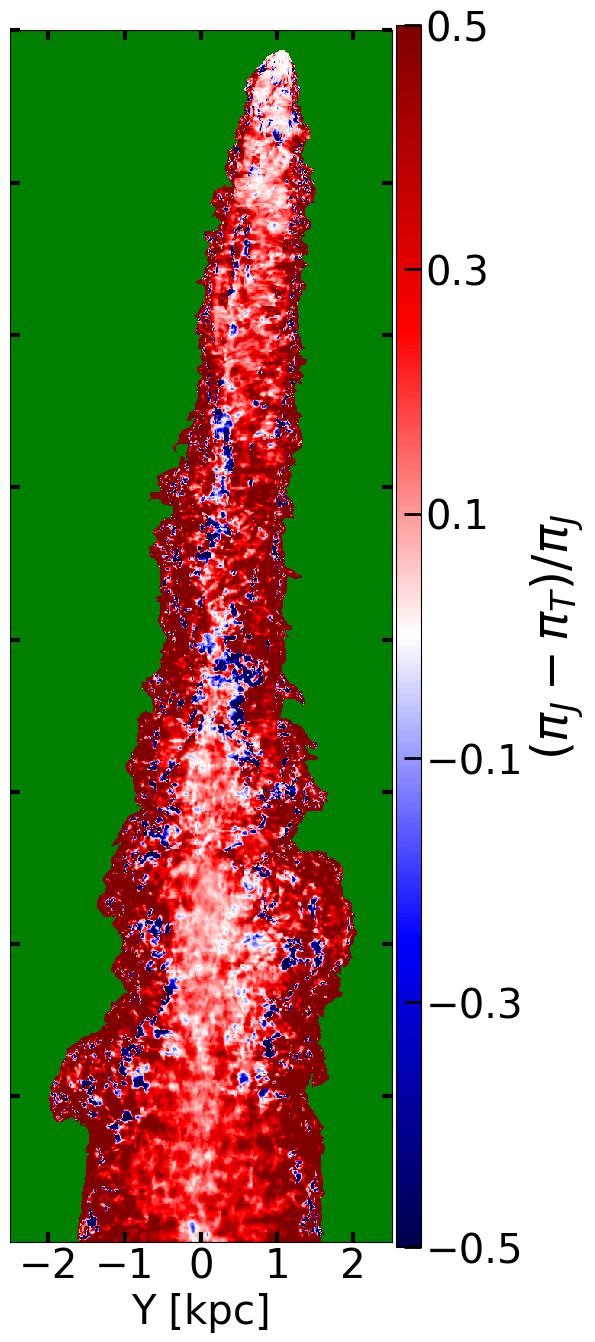}
    \end{tabular}}
     \caption{\textbf{Top:} Logarithmic total synchrotron flux, \textbf{Middle:} Polarization map ($\pi_T$), \textbf{Bottom:} Fractional change in polarization ($\Delta\pi$) for $\mathrm{J44\_L500a}$ at $\theta_I=90^\circ$ image plane for different times. \textbf{F1'} shows filamentary structures inside the cocoon at $t=293~$kyr and 440~kyr, respectively.}
\label{fig:lbox_44}
\end{figure*}

\begin{figure*}
\centerline{
\def\arraystretch{1.0}
\setlength{\tabcolsep}{0.0pt}
\begin{tabular}{lcr}
\hspace{-0.22cm}
    \includegraphics[height=9cm,keepaspectratio]{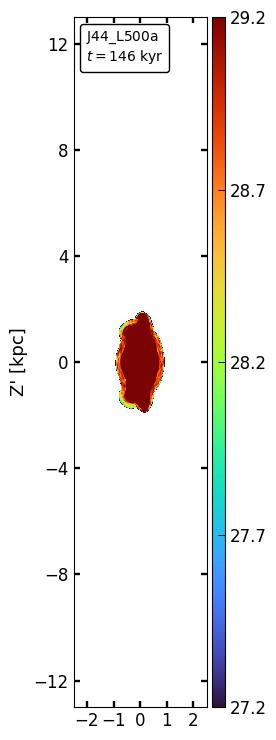}
    \hspace{0.1cm}
    \includegraphics[height=9cm,keepaspectratio]{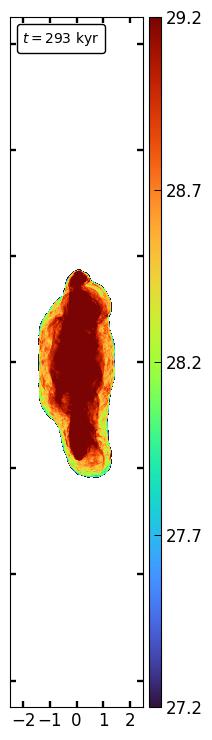}
    \hspace{0.1cm}
    \includegraphics[height=9cm,keepaspectratio]{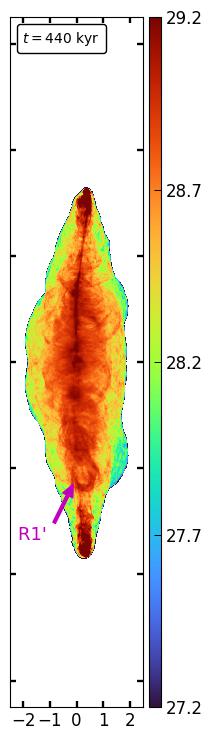}
    \hspace{0.1cm}
    \includegraphics[height=9cm,keepaspectratio]{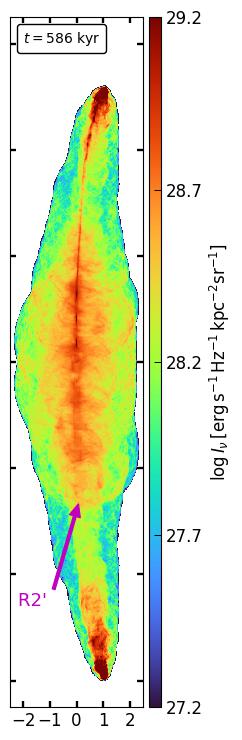}
    \end{tabular}}

    \centerline{
\def\arraystretch{1.0}
\setlength{\tabcolsep}{0.0pt}
\begin{tabular}{lcr}
    \includegraphics[height=9.25cm,keepaspectratio]{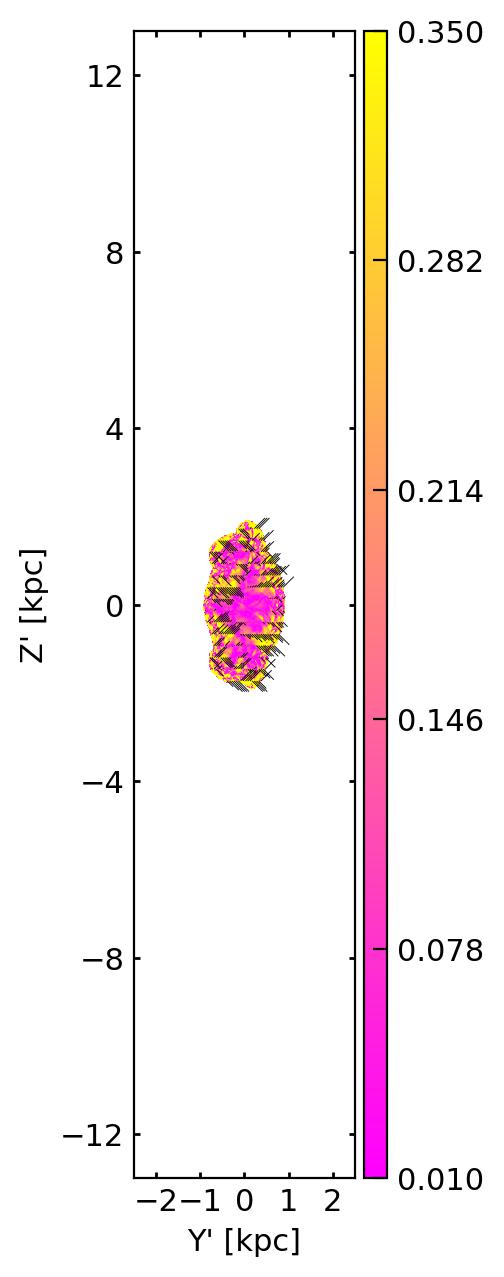}
    \includegraphics[height=9.25cm,keepaspectratio]{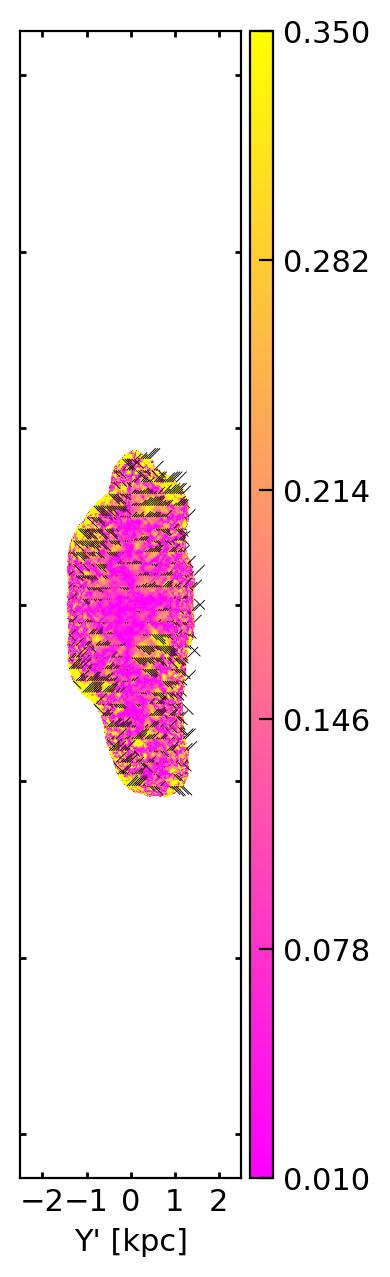}
    \includegraphics[height=9.25cm,keepaspectratio]{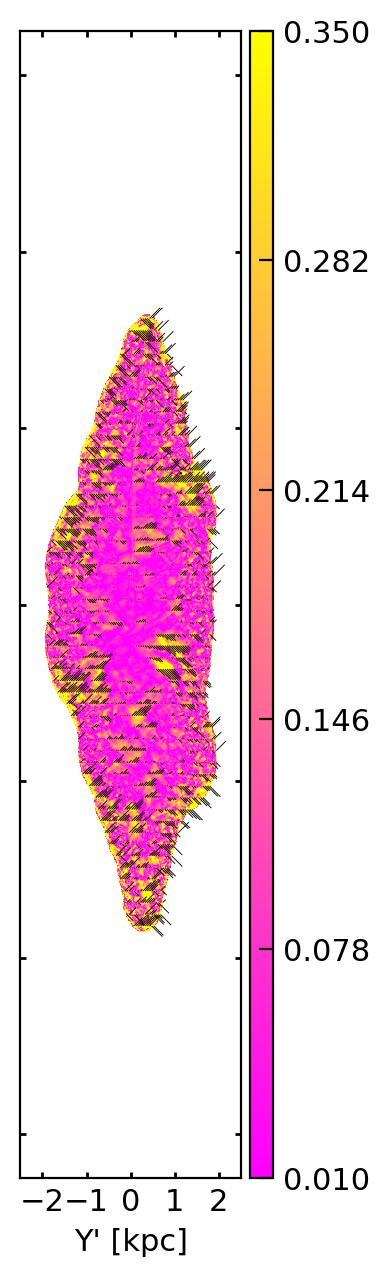}
    \includegraphics[height=9.25cm,keepaspectratio]{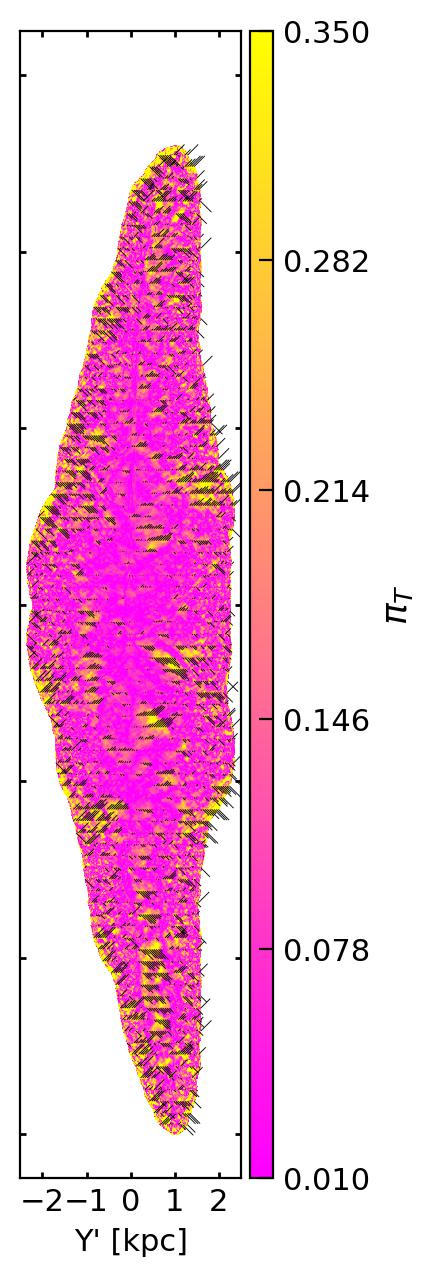}
    \end{tabular}}
     \caption{Logarithmic total synchrotron flux and polarization map ($\pi_T$) for $\mathrm{J44\_L500a}$ at $\theta_I=45^\circ$ image plane for different times. \textbf{R1'} and \textbf{R2'} show some of the arc-like features in the simulation.}
\label{fig:lbox_44_45deg}
\end{figure*}

\begin{figure*}
\centerline{
\def\arraystretch{1.0}
\setlength{\tabcolsep}{0.0pt}
\begin{tabular}{lcr}
    \includegraphics[width=0.3\linewidth,height=0.3\linewidth]
    {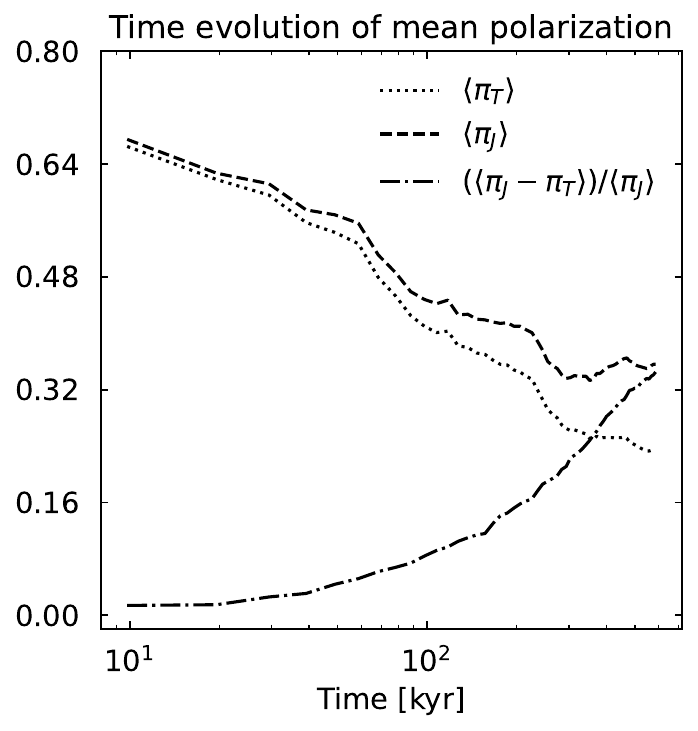}
     \includegraphics[width=0.3\linewidth,height=0.3\linewidth]
    {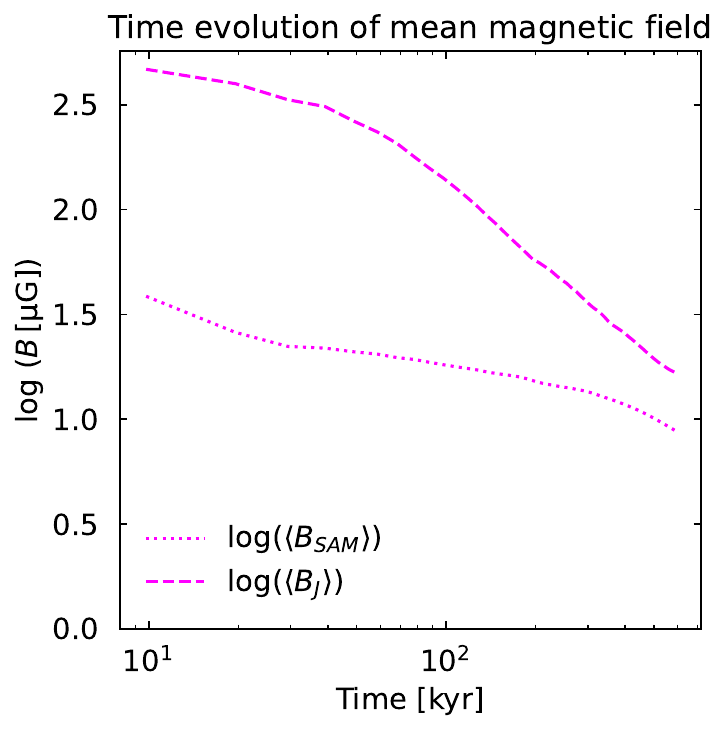}
    \includegraphics[width=0.3\linewidth,height=0.3\linewidth]
    {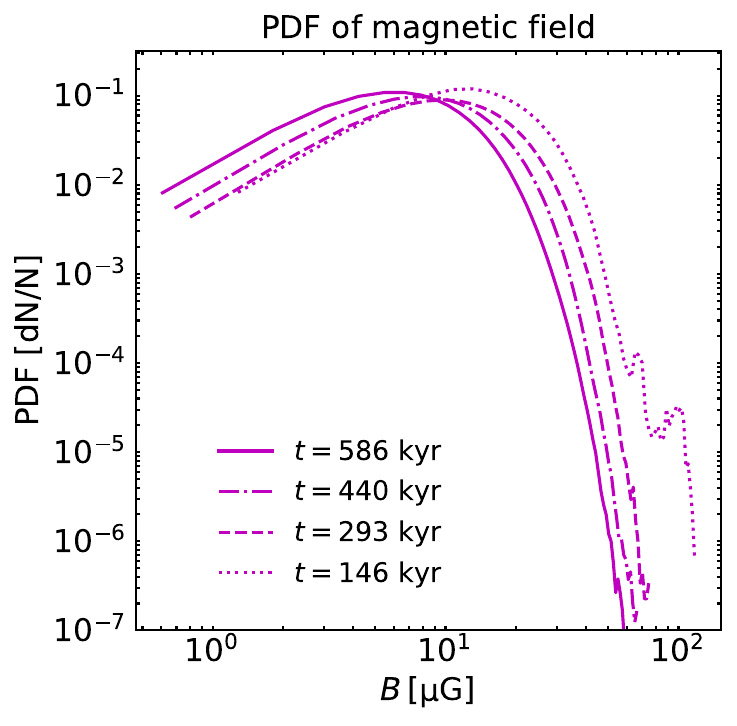}
    \end{tabular}}
    \caption{Time evolution of different measures for $\mathrm{J44\_L500a}$. \textbf{Left:} Mean of polarization fractions ($\pi_T~\mathrm{and}~\pi_J)$ and fractional change in mean polarizations $(\langle\pi_J-\pi_T\rangle)/\langle\pi_J\rangle$ at $\theta_I=90^\circ$ image plane. \textbf{Middle:} Mean magnetic field in the SAM and the jet's cocoon. \textbf{Right:} PDF of the magnetic field in the SAM at different times.}
\label{fig:pol_lbox}
\end{figure*}

\subsubsection{Depolarization effect of the SAM}
\label{sec:SAM _depolarization}

In the fourth row of Fig.~\ref{fig:inc_42}, we show the maps for the fractional change in polarization ($\Delta\pi$) which accounts for the contribution of the magnetized SAM in affecting the observed polarization from the cocoon for L~=~500~pc cases. One can notice that in all the cases the polarization fraction from the cocoon is lowered due to the presence of the turbulent magnetic fields in SAM. Similar lowering of the polarization can also be seen in the right panels of Fig.~\ref{fig:diff_len_43} in the paper, and Fig.~\ref{fig:diff_len_44} and~\ref{fig:diff_len_45} in Appendix~\ref{app:figures}. In this section, we attempt to quantify the depolarization effect from the SAM using some diagnostics. We show histograms with mean of the polarization fractions i.e. $\langle \pi_T \rangle$ (red) and $\langle \pi_J \rangle$ (green), and the fractional change in mean polarization (blue) in Fig.~\ref{fig:hist_pol}, estimated at the $\theta_I=90^\circ$ and $45^\circ$ image plane for different simulations. Three consecutive sets of the bars represent these values at correlation lengths of 100~pc, 500~pc, and 1000~pc, respectively, for a fixed mechanical power of the jet. All the calculations are done only for the regions with $\pi_J > 0$ on the image plane.
We explain our findings below.
\begin{itemize}
    \item \textbf{Dependence on the jet power:} Our analysis shows that the SAM results in a lowering of the net polarization fraction observed from the jet cocoon. For instance, Fig.~\ref{fig:hist_pol} shows that $\pi_J$ is higher than $\pi_T$ for all cases, clearly showing net depolarization from the SAM. As the jet power is decreased, the distinction between $\pi_T$ and $\pi_J$ grows significantly, resulting in an enhanced fractional change in mean polarization values.
 The magnetic fields in the cocoon undergo decay with the expansion and can become comparable to the compressed fields in the SAM for the low-power jets, as shown in Fig.~\ref{fig:dist_Bj}. Therefore, the magnetized SAM can contribute significantly to cancelling out the polarized emission. At a few sites, the compressed fields in the SAM can also lead to higher polarized emission (depending on the LOS) than what is obtained from the jet only (see e.g. Fig.~\ref{fig:inc_42}). Nevertheless, the depolarizing effect due to the vector cancellation of Stokes emissivities (see Eq.~\ref{eq:QU}) caused by the randomness in the magnetic polarity of the external field mostly dominates. Contrarily for the high-power jets, especially for $\mathrm{J45}$, the jet's magnetic field is very strong when compared to the compressed field ahead of the cocoon (see Fig.~\ref{fig:Bmag}). Thus, the contribution of the SAM in lowering the polarization from the cocoon is negligible.


\item \textbf{Dependence on correlation length of external magnetic field:} 

In our study, we find that the depolarization effect from the SAM gets strengthened as the correlation length of the external magnetic field is increased. From Fig.~\ref{fig:hist_pol}, it can be seen that the fractional change in mean polarization values is comparatively smaller for $\mathrm{L=100~pc}$ case than the large scale fields ($\mathrm{L=500~and~1000~pc}$). 
This can also be seen in the right panels of Fig.~\ref{fig:diff_len_43} in the paper and Fig.~\ref{fig:diff_len_44} and~\ref{fig:diff_len_45} in Appendix~\ref{app:figures}, where more reddening of the fractional change in polarization for $\mathrm{L=1000~pc}$ indicates stronger SAM depolarization when compared to $\mathrm{L=100~pc}$. Among large-scale fields in the SAM, high polarization fractional change for $\mathrm{L=1000~pc}$ case can be seen particularly for the J43 and J44 in Fig.~\ref{fig:hist_pol}. Due to the strong magnetic field of the cocoon, the effects are less discernible for J45. As discussed in Appendix~\ref{no_jet_test}, the magnetic fields within the SAM decay rapidly with the adiabatic expansion for the L~=~100~pc, thereby reducing its depolarization effect on the emission from the jet. Thus, in conclusion, we find that the large-scale fields in the SAM lead to more depolarization of the radiation from the jet cocoon and small-scale fields with scale $\approx 100~$pc decay fast leading to less depolarization.

\end{itemize}

\subsection{Time evolution of jet, SAM and depolarization}

In this section, we discuss the time evolution of the synchrotron emission and polarization in simulation $\mathrm{J44\_L500a}$ in Table~\ref{tab:sim_table}, where a jet of power $10^{44}\,\ergs$ is launched in a turbulent magnetic field correlated at scales of 500~pc. The emission and polarization maps at the $\theta_I=90^\circ$ and $45^\circ$ image plane for different times are shown in Fig.~\ref{fig:lbox_44} and~\ref{fig:lbox_44_45deg}, respectively. The regions inside the forward shock are selected using Eq.~\ref{eq:forward}, but with an `or' condition for the tracer to include regions that exhibit low pressure ($<4P_0$) at late times. The SAM is identified by selecting cells with $\mathrm{tr1}\leq10^{-7}$ from the regions inside the forward shock.

The synchrotron maps in Fig.~\ref{fig:lbox_44} and~\ref{fig:lbox_44_45deg} show that the jet appears compact at earlier times, and later develops a spine with a clear hotspot. Some elongated high-emission filaments, for example, see e.g. F1' in Fig.~\ref{fig:lbox_44}, are also visible in the cocoon. High-emission `arc-like' structures (e.g. see R1' and R2' in Fig.~\ref{fig:lbox_44_45deg}), similar to Fig.~\ref{fig:inc_43}, can also be seen. In the middle row of Fig.~\ref{fig:lbox_44}, the injected toroidal field of the jet causes high polarization in the lower regions of the cocoon. The bright yellow patch extending to $\sim 1~\mathrm{kpc}$ from the jet's injection zone at $t=146~$kyr indicates that the depolarization from the SAM is not so prominent at earlier times. This, however, gets diluted due to depolarization by the random fields in the SAM as the jet enlarges with time, which can also be inferred from the increased reddening in the plots in the bottom row. As shown in Appendix~\ref{appendix:forward_shock}, the backflows and shearing of fields in the cocoon also lead to low polarization values in the cocoon. Consequently, the pinkish regions in the upper and lateral regions of the cocoon originate from a combination of internal and external depolarization, as discussed in previous sections. Nonetheless, the bottom row shows that the polarization from the jet's cocoon is indeed lowered by the magnetized SAM.

At $\theta_I=45^\circ$, a large region of the jet exhibits low polarization values ($\sim 1\%$), as can be seen in Fig.~\ref{fig:lbox_44_45deg}. The low polarization regions also become more pronounced visually due to different lengths along the axes on the image plane, in contrast to previous runs in Fig.~\ref{fig:inc_43}, where the length along both axes is the same. The high values mainly occur at the edges of the forward shock and along the jet's spine, and some patches of $\sim$kpc size can also be seen. Similar behavior also occurs in Fig.~\ref{fig:pol_lbox}, where we show the time evolution of the mean of the polarization fractions $\pi_T$ and $\pi_J$ and the fractional change in mean polarization at $\theta_I=90^\circ$ image plane in dashed, dotted, and dash-dotted curves, respectively. The calculations are done for the regions where $\pi_J>0$ on the image plane. The plot clearly indicates that the presence of the turbulent magnetic field in the SAM progressively lowers the net polarization fraction caused by the jet region.

 It should be noted that while the total magnetic field energy of the cocoon and the SAM increases over time as the jet expands, the local values of the magnetic fields can actually decrease. This trend is evident in the middle panel of Fig.~\ref{fig:pol_lbox}, which illustrates that the magnetic fields within the cocoon can decay more rapidly compared to those in the SAM. Consequently, as the magnetic field strengths in the cocoon and SAM become comparable, the depolarization from the SAM strengthens. Simultaneously, the internal depolarization also gains momentum due to an increase in the cocoon's poloidal component, as discussed in Sec.~\ref{sec_pol_edge}. This contributes to a decrease in $\pi_J$ values over time, which in turn can also lead to more lowering in $\pi_T$. A faster decline of $\pi_T$ than $\pi_J$ leads to a sharp increase in the fractional change in polarization with time, as can be seen in the left panel of Fig.~\ref{fig:pol_lbox}.

In the right panel of Fig.~\ref{fig:pol_lbox}, we show the distribution of the magnetic field in the SAM at different times. At $146.7~$kyr, the magnetic field in SAM reaches very high values ($\sim 100\,\mathrm{\mu G}$), which is larger than the mean value of the external magnetic field at the initial time ($\sim 10~\mathrm{\mu G})$. However, the decay of the magnetic fields caused by the enlargement of the SAM causes the distribution to shift toward the left with time. We confirm that the PDF of the magnetic fields within the jet cocoon exhibits a similar trend as the PDF of the fields in the SAM over time. Thus, the magnetic fields in both cocoon and SAM decreases with the expansion of the jet.

\section{Discussion and summary}
\label{discussion}
    In this paper, we attempt to investigate the effect of the magnetized shocked ambient medium (SAM) surrounding the jets on their observed non-thermal synchrotron emission and polarization features. Our aim is to study these different observable features during the evolutionary phase of young evolving jets. The ambient medium is equipped with a turbulent magnetic field with a mean magnitude of $\approx$ 10 $\mathrm{\mu G}$, which is set up by following the approach given in \citet{tribble91,murgia_2004}. The jets of different mechanical powers varying from $10^{42}\ergs$ to $10^{45}\ergs$ are launched in the magnetized medium with correlation lengths of $\mathrm{100~pc, 500~pc \, and \,1000~pc}$. To estimate the non-thermal radio emission, we make an assumption that the electrons contain precisely 10\% of the internal energy of the fluid. However, this assumption is not entirely accurate, as the electrons in different regions are expected to gain varying energy from their surroundings due to different acceleration mechanisms \citep[see e.g.][]{matthews_2020}. For example, the electrons along the spine of the jet and at the hotspot will be accelerated more strongly, when compared to the cocoon. Also, the acceleration of the electrons inside the SAM, which most likely originated in the external medium, may not be as efficient as in the cocoon. Our current analysis follows a simplified approach and a more intricate study including non-thermal electrons both within and outside the outflow is reserved for the future.

The text below summarizes the main findings from our study:

\begin{enumerate}
   
    \item \textbf{Jets amplifying the external magnetic field at the forward shock:}\, The forward shock of the jet compresses and amplifies the magnetic fields from the ambient medium (see Fig.~\ref{fig:Bmag}). The amplified values of the field can go up to $200\,\mathrm{\mu G}$ (Fig.~\ref{fig:dist_43_45}), which is quite higher than the mean magnitude of the initial ambient field ($\sim10\,\mathrm{\mu G}$). Therefore, one can expect such amplification to be comparatively stronger in the starburst galaxies, where magnetic fields up to $1000\,\mathrm{\mu G}$ have been observed \citep{mcbride_2014}.  However, as the SAM expands along with the adiabatic expansion of the jet cocoon, the magnetic energy of the compressed fields in the SAM decreases (see \citet{fernandez_2021}), as can be inferred from Fig.~\ref{fig:res_test2}.

   \item \textbf{Morphology of synchrotron emission from jets:} The electrons accelerated inside the jet and the magnetized SAM can give rise to high synchrotron flux. For the magnetic fields with large correlation lengths, the flux from the SAM is much higher when compared to the fields with small coherence scales (Fig.~\ref{fig:diff_len_43}, and~\ref{fig:diff_len_44} and~\ref{fig:diff_len_45} in Appendix~\ref{app:figures}). This occurs because the magnetic fields in the SAM decay slowly with the expansion for the former case (see Appendix.~\ref{no_jet_test}), and thus the total integrated emission is high. We observe the highest emission from the jet's spine and hotspots, and some filaments are also observed in the cocoon, particularly extended from the jet's head. Thus, these filaments are likely to be formed due to backflows (see Fig.~\ref{fig:inc_42} and~\ref{fig:inc_43}), as the magnetic fields lie along them and display high polarization values of $\gtrsim 50\%$. Such high-polarized filaments originating from the hotspot with fields aligned along their ridge lines have been observed in 3C~390.3 \citep{leahy_1995} and 3C~334, 3C~336 and 3C~351 \citep{bridle_1994}, and also in simulations by \citet{huarte_2011a}.
   
   The synchrotron emission morphology varies depending on the viewing angle of the observer with respect to the jet's axis. At lower inclinations, the Doppler asymmetries lead to higher flux from the approaching jet when compared to the receding one, although the highest emission is still observed from the hot-spots of both the jets (Fig.~\ref{fig:inc_43} and~\ref{fig:inc_44}). At $\theta_I=45^\circ$, some ``ring or arc-like'' high emission structures are observed in Fig.~\ref{fig:inc_43},~\ref{fig:P43_L1000} and~\ref{fig:lbox_44_45deg}, which are likely because of the annular shocks propagating in the backflowing plasma, as shown by simulations in \citet{saxton_2002}. The rings in J44 and J45 show a polarization fraction of up to $35\%$, and the polarization vectors indicate magnetic fields along the rings. Rings or arc-like features have been observed in Hercules A \citep{dreher_1984}, 3C~219 \citep{perley_1984} and 3C~210 \citep{kraft_2012}. However, one should note that these structures can be formed due to different scenarios \citep{gizani_2003}, such as rings in Hercules A are may be due to episodic AGN activities \citep{timmerman_2022}, although contrary views of a single episode formation have also been recently proposed \citep{perucho_2023}. Our simulations suggest that annular shocks can perhaps provide similar morphologies, as observed in large-scale sources. In Fig.~\ref{fig:inc_43} and~\ref{fig:P43_L1000}, `Hotspot complexes' can be seen for the low-power jets, where the secondary spots are formed due to a wiggling of the jet's spine (as can be inferred from Fig.~\ref{fig:prs_3d_43}). Multiple hotspots have also been observed in several FRII sources \citep[for e.g.][]{leahy_1997,mahatma_2023}, and can form due to different internal processes \citep{horton_2023}.

    \item \textbf{Orientation of magnetic fields and polarization vectors:}\, We find that the injected toroidal magnetic field in the jet produces linearly polarized radiation with polarization vectors aligned along the vertical direction in the central regions of the cocoon. Observations indicate that the magnetic field in the FRII-like sources is found to be aligned along the jet direction and FR-I shows toroidal fields \citep{bridle_1984,bridle_1994}. However, helical magnetic fields are also observed in several sources \citep{asada_2002,zamanisab_2013}, which indicate the toroidal component of the launched jet \citep[see also][]{silpa_2021a,silpa_2021b}. The patchy polarization distribution suggests a complex field configuration in jets \citep{saikia_1988}. In our study, the jets are launched with a toroidal field to conserve zero divergences at the injection zone. Nevertheless, simulations incorporating jets injected with helical, toroidal, and random fields have also been conducted \citep{Gaibler_2009,huarte_2011a,hardcastle_2014,chen_2023}. These studies found that the poloidal fields can be generated due to backflows and shearing of fields, although the strength can vary for different configurations. The dominance of the toroidal component in our jets is solely a result of the injected field, which persists until the end of the simulations. However, we anticipate that, as found by \citet{hardcastle_2014} and \citet{chen_2023}, the poloidal component will strengthen over time as the jet evolves to large scales.
    
    In our study, the magnetic field appears poloidal in regions close to the outer edges of the jet's cocoon (see $\pi_J$ maps in Fig.~\ref{fig:inc_42} and~\ref{fig:inc_43}), which is caused by the velocity shears due to turbulent backflows as well as shearing of the jet's magnetic field in the outer parts of the mixing layer. Furthermore, the compression of the external magnetic fields can also produce such alignments surrounding the radio lobes \citep[]{laing_1980}, as can also be inferred from $\pi_T$ maps in Fig.~\ref{fig:inc_42} and~\ref{fig:inc_43}.
    
    Fields aligned along the edges of the radio lobes have been seen in several sources \citep{hardcastle_1996,bridle1984,mullin_2006,subrahmanyan_2008,perlman_2020,baghel_2022}. High polarization values up to 50-70\% are observed in these regions \citep{bridle_1994} and are attributed to the backflowing plasma in the cocoon, which aligns the fields parallel to the contact surface \citep{matthews_1990}. We also find similar alignments and polarization values in regions close to the edges of the cocoon and the forward shock ($\pi_J$ as well as $\pi_T$ maps in Fig.~\ref{fig:inc_42} and~\ref{fig:inc_43}). It is likely that fields parallel to the outer edges of the radio lobes in observations are caused by the back-flows in the jet cocoon, as one may anticipate the majority of detectable emissions from these regions. Nonetheless, one cannot dismiss the possibility that some of these regions might also result from the compression of the external magnetic fields.

\item \textbf{Internal depolarization due to vector cancellation:}\,

The jets in our study are injected with the toroidal field; however, a strong poloidal component gets generated in the cocoon with time.
The turbulent backflows and the shearing of fields in the cocoon lead to the production of strong poloidal fields in regions near the contact surface and the jet's head (Fig.~\ref{fig:marks3}). Additionally, the poloidal fields introduced into the cocoon from the SAM via mixing at the contact surface are further amplified due to the turbulence. Such realignments lead to vector cancellation of the orthogonal (poloidal and toroidal) polarization components along the LOS \citep{swain_1998,hardcastle_2014}. In the polarization maps for the cocoon ($\pi_J$), we observe low polarization in the regions where the configuration of the magnetic field appears to be transitioning from toroidal to poloidal (see Fig.~\ref{fig:inc_42} and~\ref{fig:inc_43}).

   \item \textbf{Depolarization from the swept-up magnetic fields:}\,
    We find that the random magnetic polarity of the fields in the SAM around the radio lobes causes depolarization of the synchrotron radiation emitted by the jet (see bottom row in Fig.~\ref{fig:inc_42}). 
    In our study, this effect is weak for the high-power jets ($\mathrm{J44}$ and $\mathrm{J45}$)  because the magnetic field of these jets is quite higher than the compressed fields in the SAM (see Fig.~\ref{fig:dist_Bj}). Contrarily, as inferred from Fig.~\ref{fig:hist_pol}, the low-power jets, which are launched with a low magnetic field, develop magnetic strengths comparable to that in the SAM, resulting in a higher contribution of SAM depolarization. Such features of external depolarization can be linked to real systems, such as the non-detection of diffuse polarized emission from lobes in B2~0258+35 \citep{adebahr_2019} can be attributed to the cancellation by the foreground screen (SAM and ICM). The radio-lobes exhibit weak magnetic fields ($\mathrm{\sim 1.1~\mu G})$ as determined through equipartition, which is similar to the estimated values for the fields in the ICM \citep{Guidetti_2008,guidetti_2010}. Contrarily, \citet{knuettel_2019} found polarization values of up to $\sim$40\% in the lobes of 3C~277.3, where magnetic fields of $\mathrm{50~\mu G}$ are observed \citep{van_1985}.

    \item \textbf{Depolarization from the fields of different correlation lengths:}\,
    We find that the magnetic fields with large correlation lengths in the SAM lead to more depolarization when compared to the small-scale fields, as inferred from Fig.~\ref{fig:diff_len_43} and~\ref{fig:hist_pol}. The depolarization is prominent when the magnetic field strengths in both the SAM and the cocoon are comparable, which is particularly the case for large-scale fields (see Fig.~\ref{fig:dist_Bj}). This happens because the magnetic fields in the SAM undergo decay with expansion (see Appendix~\ref{no_jet_test}), and small-scale fields ($\mathrm{L=100~pc}$) decay faster when compared to large-scale fields. The low field strengths in the SAM are also aided by the lowering of the fields in the external medium with time (Appendix.~\ref{no_jet_test}). Contrarily, the decay in the large-scale fields ($\mathrm{L=500}$ and $\mathrm{1000~pc}$) in the SAM is gradual, thereby causing more depolarization of the radiation from the cocoon.

    \item \textbf{Time evolution of jet, SAM, and depolarization:} As the jet progresses in an external turbulent magnetic field, it injects kinetic and thermal energy into the surrounding medium. However, as the cocoon expands, the magnetic energy density of the jet gradually decreases over time. This leads to a lowering in the observed synchrotron flux although the emitting region is extended to large scales (Fig.~\ref{fig:lbox_44} and~\ref{fig:lbox_44_45deg}); nonetheless, the highest emission can be seen along the spine of the jet and the hot-spots. The increase in the poloidal component in the cocoon by virtue of turbulent backflows and shearing of fields leads to low $\pi_J$ values with time, as also found by \citet{hardcastle_2014}. The adiabatic expansion of the cocoon leads to decay in the magnetic field of the jet. Hence, the strength of the local magnetic fields in the jet dissipates with time and becomes comparable to that in its surroundings, which strengthens the depolarizing effect of the magnetized SAM (Fig.~\ref{fig:pol_lbox}). However, one should note that younger jets are also often situated in dense regions \citep{Odea_1998} where Faraday rotation by  the intervening ISM can also lead to significant depolarization. Our primary objective is to examine the progression of the jet in a magnetized environment and demonstrate that the depolarization caused by the SAM intensifies as the jet expands and its magnetic fields weaken compared to earlier stages.

\item \textbf{Kink instabilities in low-power jets:}
As shown in previous studies \citep{bromberg_2016,tchekhovskoy_2016,dipanjan_2020}, we also find that low-power jets are more susceptible to kink mode instabilities when compared to high-power ones. The kinks lead to slow propagation of the jet and widen the cocoon when compared to the high-power ones (see Sec.~\ref{sec:synch_results}). Moreover, frequent kinks can wiggle the jet's spine in different directions, leading to the formation of multiple hotspots (see Fig.~\ref{fig:P43_L1000}). Such kinks can also change the viewing angle of the observer with respect to the jet over time (see Fig.~\ref{fig:kinks}). Reorientation of the jets has been observed in the low-power FRII radio jets by \citet{thorpe_2002}, which is attributed to a binary black-hole merger or acquisition of a small companion galaxy. Intriguingly, such a change in the direction of the restarted jet has transformed PBC~J2333.9-2343 from a Giant Radio Galaxy into a blazar \citep{garcia_2017}. Our study shows that kink modes can also cause such reorientations of the jet axis.
\end{enumerate}

Thus, our findings indicate that the dynamics and emission characteristics of the jets are strongly influenced by various factors, namely, the power and the magnetic properties of the jet and the external medium. In this paper, we initiated jets within a hot halo that exhibited a radial density decline while being influenced by an external gravitational field. However, in a realistic scenario, the gas in the ISM exhibits a complex fractal gas distribution \citep[see for e.g.][]{wagner12,dipanjan2018}. As a result, it is anticipated that young jets actively interact with the surrounding gas \citep{Breugel_1984,Odea_1998}, thereby influencing their dynamics and observable characteristics. In this work, we have focused on a simplified scenario, and the investigation for a fractal gas distribution is reserved for a later study.

\section*{Acknowledgements}
We thank the anonymous referee for his/her valuable feedback and insightful suggestions, which have contributed to the refinement of this paper. We acknowledge support by the Accordo Quadro INAF-CINECA 2017 and the `Pegasus' HPC facility at IUCAA, Pune, India, for the availability of high-performance computing resources. The authors acknowledge the assistance of the INDO-ITALIAN mobility grant (INT/Italy/P-37/2022(ER)(G)). MM thanks Ankush Mandal for useful discussions throughout the project.

\section*{DATA AVAILABILITY}
The simulations generated for this work will be shared upon a reasonable request to the corresponding authors.


\bibliographystyle{mnras}
\bibliography{manuscript} 

\begin{thebibliography}{}
\makeatletter
\relax
\def\mn@urlcharsother{\let\do\@makeother \do\$\do\&\do\#\do\^\do\_\do\%\do\~}
\def\mn@doi{\begingroup\mn@urlcharsother \@ifnextchar [ {\mn@doi@}
  {\mn@doi@[]}}
\def\mn@doi@[#1]#2{\def\@tempa{#1}\ifx\@tempa\@empty \href
  {http://dx.doi.org/#2} {doi:#2}\else \href {http://dx.doi.org/#2} {#1}\fi
  \endgroup}
\def\mn@eprint#1#2{\mn@eprint@#1:#2::\@nil}
\def\mn@eprint@arXiv#1{\href {http://arxiv.org/abs/#1} {{\tt arXiv:#1}}}
\def\mn@eprint@dblp#1{\href {http://dblp.uni-trier.de/rec/bibtex/#1.xml}
  {dblp:#1}}
\def\mn@eprint@#1:#2:#3:#4\@nil{\def\@tempa {#1}\def\@tempb {#2}\def\@tempc
  {#3}\ifx \@tempc \@empty \let \@tempc \@tempb \let \@tempb \@tempa \fi \ifx
  \@tempb \@empty \def\@tempb {arXiv}\fi \@ifundefined
  {mn@eprint@\@tempb}{\@tempb:\@tempc}{\expandafter \expandafter \csname
  mn@eprint@\@tempb\endcsname \expandafter{\@tempc}}}

\bibitem[\protect\citeauthoryear{{Adebahr}, {Brienza}  \& {Morganti}}{{Adebahr}
  et~al.}{2019}]{adebahr_2019}
{Adebahr} B.,  {Brienza} M.,   {Morganti} R.,  2019, \mn@doi [\aap]
  {10.1051/0004-6361/201833988}, \href
  {https://ui.adsabs.harvard.edu/abs/2019A&A...622A.209A} {622, A209}

\bibitem[\protect\citeauthoryear{{Anderson}, {Gaensler}, {Heald}, {O'Sullivan},
  {Kaczmarek}  \& {Feain}}{{Anderson} et~al.}{2018}]{anderson_2018}
{Anderson} C.~S.,  {Gaensler} B.~M.,  {Heald} G.~H.,  {O'Sullivan} S.~P.,
  {Kaczmarek} J.~F.,   {Feain} I.~J.,  2018, \mn@doi [\apj]
  {10.3847/1538-4357/aaaec0}, \href
  {https://ui.adsabs.harvard.edu/abs/2018ApJ...855...41A} {855, 41}

\bibitem[\protect\citeauthoryear{{Asada}, {Inoue}, {Uchida}, {Kameno},
  {Fujisawa}, {Iguchi}  \& {Mutoh}}{{Asada} et~al.}{2002}]{asada_2002}
{Asada} K.,  {Inoue} M.,  {Uchida} Y.,  {Kameno} S.,  {Fujisawa} K.,  {Iguchi}
  S.,   {Mutoh} M.,  2002, \mn@doi [\pasj] {10.1093/pasj/54.3.L39}, \href
  {https://ui.adsabs.harvard.edu/abs/2002PASJ...54L..39A} {54, L39}

\bibitem[\protect\citeauthoryear{{Baghel}, {Kharb}, {Silpa}, {Ho}  \&
  {Harrison}}{{Baghel} et~al.}{2022}]{baghel_2022}
{Baghel} J.,  {Kharb} P.,  {Silpa} S.,  {Ho} L.~C.,   {Harrison} C.~M.,  2022,
  arXiv e-prints, \href {https://ui.adsabs.harvard.edu/abs/2022arXiv221207061B}
  {p. arXiv:2212.07061}

\bibitem[\protect\citeauthoryear{{Balsara} \& {Spicer}}{{Balsara} \&
  {Spicer}}{1999}]{balsara_1999}
{Balsara} D.~S.,  {Spicer} D.~S.,  1999, \mn@doi [Journal of Computational
  Physics] {10.1006/jcph.1998.6153}, \href
  {https://ui.adsabs.harvard.edu/abs/1999JCoPh.149..270B} {149, 270}

\bibitem[\protect\citeauthoryear{{Bell} \& {Comeau}}{{Bell} \&
  {Comeau}}{2013}]{bell_2013}
{Bell} M.~B.,  {Comeau} S.~P.,  2013, \mn@doi [\apss]
  {10.1007/s10509-012-1310-4}, \href
  {https://ui.adsabs.harvard.edu/abs/2013Ap&SS.344..205B} {344, 205}

\bibitem[\protect\citeauthoryear{{Blandford} \& {Payne}}{{Blandford} \&
  {Payne}}{1982}]{blandford_1982}
{Blandford} R.~D.,  {Payne} D.~G.,  1982, \mn@doi [\mnras]
  {10.1093/mnras/199.4.883}, \href
  {https://ui.adsabs.harvard.edu/abs/1982MNRAS.199..883B} {199, 883}

\bibitem[\protect\citeauthoryear{{Blandford} \& {Znajek}}{{Blandford} \&
  {Znajek}}{1977}]{blandford_1977}
{Blandford} R.~D.,  {Znajek} R.~L.,  1977, \mn@doi [\mnras]
  {10.1093/mnras/179.3.433}, \href
  {https://ui.adsabs.harvard.edu/abs/1977MNRAS.179..433B} {179, 433}

\bibitem[\protect\citeauthoryear{{Bless}}{{Bless}}{1962}]{bless_1962}
{Bless} R.~C.,  1962, \mn@doi [\apj] {10.1086/147257}, \href
  {https://ui.adsabs.harvard.edu/abs/1962ApJ...135..187B} {135, 187}

\bibitem[\protect\citeauthoryear{{Bodo}, {Cattaneo}, {Ferrari}, {Mignone}  \&
  {Rossi}}{{Bodo} et~al.}{2011}]{Bodo_2011}
{Bodo} G.,  {Cattaneo} F.,  {Ferrari} A.,  {Mignone} A.,   {Rossi} P.,  2011,
  \mn@doi [\apj] {10.1088/0004-637X/739/2/82}, \href
  {https://ui.adsabs.harvard.edu/abs/2011ApJ...739...82B} {739, 82}

\bibitem[\protect\citeauthoryear{{Bodo}, {Mamatsashvili}, {Rossi}  \&
  {Mignone}}{{Bodo} et~al.}{2013}]{bodo_2013}
{Bodo} G.,  {Mamatsashvili} G.,  {Rossi} P.,   {Mignone} A.,  2013, \mn@doi
  [\mnras] {10.1093/mnras/stt1225}, \href
  {https://ui.adsabs.harvard.edu/abs/2013MNRAS.434.3030B} {434, 3030}

\bibitem[\protect\citeauthoryear{{Bodo}, {Tavecchio}  \& {Sironi}}{{Bodo}
  et~al.}{2021}]{bodo_2021}
{Bodo} G.,  {Tavecchio} F.,   {Sironi} L.,  2021, \mn@doi [\mnras]
  {10.1093/mnras/staa3620}, \href
  {https://ui.adsabs.harvard.edu/abs/2021MNRAS.501.2836B} {501, 2836}

\bibitem[\protect\citeauthoryear{{B{\"o}hringer}, {Chon}, {Ellis}, {Barrena}
  \& {Laporte}}{{B{\"o}hringer} et~al.}{2022}]{bohringer_2022}
{B{\"o}hringer} H.,  {Chon} G.,  {Ellis} R.~S.,  {Barrena} R.,   {Laporte} N.,
  2022, \mn@doi [\aap] {10.1051/0004-6361/202243424}, \href
  {https://ui.adsabs.harvard.edu/abs/2022A&A...664A..57B} {664, A57}

\bibitem[\protect\citeauthoryear{{Bovino}, {Schleicher}  \& {Schober}}{{Bovino}
  et~al.}{2013}]{bovino_2013}
{Bovino} S.,  {Schleicher} D.~R.~G.,   {Schober} J.,  2013, \mn@doi [New
  Journal of Physics] {10.1088/1367-2630/15/1/013055}, \href
  {https://ui.adsabs.harvard.edu/abs/2013NJPh...15a3055B} {15, 013055}

\bibitem[\protect\citeauthoryear{{Bridle}}{{Bridle}}{1984}]{bridle1984}
{Bridle} A.~H.,  1984, \mn@doi [\aj] {10.1086/113593}, \href
  {https://ui.adsabs.harvard.edu/abs/1984AJ.....89..979B} {89, 979}

\bibitem[\protect\citeauthoryear{{Bridle} \& {Perley}}{{Bridle} \&
  {Perley}}{1984}]{bridle_1984}
{Bridle} A.~H.,  {Perley} R.~A.,  1984, \mn@doi [\araa]
  {10.1146/annurev.aa.22.090184.001535}, \href
  {https://ui.adsabs.harvard.edu/abs/1984ARA&A..22..319B} {22, 319}

\bibitem[\protect\citeauthoryear{{Bridle}, {Hough}, {Lonsdale}, {Burns}  \&
  {Laing}}{{Bridle} et~al.}{1994}]{bridle_1994}
{Bridle} A.~H.,  {Hough} D.~H.,  {Lonsdale} C.~J.,  {Burns} J.~O.,   {Laing}
  R.~A.,  1994, \mn@doi [\aj] {10.1086/117112}, \href
  {https://ui.adsabs.harvard.edu/abs/1994AJ....108..766B} {108, 766}

\bibitem[\protect\citeauthoryear{{Bromberg} \& {Tchekhovskoy}}{{Bromberg} \&
  {Tchekhovskoy}}{2016}]{bromberg_2016}
{Bromberg} O.,  {Tchekhovskoy} A.,  2016, \mn@doi [\mnras]
  {10.1093/mnras/stv2591}, \href
  {https://ui.adsabs.harvard.edu/abs/2016MNRAS.456.1739B} {456, 1739}

\bibitem[\protect\citeauthoryear{{Burbidge}}{{Burbidge}}{1956}]{burbidge_1956}
{Burbidge} G.~R.,  1956, \mn@doi [\apj] {10.1086/146237}, \href
  {https://ui.adsabs.harvard.edu/abs/1956ApJ...124..416B} {124, 416}

\bibitem[\protect\citeauthoryear{{Burn}}{{Burn}}{1966}]{burn_1966}
{Burn} B.~J.,  1966, \mn@doi [MNRAS] {10.1093/mnras/133.1.67}, \href
  {https://ui.adsabs.harvard.edu/abs/1966MNRAS.133...67B} {133, 67}

\bibitem[\protect\citeauthoryear{{Cantwell} et~al.,}{{Cantwell}
  et~al.}{2020}]{cantwell_2020}
{Cantwell} T.~M.,  et~al., 2020, \mn@doi [\mnras] {10.1093/mnras/staa1160},
  \href {https://ui.adsabs.harvard.edu/abs/2020MNRAS.495..143C} {495, 143}

\bibitem[\protect\citeauthoryear{{Chatterjee}, {Liska}, {Tchekhovskoy}  \&
  {Markoff}}{{Chatterjee} et~al.}{2019}]{chatterjee_2019}
{Chatterjee} K.,  {Liska} M.,  {Tchekhovskoy} A.,   {Markoff} S.~B.,  2019,
  \mn@doi [\mnras] {10.1093/mnras/stz2626}, \href
  {https://ui.adsabs.harvard.edu/abs/2019MNRAS.490.2200C} {490, 2200}

\bibitem[\protect\citeauthoryear{{Chen}, {Heinz}  \& {Hooper}}{{Chen}
  et~al.}{2023}]{chen_2023}
{Chen} Y.-H.,  {Heinz} S.,   {Hooper} E.,  2023, \mn@doi [\mnras]
  {10.1093/mnras/stad1074}, \href
  {https://ui.adsabs.harvard.edu/abs/2023MNRAS.522.2850C} {522, 2850}

\bibitem[\protect\citeauthoryear{{Chy{\.z}y}, {We{\.z}gowiec}, {Beck}  \&
  {Bomans}}{{Chy{\.z}y} et~al.}{2011}]{chyzy_2011}
{Chy{\.z}y} K.~T.,  {We{\.z}gowiec} M.,  {Beck} R.,   {Bomans} D.~J.,  2011,
  \mn@doi [\aap] {10.1051/0004-6361/201015393}, \href
  {https://ui.adsabs.harvard.edu/abs/2011A&A...529A..94C} {529, A94}

\bibitem[\protect\citeauthoryear{{Croston} et~al.,}{{Croston}
  et~al.}{2008}]{Croston_2008}
{Croston} J.~H.,  et~al., 2008, \mn@doi [\aap] {10.1051/0004-6361:20079154},
  \href {https://ui.adsabs.harvard.edu/abs/2008A&A...487..431C} {487, 431}

\bibitem[\protect\citeauthoryear{{Del Zanna}, {Volpi}, {Amato}  \&
  {Bucciantini}}{{Del Zanna} et~al.}{2006}]{Zanna_2006}
{Del Zanna} L.,  {Volpi} D.,  {Amato} E.,   {Bucciantini} N.,  2006, \mn@doi
  [\aap] {10.1051/0004-6361:20064858}, \href
  {https://ui.adsabs.harvard.edu/abs/2006A&A...453..621D} {453, 621}

\bibitem[\protect\citeauthoryear{{Dennett-Thorpe}, {Scheuer}, {Laing},
  {Bridle}, {Pooley}  \& {Reich}}{{Dennett-Thorpe} et~al.}{2002}]{thorpe_2002}
{Dennett-Thorpe} J.,  {Scheuer} P.~A.~G.,  {Laing} R.~A.,  {Bridle} A.~H.,
  {Pooley} G.~G.,   {Reich} W.,  2002, \mn@doi [\mnras]
  {10.1046/j.1365-8711.2002.05106.x}, \href
  {https://ui.adsabs.harvard.edu/abs/2002MNRAS.330..609D} {330, 609}

\bibitem[\protect\citeauthoryear{{Dom{\'\i}nguez-Fern{\'a}ndez}
  et~al.,}{{Dom{\'\i}nguez-Fern{\'a}ndez} et~al.}{2021}]{fernandez_2021}
{Dom{\'\i}nguez-Fern{\'a}ndez} P.,  et~al., 2021, \mn@doi [\mnras]
  {10.1093/mnras/stab2353}, \href
  {https://ui.adsabs.harvard.edu/abs/2021MNRAS.507.2714D} {507, 2714}

\bibitem[\protect\citeauthoryear{{Dreher} \& {Feigelson}}{{Dreher} \&
  {Feigelson}}{1984}]{dreher_1984}
{Dreher} J.~W.,  {Feigelson} E.~D.,  1984, \mn@doi [\nat] {10.1038/308043a0},
  \href {https://ui.adsabs.harvard.edu/abs/1984Natur.308...43D} {308, 43}

\bibitem[\protect\citeauthoryear{{Federrath}}{{Federrath}}{2016}]{federrath_2016}
{Federrath} C.,  2016, \mn@doi [Journal of Plasma Physics]
  {10.1017/S0022377816001069}, \href
  {https://ui.adsabs.harvard.edu/abs/2016JPlPh..82f5301F} {82, 535820601}

\bibitem[\protect\citeauthoryear{{Gaibler}, {Krause}  \& {Camenzind}}{{Gaibler}
  et~al.}{2009}]{Gaibler_2009}
{Gaibler} V.,  {Krause} M.,   {Camenzind} M.,  2009, \mn@doi [\mnras]
  {10.1111/j.1365-2966.2009.15625.x}, \href
  {https://ui.adsabs.harvard.edu/abs/2009MNRAS.400.1785G} {400, 1785}

\bibitem[\protect\citeauthoryear{{Gardiner} \& {Stone}}{{Gardiner} \&
  {Stone}}{2005}]{gardiner_2005}
{Gardiner} T.~A.,  {Stone} J.~M.,  2005, \mn@doi [Journal of Computational
  Physics] {10.1016/j.jcp.2004.11.016}, \href
  {https://ui.adsabs.harvard.edu/abs/2005JCoPh.205..509G} {205, 509}

\bibitem[\protect\citeauthoryear{{Garrington}, {Leahy}, {Conway}  \&
  {Laing}}{{Garrington} et~al.}{1988}]{garrington_1988}
{Garrington} S.~T.,  {Leahy} J.~P.,  {Conway} R.~G.,   {Laing} R.~A.,  1988,
  \mn@doi [\nat] {10.1038/331147a0}, \href
  {https://ui.adsabs.harvard.edu/abs/1988Natur.331..147G} {331, 147}

\bibitem[\protect\citeauthoryear{{Ginzburg} \& {Syrovatskii}}{{Ginzburg} \&
  {Syrovatskii}}{1965}]{Ginzburg_1965}
{Ginzburg} V.~L.,  {Syrovatskii} S.~I.,  1965, \mn@doi [\araa]
  {10.1146/annurev.aa.03.090165.001501}, \href
  {https://ui.adsabs.harvard.edu/abs/1965ARA&A...3..297G} {3, 297}

\bibitem[\protect\citeauthoryear{{Gizani} \& {Leahy}}{{Gizani} \&
  {Leahy}}{2003}]{gizani_2003}
{Gizani} N. A.~B.,  {Leahy} J.~P.,  2003, \mn@doi [\mnras]
  {10.1046/j.1365-8711.2003.06469.x}, \href
  {https://ui.adsabs.harvard.edu/abs/2003MNRAS.342..399G} {342, 399}

\bibitem[\protect\citeauthoryear{{Goodlet}, {Kaiser}, {Best}  \&
  {Dennett-Thorpe}}{{Goodlet} et~al.}{2004}]{goodlet_2004}
{Goodlet} J.~A.,  {Kaiser} C.~R.,  {Best} P.~N.,   {Dennett-Thorpe} J.,  2004,
  \mn@doi [\mnras] {10.1111/j.1365-2966.2004.07225.x}, \href
  {https://ui.adsabs.harvard.edu/abs/2004MNRAS.347..508G} {347, 508}

\bibitem[\protect\citeauthoryear{{Govoni}, {Murgia}, {Feretti}, {Giovannini},
  {Dolag}  \& {Taylor}}{{Govoni} et~al.}{2006}]{govoni_2006}
{Govoni} F.,  {Murgia} M.,  {Feretti} L.,  {Giovannini} G.,  {Dolag} K.,
  {Taylor} G.~B.,  2006, \mn@doi [\aap] {10.1051/0004-6361:20065964}, \href
  {https://ui.adsabs.harvard.edu/abs/2006A&A...460..425G} {460, 425}

\bibitem[\protect\citeauthoryear{{Guidetti}, {Murgia}, {Govoni}, {Parma},
  {Gregorini}, {de Ruiter}, {Cameron}  \& {Fanti}}{{Guidetti}
  et~al.}{2008}]{Guidetti_2008}
{Guidetti} D.,  {Murgia} M.,  {Govoni} F.,  {Parma} P.,  {Gregorini} L.,  {de
  Ruiter} H.~R.,  {Cameron} R.~A.,   {Fanti} R.,  2008, \mn@doi [\aap]
  {10.1051/0004-6361:20078576}, \href
  {https://ui.adsabs.harvard.edu/abs/2008A&A...483..699G} {483, 699}

\bibitem[\protect\citeauthoryear{{Guidetti}, {Laing}, {Murgia}, {Govoni},
  {Gregorini}  \& {Parma}}{{Guidetti} et~al.}{2010}]{guidetti_2010}
{Guidetti} D.,  {Laing} R.~A.,  {Murgia} M.,  {Govoni} F.,  {Gregorini} L.,
  {Parma} P.,  2010, \mn@doi [\aap] {10.1051/0004-6361/200913872}, \href
  {https://ui.adsabs.harvard.edu/abs/2010A&A...514A..50G} {514, A50}

\bibitem[\protect\citeauthoryear{{Guidetti}, {Laing}, {Bridle}, {Parma}  \&
  {Gregorini}}{{Guidetti} et~al.}{2011}]{guidetti_2011}
{Guidetti} D.,  {Laing} R.~A.,  {Bridle} A.~H.,  {Parma} P.,   {Gregorini} L.,
  2011, \mn@doi [\mnras] {10.1111/j.1365-2966.2011.18321.x}, \href
  {https://ui.adsabs.harvard.edu/abs/2011MNRAS.413.2525G} {413, 2525}

\bibitem[\protect\citeauthoryear{{Hammond}, {Robishaw}  \&
  {Gaensler}}{{Hammond} et~al.}{2012}]{hammond_2012}
{Hammond} A.~M.,  {Robishaw} T.,   {Gaensler} B.~M.,  2012, arXiv e-prints,
  \href {https://ui.adsabs.harvard.edu/abs/2012arXiv1209.1438H} {p.
  arXiv:1209.1438}

\bibitem[\protect\citeauthoryear{{Hardcastle} \& {Krause}}{{Hardcastle} \&
  {Krause}}{2014}]{hardcastle_2014}
{Hardcastle} M.~J.,  {Krause} M.~G.~H.,  2014, \mn@doi [\mnras]
  {10.1093/mnras/stu1229}, \href
  {https://ui.adsabs.harvard.edu/abs/2014MNRAS.443.1482H} {443, 1482}

\bibitem[\protect\citeauthoryear{{Hardcastle}, {Alexander}, {Pooley}  \&
  {Riley}}{{Hardcastle} et~al.}{1996}]{hardcastle_1996}
{Hardcastle} M.~J.,  {Alexander} P.,  {Pooley} G.~G.,   {Riley} J.~M.,  1996,
  \mn@doi [\mnras] {10.1093/mnras/278.1.273}, \href
  {https://ui.adsabs.harvard.edu/abs/1996MNRAS.278..273H} {278, 273}

\bibitem[\protect\citeauthoryear{{Hern{\'a}ndez-Garc{\'\i}a}
  et~al.,}{{Hern{\'a}ndez-Garc{\'\i}a} et~al.}{2017}]{garcia_2017}
{Hern{\'a}ndez-Garc{\'\i}a} L.,  et~al., 2017, \mn@doi [\aap]
  {10.1051/0004-6361/201730530}, \href
  {https://ui.adsabs.harvard.edu/abs/2017A&A...603A.131H} {603, A131}

\bibitem[\protect\citeauthoryear{{Hernquist}}{{Hernquist}}{1990}]{Hernquist_1990}
{Hernquist} L.,  1990, \mn@doi [\apj] {10.1086/168845}, \href
  {https://ui.adsabs.harvard.edu/abs/1990ApJ...356..359H} {356, 359}

\bibitem[\protect\citeauthoryear{{Horton}, {Krause}  \& {Hardcastle}}{{Horton}
  et~al.}{2023}]{horton_2023}
{Horton} M.~A.,  {Krause} M. G.~H.,   {Hardcastle} M.~J.,  2023, arXiv
  e-prints, \href {https://ui.adsabs.harvard.edu/abs/2023arXiv230214023H} {p.
  arXiv:2302.14023}

\bibitem[\protect\citeauthoryear{{Hovatta}, {Valtaoja}, {Tornikoski}  \&
  {L{\"a}hteenm{\"a}ki}}{{Hovatta} et~al.}{2009}]{hovatta_2009}
{Hovatta} T.,  {Valtaoja} E.,  {Tornikoski} M.,   {L{\"a}hteenm{\"a}ki} A.,
  2009, \mn@doi [\aap] {10.1051/0004-6361:200811150}, \href
  {https://ui.adsabs.harvard.edu/abs/2009A&A...494..527H} {494, 527}

\bibitem[\protect\citeauthoryear{{Huarte-Espinosa}, {Krause}  \&
  {Alexander}}{{Huarte-Espinosa} et~al.}{2011a}]{huarte_2011a}
{Huarte-Espinosa} M.,  {Krause} M.,   {Alexander} P.,  2011a, \mn@doi [\mnras]
  {10.1111/j.1365-2966.2011.19271.x}, \href
  {https://ui.adsabs.harvard.edu/abs/2011MNRAS.417..382H} {417, 382}

\bibitem[\protect\citeauthoryear{{Huarte-Espinosa}, {Krause}  \&
  {Alexander}}{{Huarte-Espinosa} et~al.}{2011b}]{huarte_2011b}
{Huarte-Espinosa} M.,  {Krause} M.,   {Alexander} P.,  2011b, \mn@doi [\mnras]
  {10.1111/j.1365-2966.2011.19545.x}, \href
  {https://ui.adsabs.harvard.edu/abs/2011MNRAS.418.1621H} {418, 1621}

\bibitem[\protect\citeauthoryear{{Jarvis} et~al.,}{{Jarvis}
  et~al.}{2019}]{jarvis_2019}
{Jarvis} M.~E.,  et~al., 2019, \mn@doi [\mnras] {10.1093/mnras/stz556}, \href
  {https://ui.adsabs.harvard.edu/abs/2019MNRAS.485.2710J} {485, 2710}

\bibitem[\protect\citeauthoryear{{Jarvis} et~al.,}{{Jarvis}
  et~al.}{2021}]{jarvis_2021}
{Jarvis} M.~E.,  et~al., 2021, \mn@doi [\mnras] {10.1093/mnras/stab549}, \href
  {https://ui.adsabs.harvard.edu/abs/2021MNRAS.503.1780J} {503, 1780}

\bibitem[\protect\citeauthoryear{{Kaiser} \& {Alexander}}{{Kaiser} \&
  {Alexander}}{1997}]{kaiser_1997}
{Kaiser} C.~R.,  {Alexander} P.,  1997, \mn@doi [\mnras]
  {10.1093/mnras/286.1.215}, \href
  {https://ui.adsabs.harvard.edu/abs/1997MNRAS.286..215K} {286, 215}

\bibitem[\protect\citeauthoryear{{Kepley}, {M{\"u}hle}, {Everett}, {Zweibel},
  {Wilcots}  \& {Klein}}{{Kepley} et~al.}{2010}]{kepley_2010}
{Kepley} A.~A.,  {M{\"u}hle} S.,  {Everett} J.,  {Zweibel} E.~G.,  {Wilcots}
  E.~M.,   {Klein} U.,  2010, \mn@doi [\apj] {10.1088/0004-637X/712/1/536},
  \href {https://ui.adsabs.harvard.edu/abs/2010ApJ...712..536K} {712, 536}

\bibitem[\protect\citeauthoryear{{Kharb}, {O'Dea}, {Baum}, {Daly}, {Mory},
  {Donahue}  \& {Guerra}}{{Kharb} et~al.}{2008}]{kharb_2008}
{Kharb} P.,  {O'Dea} C.~P.,  {Baum} S.~A.,  {Daly} R.~A.,  {Mory} M.~P.,
  {Donahue} M.,   {Guerra} E.~J.,  2008, \mn@doi [\apjs] {10.1086/520840},
  \href {https://ui.adsabs.harvard.edu/abs/2008ApJS..174...74K} {174, 74}

\bibitem[\protect\citeauthoryear{{Knuettel}, {O'Sullivan}, {Curiel}  \&
  {Emonts}}{{Knuettel} et~al.}{2019}]{knuettel_2019}
{Knuettel} S.,  {O'Sullivan} S.~P.,  {Curiel} S.,   {Emonts} B.~H.~C.,  2019,
  \mn@doi [\mnras] {10.1093/mnras/sty3018}, \href
  {https://ui.adsabs.harvard.edu/abs/2019MNRAS.482.4606K} {482, 4606}

\bibitem[\protect\citeauthoryear{{Kollgaard}, {Wardle}  \&
  {Roberts}}{{Kollgaard} et~al.}{1990}]{kollgaard_1990}
{Kollgaard} R.~I.,  {Wardle} J.~F.~C.,   {Roberts} D.~H.,  1990, \mn@doi [\aj]
  {10.1086/115579}, \href
  {https://ui.adsabs.harvard.edu/abs/1990AJ....100.1057K} {100, 1057}

\bibitem[\protect\citeauthoryear{{Komissarov} \& {Falle}}{{Komissarov} \&
  {Falle}}{1998}]{komissarov_1998}
{Komissarov} S.~S.,  {Falle} S.~A.~E.~G.,  1998, \mn@doi [\mnras]
  {10.1046/j.1365-8711.1998.01547.x}, \href
  {https://ui.adsabs.harvard.edu/abs/1998MNRAS.297.1087K} {297, 1087}

\bibitem[\protect\citeauthoryear{{Kondapally} et~al.,}{{Kondapally}
  et~al.}{2023}]{kondapally_2023}
{Kondapally} R.,  et~al., 2023, \mn@doi [\mnras] {10.1093/mnras/stad1813},
  \href {https://ui.adsabs.harvard.edu/abs/2023MNRAS.523.5292K} {523, 5292}

\bibitem[\protect\citeauthoryear{{Kraft} et~al.,}{{Kraft}
  et~al.}{2012}]{kraft_2012}
{Kraft} R.~P.,  et~al., 2012, \mn@doi [\apj] {10.1088/0004-637X/749/1/19},
  \href {https://ui.adsabs.harvard.edu/abs/2012ApJ...749...19K} {749, 19}

\bibitem[\protect\citeauthoryear{{Laing}}{{Laing}}{1980}]{laing_1980}
{Laing} R.~A.,  1980, \mn@doi [\mnras] {10.1093/mnras/193.3.439}, \href
  {https://ui.adsabs.harvard.edu/abs/1980MNRAS.193..439L} {193, 439}

\bibitem[\protect\citeauthoryear{{Laing}}{{Laing}}{1988}]{laing_1988}
{Laing} R.~A.,  1988, \mn@doi [\nat] {10.1038/331149a0}, \href
  {https://ui.adsabs.harvard.edu/abs/1988Natur.331..149L} {331, 149}

\bibitem[\protect\citeauthoryear{{Laing}, {Bridle}, {Parma}  \&
  {Murgia}}{{Laing} et~al.}{2008}]{laing_2008}
{Laing} R.~A.,  {Bridle} A.~H.,  {Parma} P.,   {Murgia} M.,  2008, \mn@doi
  [\mnras] {10.1111/j.1365-2966.2008.13895.x}, \href
  {https://ui.adsabs.harvard.edu/abs/2008MNRAS.391..521L} {391, 521}

\bibitem[\protect\citeauthoryear{{Laor}}{{Laor}}{2000}]{laor_2000}
{Laor} A.,  2000, \mn@doi [\apjl] {10.1086/317280}, \href
  {https://ui.adsabs.harvard.edu/abs/2000ApJ...543L.111L} {543, L111}

\bibitem[\protect\citeauthoryear{{Leahy} \& {Perley}}{{Leahy} \&
  {Perley}}{1995}]{leahy_1995}
{Leahy} J.~P.,  {Perley} R.~A.,  1995, \mn@doi [\mnras]
  {10.1093/mnras/277.3.1097}, \href
  {https://ui.adsabs.harvard.edu/abs/1995MNRAS.277.1097L} {277, 1097}

\bibitem[\protect\citeauthoryear{{Leahy}, {Pooley}  \& {Riley}}{{Leahy}
  et~al.}{1986}]{leahy_1986}
{Leahy} J.~P.,  {Pooley} G.~G.,   {Riley} J.~M.,  1986, \mn@doi [\mnras]
  {10.1093/mnras/222.4.753}, \href
  {https://ui.adsabs.harvard.edu/abs/1986MNRAS.222..753L} {222, 753}

\bibitem[\protect\citeauthoryear{{Leahy}, {Black}, {Dennett-Thorpe},
  {Hardcastle}, {Komissarov}, {Perley}, {Riley}  \& {Scheuer}}{{Leahy}
  et~al.}{1997}]{leahy_1997}
{Leahy} J.~P.,  {Black} A.~R.~S.,  {Dennett-Thorpe} J.,  {Hardcastle} M.~J.,
  {Komissarov} S.,  {Perley} R.~A.,  {Riley} J.~M.,   {Scheuer} P.~A.~G.,
  1997, \mn@doi [\mnras] {10.1093/mnras/291.1.20}, \href
  {https://ui.adsabs.harvard.edu/abs/1997MNRAS.291...20L} {291, 20}

\bibitem[\protect\citeauthoryear{{Livingston}, {McClure-Griffiths}, {Mao},
  {Ma}, {Gaensler}, {Heald}  \& {Seta}}{{Livingston}
  et~al.}{2022}]{livingston_2022}
{Livingston} J.~D.,  {McClure-Griffiths} N.~M.,  {Mao} S.~A.,  {Ma} Y.~K.,
  {Gaensler} B.~M.,  {Heald} G.,   {Seta} A.,  2022, \mn@doi [\mnras]
  {10.1093/mnras/stab3375}, \href
  {https://ui.adsabs.harvard.edu/abs/2022MNRAS.510..260L} {510, 260}

\bibitem[\protect\citeauthoryear{{Lyutikov}, {Pariev}  \&
  {Blandford}}{{Lyutikov} et~al.}{2003}]{lyutikov_2003}
{Lyutikov} M.,  {Pariev} V.~I.,   {Blandford} R.~D.,  2003, \mn@doi [\apj]
  {10.1086/378497}, \href
  {https://ui.adsabs.harvard.edu/abs/2003ApJ...597..998L} {597, 998}

\bibitem[\protect\citeauthoryear{{Mahatma}, {Basu}, {Hardcastle}, {Morabito}
  \& {van Weeren}}{{Mahatma} et~al.}{2023}]{mahatma_2023}
{Mahatma} V.~H.,  {Basu} A.,  {Hardcastle} M.~J.,  {Morabito} L.~K.,   {van
  Weeren} R.~J.,  2023, \mn@doi [arXiv e-prints] {10.48550/arXiv.2302.01357},
  \href {https://ui.adsabs.harvard.edu/abs/2023arXiv230201357M} {p.
  arXiv:2302.01357}

\bibitem[\protect\citeauthoryear{{Mart{\'\i}}, {M{\"u}ller}, {Font},
  {Ib{\'a}{\~n}ez}  \& {Marquina}}{{Mart{\'\i}} et~al.}{1997}]{marti_1997}
{Mart{\'\i}} J.~M.,  {M{\"u}ller} E.,  {Font} J.~A.,  {Ib{\'a}{\~n}ez}
  J.~M.~Z.,   {Marquina} A.,  1997, \mn@doi [\apj] {10.1086/303842}, \href
  {https://ui.adsabs.harvard.edu/abs/1997ApJ...479..151M} {479, 151}

\bibitem[\protect\citeauthoryear{{Massaglia}, {Bodo}, {Rossi}, {Capetti}  \&
  {Mignone}}{{Massaglia} et~al.}{2022}]{massaglia_2022}
{Massaglia} S.,  {Bodo} G.,  {Rossi} P.,  {Capetti} A.,   {Mignone} A.,  2022,
  \mn@doi [\aap] {10.1051/0004-6361/202038724}, \href
  {https://ui.adsabs.harvard.edu/abs/2022A&A...659A.139M} {659, A139}

\bibitem[\protect\citeauthoryear{{Matthews} \& {Scheuer}}{{Matthews} \&
  {Scheuer}}{1990}]{matthews_1990}
{Matthews} A.~P.,  {Scheuer} P.~A.~G.,  1990, \mn@doi [MNRAS]
  {10.1093/mnras/242.4.623}, \href
  {https://ui.adsabs.harvard.edu/abs/1990MNRAS.242..623M} {242, 623}

\bibitem[\protect\citeauthoryear{{Matthews}, {Bell}  \& {Blundell}}{{Matthews}
  et~al.}{2020}]{matthews_2020}
{Matthews} J.~H.,  {Bell} A.~R.,   {Blundell} K.~M.,  2020, \mn@doi [\nar]
  {10.1016/j.newar.2020.101543}, \href
  {https://ui.adsabs.harvard.edu/abs/2020NewAR..8901543M} {89, 101543}

\bibitem[\protect\citeauthoryear{{McBride}, {Quataert}, {Heiles}  \&
  {Bauermeister}}{{McBride} et~al.}{2014}]{mcbride_2014}
{McBride} J.,  {Quataert} E.,  {Heiles} C.,   {Bauermeister} A.,  2014, \mn@doi
  [\apj] {10.1088/0004-637X/780/2/182}, \href
  {https://ui.adsabs.harvard.edu/abs/2014ApJ...780..182M} {780, 182}

\bibitem[\protect\citeauthoryear{{Meenakshi} et~al.,}{{Meenakshi}
  et~al.}{2022}]{meenakshi_2022}
{Meenakshi} M.,  et~al., 2022, \mn@doi [\mnras] {10.1093/mnras/stac2251}, \href
  {https://ui.adsabs.harvard.edu/abs/2022MNRAS.516..766M} {516, 766}

\bibitem[\protect\citeauthoryear{{Mignone} \& {Del Zanna}}{{Mignone} \& {Del
  Zanna}}{2021}]{mignone_2021}
{Mignone} A.,  {Del Zanna} L.,  2021, \mn@doi [Journal of Computational
  Physics] {10.1016/j.jcp.2020.109748}, \href
  {https://ui.adsabs.harvard.edu/abs/2021JCoPh.42409748M} {424, 109748}

\bibitem[\protect\citeauthoryear{{Mignone}, {Plewa}  \& {Bodo}}{{Mignone}
  et~al.}{2005}]{mignone_2005}
{Mignone} A.,  {Plewa} T.,   {Bodo} G.,  2005, \mn@doi [\apjs]
  {10.1086/430905}, \href
  {https://ui.adsabs.harvard.edu/abs/2005ApJS..160..199M} {160, 199}

\bibitem[\protect\citeauthoryear{{Mignone}, {Bodo}, {Massaglia}, {Matsakos},
  {Tesileanu}, {Zanni}  \& {Ferrari}}{{Mignone} et~al.}{2007}]{mignone_2007}
{Mignone} A.,  {Bodo} G.,  {Massaglia} S.,  {Matsakos} T.,  {Tesileanu} O.,
  {Zanni} C.,   {Ferrari} A.,  2007, \mn@doi [\apjs] {10.1086/513316}, \href
  {https://ui.adsabs.harvard.edu/abs/2007ApJS..170..228M} {170, 228}

\bibitem[\protect\citeauthoryear{{Mignone}, {Ugliano}  \& {Bodo}}{{Mignone}
  et~al.}{2009}]{mignone_2009}
{Mignone} A.,  {Ugliano} M.,   {Bodo} G.,  2009, \mn@doi [\mnras]
  {10.1111/j.1365-2966.2008.14221.x}, \href
  {https://ui.adsabs.harvard.edu/abs/2009MNRAS.393.1141M} {393, 1141}

\bibitem[\protect\citeauthoryear{{Mignone}, {Rossi}, {Bodo}, {Ferrari}  \&
  {Massaglia}}{{Mignone} et~al.}{2010}]{mignone_2010}
{Mignone} A.,  {Rossi} P.,  {Bodo} G.,  {Ferrari} A.,   {Massaglia} S.,  2010,
  \mn@doi [\mnras] {10.1111/j.1365-2966.2009.15642.x}, \href
  {https://ui.adsabs.harvard.edu/abs/2010MNRAS.402....7M} {402, 7}

\bibitem[\protect\citeauthoryear{{Morganti}, {Parma}, {Capetti}, {Fanti}, {de
  Ruiter}  \& {Prandoni}}{{Morganti} et~al.}{1997a}]{morganti_1997}
{Morganti} R.,  {Parma} P.,  {Capetti} A.,  {Fanti} R.,  {de Ruiter} H.~R.,
  {Prandoni} I.,  1997a, \mn@doi [\aaps] {10.1051/aas:1997269}, \href
  {https://ui.adsabs.harvard.edu/abs/1997A&AS..126..335M} {126, 335}

\bibitem[\protect\citeauthoryear{{Morganti}, {Parma}, {Capetti}, {Fanti}  \&
  {de Ruiter}}{{Morganti} et~al.}{1997b}]{morganti_1997a}
{Morganti} R.,  {Parma} P.,  {Capetti} A.,  {Fanti} R.,   {de Ruiter} H.~R.,
  1997b, \aap, \href {https://ui.adsabs.harvard.edu/abs/1997A&A...326..919M}
  {326, 919}

\bibitem[\protect\citeauthoryear{{Mukherjee}, {Bicknell}, {Wagner},
  {Sutherland}  \& {Silk}}{{Mukherjee} et~al.}{2018}]{dipanjan2018}
{Mukherjee} D.,  {Bicknell} G.~V.,  {Wagner} A. e.~Y.,  {Sutherland} R.~S.,
  {Silk} J.,  2018, \mn@doi [\mnras] {10.1093/mnras/sty1776}, \href
  {https://ui.adsabs.harvard.edu/abs/2018MNRAS.479.5544M} {479, 5544}

\bibitem[\protect\citeauthoryear{{Mukherjee}, {Bodo}, {Mignone}, {Rossi}  \&
  {Vaidya}}{{Mukherjee} et~al.}{2020}]{dipanjan_2020}
{Mukherjee} D.,  {Bodo} G.,  {Mignone} A.,  {Rossi} P.,   {Vaidya} B.,  2020,
  \mn@doi [\mnras] {10.1093/mnras/staa2934}, \href
  {https://ui.adsabs.harvard.edu/abs/2020MNRAS.499..681M} {499, 681}

\bibitem[\protect\citeauthoryear{{Mullin}, {Hardcastle}  \& {Riley}}{{Mullin}
  et~al.}{2006}]{mullin_2006}
{Mullin} L.~M.,  {Hardcastle} M.~J.,   {Riley} J.~M.,  2006, \mn@doi [\mnras]
  {10.1111/j.1365-2966.2006.10763.x}, \href
  {https://ui.adsabs.harvard.edu/abs/2006MNRAS.372..113M} {372, 113}

\bibitem[\protect\citeauthoryear{{Murgia}, {Govoni}, {Feretti}, {Giovannini},
  {Dallacasa}, {Fanti}, {Taylor}  \& {Dolag}}{{Murgia}
  et~al.}{2004}]{murgia_2004}
{Murgia} M.,  {Govoni} F.,  {Feretti} L.,  {Giovannini} G.,  {Dallacasa} D.,
  {Fanti} R.,  {Taylor} G.~B.,   {Dolag} K.,  2004, \mn@doi [\aap]
  {10.1051/0004-6361:20040191}, \href
  {https://ui.adsabs.harvard.edu/abs/2004A&A...424..429M} {424, 429}

\bibitem[\protect\citeauthoryear{{Navarro}, {Frenk}  \& {White}}{{Navarro}
  et~al.}{1996}]{NFW_1996}
{Navarro} J.~F.,  {Frenk} C.~S.,   {White} S. D.~M.,  1996, \mn@doi [\apj]
  {10.1086/177173}, \href
  {https://ui.adsabs.harvard.edu/abs/1996ApJ...462..563N} {462, 563}

\bibitem[\protect\citeauthoryear{{O'Dea}}{{O'Dea}}{1998}]{Odea_1998}
{O'Dea} C.~P.,  1998, \mn@doi [\pasp] {10.1086/316162}, \href
  {https://ui.adsabs.harvard.edu/abs/1998PASP..110..493O} {110, 493}

\bibitem[\protect\citeauthoryear{{O'Dea} \& {Saikia}}{{O'Dea} \&
  {Saikia}}{2021}]{odea_2021}
{O'Dea} C.~P.,  {Saikia} D.~J.,  2021, \mn@doi [\aapr]
  {10.1007/s00159-021-00131-w}, \href
  {https://ui.adsabs.harvard.edu/abs/2021A&ARv..29....3O} {29, 3}

\bibitem[\protect\citeauthoryear{{O'Sullivan} et~al.,}{{O'Sullivan}
  et~al.}{2013}]{Sullivan_2013}
{O'Sullivan} S.~P.,  et~al., 2013, \mn@doi [\apj]
  {10.1088/0004-637X/764/2/162}, \href
  {https://ui.adsabs.harvard.edu/abs/2013ApJ...764..162O} {764, 162}

\bibitem[\protect\citeauthoryear{{Perley}, {Dreher}  \& {Cowan}}{{Perley}
  et~al.}{1984}]{perley_1984}
{Perley} R.~A.,  {Dreher} J.~W.,   {Cowan} J.~J.,  1984, \mn@doi [\apjl]
  {10.1086/184360}, \href
  {https://ui.adsabs.harvard.edu/abs/1984ApJ...285L..35P} {285, L35}

\bibitem[\protect\citeauthoryear{{Perlman}, {Clautice}, {Avachat}, {Cara},
  {Sparks}, {Georganopoulos}  \& {Meyer}}{{Perlman}
  et~al.}{2020}]{perlman_2020}
{Perlman} E.~S.,  {Clautice} D.,  {Avachat} S.,  {Cara} M.,  {Sparks} W.~B.,
  {Georganopoulos} M.,   {Meyer} E.,  2020, \mn@doi [Galaxies]
  {10.3390/galaxies8040071}, \href
  {https://ui.adsabs.harvard.edu/abs/2020Galax...8...71P} {8, 71}

\bibitem[\protect\citeauthoryear{{Perucho}, {Mart{\'\i}}, {Quilis}  \&
  {Ricciardelli}}{{Perucho} et~al.}{2014}]{perucho_2014}
{Perucho} M.,  {Mart{\'\i}} J.-M.,  {Quilis} V.,   {Ricciardelli} E.,  2014,
  \mn@doi [\mnras] {10.1093/mnras/stu1828}, \href
  {https://ui.adsabs.harvard.edu/abs/2014MNRAS.445.1462P} {445, 1462}

\bibitem[\protect\citeauthoryear{{Perucho}, {Mart{\'\i}}  \&
  {Quilis}}{{Perucho} et~al.}{2019}]{perucho_2019}
{Perucho} M.,  {Mart{\'\i}} J.-M.,   {Quilis} V.,  2019, \mn@doi [\mnras]
  {10.1093/mnras/sty2912}, \href
  {https://ui.adsabs.harvard.edu/abs/2019MNRAS.482.3718P} {482, 3718}

\bibitem[\protect\citeauthoryear{{Perucho}, {Mart{\'\i}}  \&
  {Quilis}}{{Perucho} et~al.}{2022}]{perucho_2022}
{Perucho} M.,  {Mart{\'\i}} J.-M.,   {Quilis} V.,  2022, \mn@doi [\mnras]
  {10.1093/mnras/stab3560}, \href
  {https://ui.adsabs.harvard.edu/abs/2022MNRAS.510.2084P} {510, 2084}

\bibitem[\protect\citeauthoryear{{Perucho}, {L{\'o}pez-Miralles}, {Gizani},
  {Mart{\'\i}}  \& {Boccardi}}{{Perucho} et~al.}{2023}]{perucho_2023}
{Perucho} M.,  {L{\'o}pez-Miralles} J.,  {Gizani} N. A.~B.,  {Mart{\'\i}}
  J.~M.,   {Boccardi} B.,  2023, \mn@doi [\mnras] {10.1093/mnras/stad1640},
  \href {https://ui.adsabs.harvard.edu/abs/2023MNRAS.523.3583P} {523, 3583}

\bibitem[\protect\citeauthoryear{{Planck Collaboration}}{{Planck
  Collaboration}}{2016}]{planck_2015}
{Planck Collaboration} 2016, \mn@doi [A\&A] {10.1051/0004-6361/201525830}, 594,
  A13

\bibitem[\protect\citeauthoryear{Pushkarev, Kovalev, Lister, Savolainen, Aller,
  Aller  \& Hodge}{Pushkarev et~al.}{2017}]{pushkarev_2017}
Pushkarev A.~B.,  Kovalev Y.~Y.,  Lister M.~L.,  Savolainen T.,  Aller M.~F.,
  Aller H.~D.,   Hodge M.~A.,  2017, Galaxies, 5

\bibitem[\protect\citeauthoryear{{Rossi}, {Mignone}, {Bodo}, {Massaglia}  \&
  {Ferrari}}{{Rossi} et~al.}{2008}]{rossi_2008}
{Rossi} P.,  {Mignone} A.,  {Bodo} G.,  {Massaglia} S.,   {Ferrari} A.,  2008,
  \mn@doi [\aap] {10.1051/0004-6361:200809687}, \href
  {https://ui.adsabs.harvard.edu/abs/2008A&A...488..795R} {488, 795}

\bibitem[\protect\citeauthoryear{{Roy}, {Pramesh Rao}  \& {Subrahmanyan}}{{Roy}
  et~al.}{2003}]{roy_2003}
{Roy} S.,  {Pramesh Rao} A.,   {Subrahmanyan} R.,  2003, \mn@doi [Astronomische
  Nachrichten Supplement] {10.1002/asna.200385115}, \href
  {https://ui.adsabs.harvard.edu/abs/2003ANS...324...41R} {324, 41}

\bibitem[\protect\citeauthoryear{{Rybicki} \& {Lightman}}{{Rybicki} \&
  {Lightman}}{1979}]{rybicki_1979}
{Rybicki} G.~B.,  {Lightman} A.~P.,  1979, {Radiative processes in
  astrophysics}.
John Wiley \& Sons, Inc.

\bibitem[\protect\citeauthoryear{{Saikia}}{{Saikia}}{2022}]{Saikia_2022}
{Saikia} D.~J.,  2022, \mn@doi [Journal of Astrophysics and Astronomy]
  {10.1007/s12036-022-09863-2}, \href
  {https://ui.adsabs.harvard.edu/abs/2022JApA...43...97S} {43, 97}

\bibitem[\protect\citeauthoryear{{Saikia} \& {Salter}}{{Saikia} \&
  {Salter}}{1988}]{saikia_1988}
{Saikia} D.~J.,  {Salter} C.~J.,  1988, \mn@doi [\araa]
  {10.1146/annurev.aa.26.090188.000521}, \href
  {https://ui.adsabs.harvard.edu/abs/1988ARA&A..26...93S} {26, 93}

\bibitem[\protect\citeauthoryear{{Saxton}, {Bicknell}  \&
  {Sutherland}}{{Saxton} et~al.}{2002}]{saxton_2002}
{Saxton} C.~J.,  {Bicknell} G.~V.,   {Sutherland} R.~S.,  2002, \mn@doi [\apj]
  {10.1086/342679}, \href
  {https://ui.adsabs.harvard.edu/abs/2002ApJ...579..176S} {579, 176}

\bibitem[\protect\citeauthoryear{{Schekochihin}, {Cowley}, {Taylor}, {Maron}
  \& {McWilliams}}{{Schekochihin} et~al.}{2004}]{schekochinin_2004}
{Schekochihin} A.~A.,  {Cowley} S.~C.,  {Taylor} S.~F.,  {Maron} J.~L.,
  {McWilliams} J.~C.,  2004, \mn@doi [\apj] {10.1086/422547}, \href
  {https://ui.adsabs.harvard.edu/abs/2004ApJ...612..276S} {612, 276}

\bibitem[\protect\citeauthoryear{{Sebokolodi}, {Perley}, {Eilek}, {Carilli},
  {Smirnov}, {Laing}, {Greisen}  \& {Wise}}{{Sebokolodi}
  et~al.}{2020}]{lerato_2020}
{Sebokolodi} M. L.~L.,  {Perley} R.,  {Eilek} J.,  {Carilli} C.,  {Smirnov} O.,
   {Laing} R.,  {Greisen} E.~W.,   {Wise} M.,  2020, \mn@doi [\apj]
  {10.3847/1538-4357/abb80e}, \href
  {https://ui.adsabs.harvard.edu/abs/2020ApJ...903...36S} {903, 36}

\bibitem[\protect\citeauthoryear{{Seta} \& {Federrath}}{{Seta} \&
  {Federrath}}{2020}]{seta_2020}
{Seta} A.,  {Federrath} C.,  2020, \mn@doi [\mnras] {10.1093/mnras/staa2978},
  \href {https://ui.adsabs.harvard.edu/abs/2020MNRAS.499.2076S} {499, 2076}

\bibitem[\protect\citeauthoryear{{Seta}, {Shukurov}, {Wood}, {Bushby}  \&
  {Snodin}}{{Seta} et~al.}{2018}]{seta_2018}
{Seta} A.,  {Shukurov} A.,  {Wood} T.~S.,  {Bushby} P.~J.,   {Snodin} A.~P.,
  2018, \mn@doi [\mnras] {10.1093/mnras/stx260610.48550/arXiv.1708.07499},
  \href {https://ui.adsabs.harvard.edu/abs/2018MNRAS.473.4544S} {473, 4544}

\bibitem[\protect\citeauthoryear{{Shah} \& {Seta}}{{Shah} \&
  {Seta}}{2021}]{hilay_2021}
{Shah} H.,  {Seta} A.,  2021, \mn@doi [\mnras] {10.1093/mnras/stab2500}, \href
  {https://ui.adsabs.harvard.edu/abs/2021MNRAS.508.1371S} {508, 1371}

\bibitem[\protect\citeauthoryear{{Silpa}, {Kharb}, {Ho}, {Ishwara-Chandra},
  {Jarvis}  \& {Harrison}}{{Silpa} et~al.}{2020}]{silpa_2020}
{Silpa} S.,  {Kharb} P.,  {Ho} L.~C.,  {Ishwara-Chandra} C.~H.,  {Jarvis}
  M.~E.,   {Harrison} C.,  2020, \mn@doi [\mnras] {10.1093/mnras/staa2970},
  \href {https://ui.adsabs.harvard.edu/abs/2020MNRAS.499.5826S} {499, 5826}

\bibitem[\protect\citeauthoryear{{Silpa}, {Kharb}, {Harrison}, {Ho}, {Jarvis},
  {Ishwara-Chandra}  \& {Sebastian}}{{Silpa} et~al.}{2021a}]{silpa_2021a}
{Silpa} S.,  {Kharb} P.,  {Harrison} C.~M.,  {Ho} L.~C.,  {Jarvis} M.~E.,
  {Ishwara-Chandra} C.~H.,   {Sebastian} B.,  2021a, \mn@doi [\mnras]
  {10.1093/mnras/stab1870}, \href
  {https://ui.adsabs.harvard.edu/abs/2021MNRAS.507..991S} {507, 991}

\bibitem[\protect\citeauthoryear{{Silpa}, {Kharb}, {Harrison}, {Ho}, {Jarvis},
  {Ishwara-Chandra}  \& {Sebastian}}{{Silpa} et~al.}{2021b}]{silpa_2021b}
{Silpa} S.,  {Kharb} P.,  {Harrison} C.~M.,  {Ho} L.~C.,  {Jarvis} M.~E.,
  {Ishwara-Chandra} C.~H.,   {Sebastian} B.,  2021b, \mn@doi [\mnras]
  {10.1093/mnras/stab1870}, \href
  {https://ui.adsabs.harvard.edu/abs/2021MNRAS.507..991S} {507, 991}

\bibitem[\protect\citeauthoryear{{Silpa}, {Kharb}, {Harrison}, {Girdhar},
  {Mukherjee}, {Mainieri}  \& {Jarvis}}{{Silpa} et~al.}{2022}]{silpa_2022}
{Silpa} S.,  {Kharb} P.,  {Harrison} C.~M.,  {Girdhar} A.,  {Mukherjee} D.,
  {Mainieri} V.,   {Jarvis} M.~E.,  2022, \mn@doi [\mnras]
  {10.1093/mnras/stac1044}, \href
  {https://ui.adsabs.harvard.edu/abs/2022MNRAS.513.4208S} {513, 4208}

\bibitem[\protect\citeauthoryear{{Subrahmanyan}, {Saripalli}, {Safouris}  \&
  {Hunstead}}{{Subrahmanyan} et~al.}{2008}]{subrahmanyan_2008}
{Subrahmanyan} R.,  {Saripalli} L.,  {Safouris} V.,   {Hunstead} R.~W.,  2008,
  \mn@doi [\apj] {10.1086/529007}, \href
  {https://ui.adsabs.harvard.edu/abs/2008ApJ...677...63S} {677, 63}

\bibitem[\protect\citeauthoryear{{Subramanian}}{{Subramanian}}{2018}]{kandu_2018}
{Subramanian} K.,  2018, arXiv e-prints, \href
  {https://ui.adsabs.harvard.edu/abs/2018arXiv180903543S} {p. arXiv:1809.03543}

\bibitem[\protect\citeauthoryear{{Swain}, {Bridle}  \& {Baum}}{{Swain}
  et~al.}{1998}]{swain_1998}
{Swain} M.~R.,  {Bridle} A.~H.,   {Baum} S.~A.,  1998, \mn@doi [\apjl]
  {10.1086/311663}, \href
  {https://ui.adsabs.harvard.edu/abs/1998ApJ...507L..29S} {507, L29}

\bibitem[\protect\citeauthoryear{{Tchekhovskoy} \& {Bromberg}}{{Tchekhovskoy}
  \& {Bromberg}}{2016}]{tchekhovskoy_2016}
{Tchekhovskoy} A.,  {Bromberg} O.,  2016, \mn@doi [\mnras]
  {10.1093/mnrasl/slw064}, \href
  {https://ui.adsabs.harvard.edu/abs/2016MNRAS.461L..46T} {461, L46}

\bibitem[\protect\citeauthoryear{{Timmerman} et~al.,}{{Timmerman}
  et~al.}{2022}]{timmerman_2022}
{Timmerman} R.,  et~al., 2022, \mn@doi [\aap] {10.1051/0004-6361/202140880},
  \href {https://ui.adsabs.harvard.edu/abs/2022A&A...658A...5T} {658, A5}

\bibitem[\protect\citeauthoryear{{Tribble}}{{Tribble}}{1991}]{tribble91}
{Tribble} P.~C.,  1991, \mn@doi [\mnras] {10.1093/mnras/253.1.147}, \href
  {https://ui.adsabs.harvard.edu/abs/1991MNRAS.253..147T} {253, 147}

\bibitem[\protect\citeauthoryear{{Vaidya}, {Mignone}, {Bodo}, {Rossi}  \&
  {Massaglia}}{{Vaidya} et~al.}{2018}]{vaidya_2018}
{Vaidya} B.,  {Mignone} A.,  {Bodo} G.,  {Rossi} P.,   {Massaglia} S.,  2018,
  \mn@doi [\apj] {10.3847/1538-4357/aadd17}, \href
  {https://ui.adsabs.harvard.edu/abs/2018ApJ...865..144V} {865, 144}

\bibitem[\protect\citeauthoryear{{Wagner}, {Bicknell}  \& {Umemura}}{{Wagner}
  et~al.}{2012}]{wagner12}
{Wagner} A.~Y.,  {Bicknell} G.~V.,   {Umemura} M.,  2012, \mn@doi [\apj]
  {10.1088/0004-637X/757/2/136}, \href
  {https://ui.adsabs.harvard.edu/abs/2012ApJ...757..136W} {757, 136}

\bibitem[\protect\citeauthoryear{{Zamaninasab}, {Savolainen}, {Clausen-Brown},
  {Hovatta}, {Lister}, {Krichbaum}, {Kovalev}  \& {Pushkarev}}{{Zamaninasab}
  et~al.}{2013}]{zamanisab_2013}
{Zamaninasab} M.,  {Savolainen} T.,  {Clausen-Brown} E.,  {Hovatta} T.,
  {Lister} M.~L.,  {Krichbaum} T.~P.,  {Kovalev} Y.~Y.,   {Pushkarev} A.~B.,
  2013, \mn@doi [MNRAS] {10.1093/mnras/stt1816}, \href
  {https://ui.adsabs.harvard.edu/abs/2013MNRAS.436.3341Z} {436, 3341}

\bibitem[\protect\citeauthoryear{{van Breugel}, {Miley}  \& {Heckman}}{{van
  Breugel} et~al.}{1984}]{Breugel_1984}
{van Breugel} W.,  {Miley} G.,   {Heckman} T.,  1984, \mn@doi [\aj]
  {10.1086/113480}, \href
  {https://ui.adsabs.harvard.edu/abs/1984AJ.....89....5V} {89, 5}

\bibitem[\protect\citeauthoryear{{van Breugel}, {Miley}, {Heckman}, {Butcher}
  \& {Bridle}}{{van Breugel} et~al.}{1985}]{van_1985}
{van Breugel} W.,  {Miley} G.,  {Heckman} T.,  {Butcher} H.,   {Bridle} A.,
  1985, \mn@doi [\apj] {10.1086/163007}, \href
  {https://ui.adsabs.harvard.edu/abs/1985ApJ...290..496V} {290, 496}

\makeatother
\end{thebibliography}

\appendix

\section{Decrease in the magnetic field in the `no-jet' runs and the SAM}
\label{no_jet_test}

\subsection{Time evolution of `no-jet' runs}
\label{A1}
\begin{figure*}  
\centerline{
\def\arraystretch{1.0}
\setlength{\tabcolsep}{0.0pt}
\begin{tabular}{lcr}
 \includegraphics[width=0.4\linewidth]{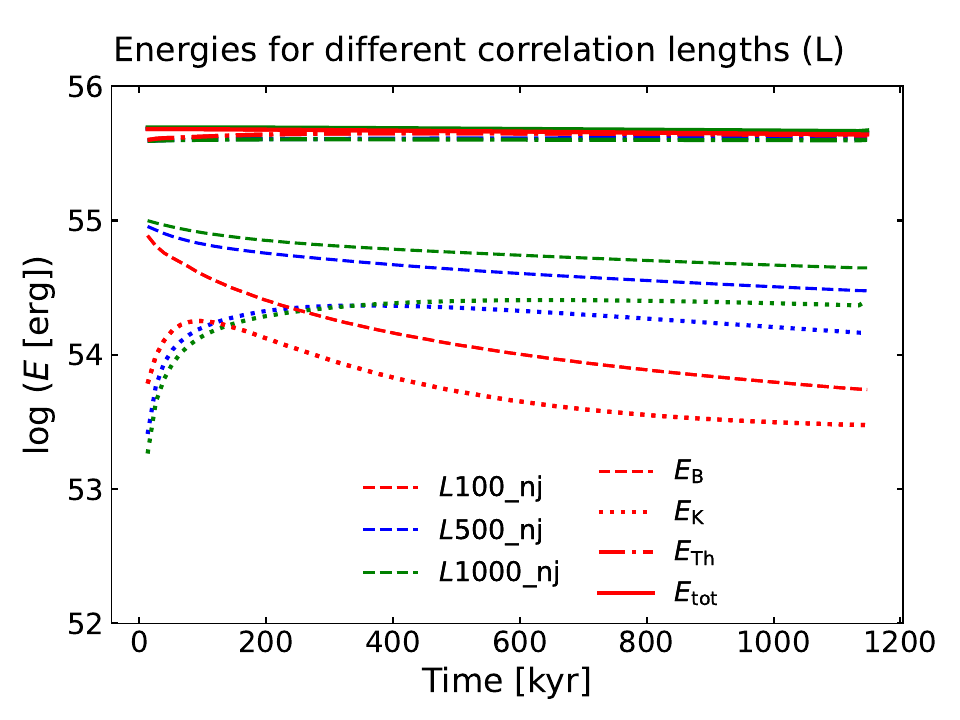} 
    \includegraphics[width=0.4\linewidth]{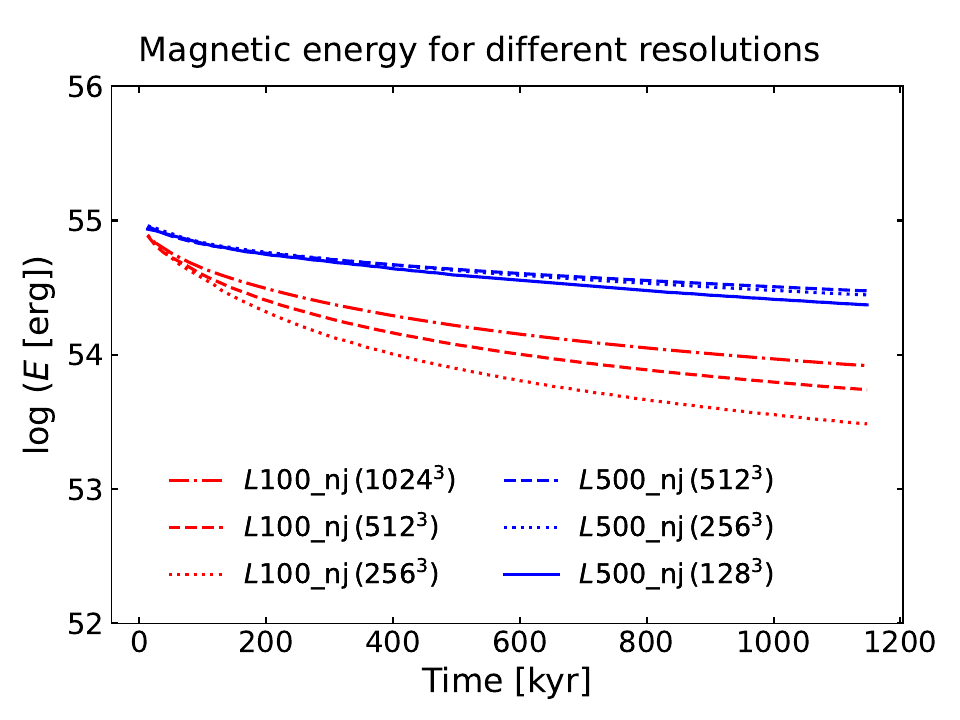} 
     \end{tabular}}    
     \caption{Left: Time evolution of the different energies: magnetic ($E_\mathrm{B}$), thermal ($E_\mathrm{Th}$), kinetic ($E_\mathrm{K}$) and total energy ($E_\mathrm{tot}$) for the `no-jet' runs at different correlation lengths. Evolution of magnetic energy for different resolutions.}
     \label{fig:no_jet}
\end{figure*}

We examine the time evolution of the turbulent magnetic field for the control or `no-jet' runs in this section.
We do not use the `Shock-flattening' in these runs, where a more diffusive Riemann solver is used in the shocked cells (see Sec.~\ref{pluto_setup}), except for $1024^3$, where the limit for the ratio of the pressure difference and minimum pressure across the consecutive zones is kept at 6 to identify the shocked cells.
In the left panel of Fig.~\ref{fig:no_jet}, we show the time evolution of the different energy measures for `no-jet' runs with different correlation lengths. It can be seen that the total energy (illustrated by the black curve) remains almost constant till the end time, with a minor decrease of up to 7\% occurring due to the imposed outflow conditions at the boundaries. The magnetic energy clearly decays with time in all cases. For a similar grid resolution, the fields with small coherence lengths decay faster, leading to a rapid lowering in the magnetic energy.

We also perform convergence studies for $\mathrm{L=100~pc}$ and $\mathrm{L=500~pc}$ cases with different resolutions to investigate the effect of numerical diffusion. The time evolution of magnetic energies is shown in the right panel of Fig.~\ref{fig:no_jet}. The convergence can be seen from $256^3$ resolution onwards for 500~pc case. The magnetic energy for the $\mathrm{L=100~pc}$ case appears to progress towards convergence as well from $512^3$ onwards. We compare the $1024^3$ run for $\mathrm{L=100~pc}$ with $256^3$ run for $\mathrm{L=500~pc}$ correlation length for analyzing the effect of numerical diffusion. The correlation length of the field in the former is resolved by around 25 cells, which is comparable to 30 cells for resolving the length in the latter. Thus the two simulations, although of different total grid size and correlation lengths, resolve the magnetic field structures by similar number of cells, and hence the magnetic filaments are expected to experience similar amounts of numerical diffusion. However, even with similar resolution for the filaments, the $\mathrm{L=100~pc}$ decays faster, suggesting that the numerical diffusion is not totally dominating in lowering the magnetic energy in the simulation domain.

We also find that the physical factors such as density, pressure, and velocity within the surrounding environment attempt to align with the turbulent magnetic field. As a consequence, gas movements occur, potentially leading to a conversion of the magnetic energy into kinetic and thermal forms. The change in total kinetic energy is around $\sim 10^{54}~\mathrm{erg}$ for the large-scale fields, which is comparable to the change in the magnetic energy. Contrarily, for L~=~100~pc, the increase in the kinetic energy is one order lower. Thus the inhomogeneous magnetic fields give rise to local motions, which in turn enhances the decay of the magnetic field. The field structures with shorter correlation length scales are then expected to decay faster as they will likely have shorter turn over time scales for the local eddies. This is likely giving rise to the more rapid decay of the $\mathrm{L=100~pc}$ scale runs, even without the jet. Such a decay in the external medium can lead to low field values at the forward shock.

\subsection{Evolution of the magnetic field in the SAM for different external magnetic fields in jetted simulations}
\label{A2}
\begin{figure}
 \centerline{
\def\arraystretch{1.0}
\setlength{\tabcolsep}{0.0pt}
\begin{tabular}{lcr}
    \includegraphics[width=0.7\linewidth]{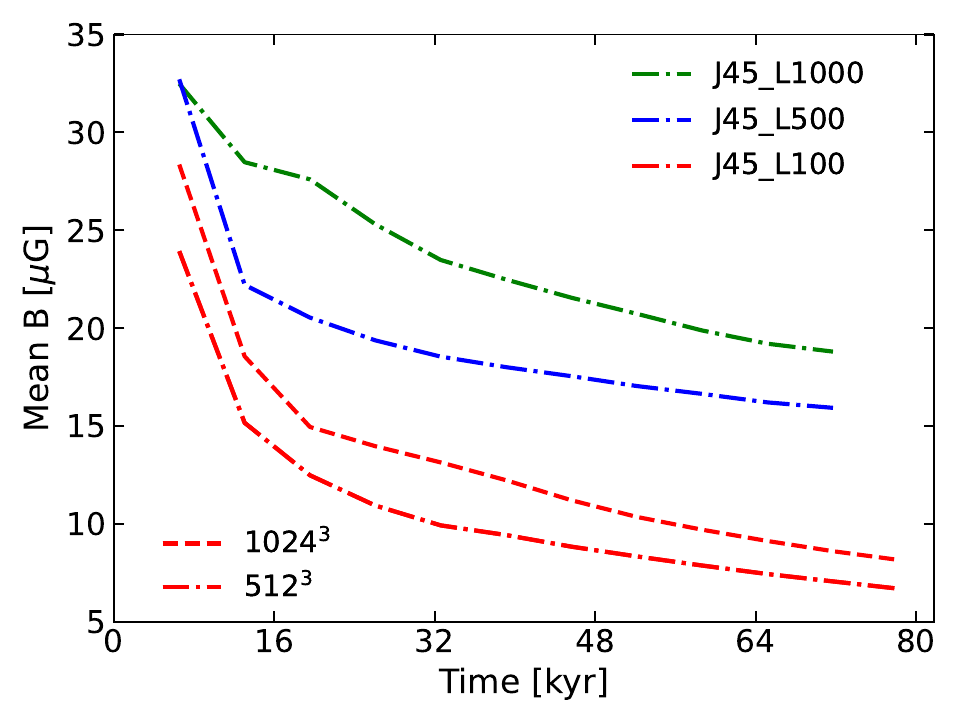}
     \end{tabular}}    
     \caption{Time evolution of the mean magnetic field in the shocked ambient medium (SAM) for jets with power $10^{45}\,\ergs$ interacting with fields at different correlation lengths and different resolutions of the simulations. The dash-dotted curves show the evolution for $\mathrm{J45\_L100}0$ (green), $\mathrm{J45\_L500}$ (blue) and $\mathrm{J45\_L100}$ (red) at the standard resolution of $512^3$. The dashed red curve shows the evolution for a resolution of $1024^3$ for $\mathrm{J45\_L100}$. 
     }
     \label{fig:res_test2}
\end{figure}

We perform a run for $\mathrm{J45\_L100}$ using a higher grid resolution of $1024^3$ than the standard run ($512^3$). We show the time evolution of the mean magnetic field in the SAM in Fig.~\ref{fig:res_test2}. In this plot, we also show the time evolution for the $\mathrm{J45\_L500}$ and $\mathrm{J45\_L1000}$ in blue and green curves, respectively. As explained in the previous section~\ref{A1}, the magnetic fields for all correlation lengths decay, with a faster decrease for the $\mathrm{L=100~pc}$ case due to plasma motions. For the jetted simulations, this decay is further enhanced due to adiabatic expansion of the magnetic fields in the SAM. Nonetheless, the mean-field values obtained for the L~=~100~pc case for a resolutions of $512^3$ and $1024^3$ are quite comparable at the end of the simulations.

Thus, we conclude from the above discussion that the low-values of the magnetic fields for $\mathrm{L=100~pc}$ cases in the SAM is aided by two causes: 1) Random motions induced due to the presence of turbulent magnetic field in the ambient medium, and 2) Adiabatic expansion of fields in the SAM.


\section{Identification of cocoon and forward shock}
\label{appendix:forward_shock}
\begin{figure}
    \centering
    \includegraphics[scale=0.5]{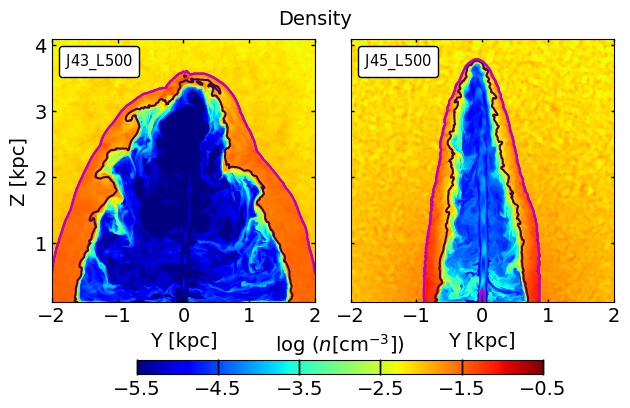}
    \caption{Density slice in the Y-Z plane for $\mathrm{J43\_L500}$ and $\mathrm{J45\_L500}$ at 834.6~kyr and 71.7~kyr, respectively. The black contour encloses the cocoon region (i.e. tr1$> 10^{-7}$ for $\pi_J$ maps), and the region inside the forward shock is enclosed by the magenta contour, which is used for the total polarization ($\pi_T$).}
    \label{fig:marks}
\end{figure}

\begin{figure}
    \centerline{
\def\arraystretch{1.0}
\setlength{\tabcolsep}{0.0pt}
\begin{tabular}{lcr}
    \includegraphics[scale=0.35]{distr_B_j45.pdf} &
    \includegraphics[scale=0.35]{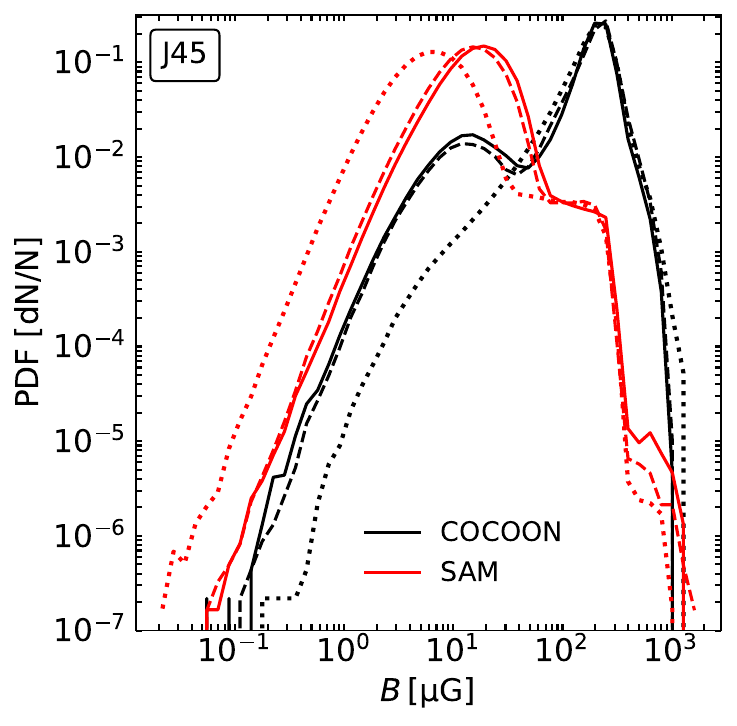} 
     \end{tabular}}
    \caption{PDF of magnetic fields in the cocoon and SAM for J45 at jet-tracer threshold of $\mathrm{tr1}>10^{-7}$ (left) and $\mathrm{tr1}>10^{-5}$ (right). Corresponding SAM is identified with $\mathrm{tr1}\leq 10^{-7}$ and $\mathrm{tr1}\leq10^{-5}$ in the left and right plots, respectively, from Eq.~\ref{eq:forward}. The cases with lower tracer threshold show high magnetic fields in the SAM, which are likely due to the field from the cocoon mixing with the SAM and being misidentified. We thus use a tracer threshold of $10^{-7}$, which give acceptable results in separating the SAM and cocoon. The solid lines correspond to cases with a field correlation length of L~=~1000 pc, the dashed lines represent 500~pc, and the dotted lines depict 100~pc.}
    \label{fig:marks2}
\end{figure}

\begin{figure}
    \centering
    \includegraphics[scale=0.4]{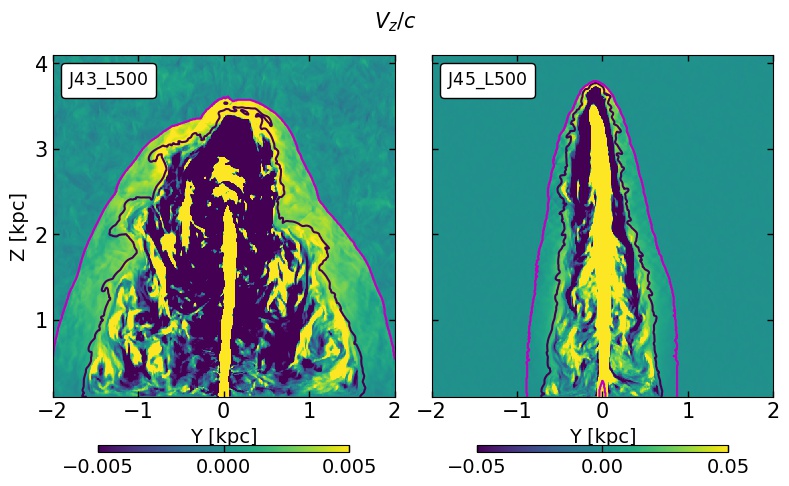} 
    \includegraphics[scale=0.4]{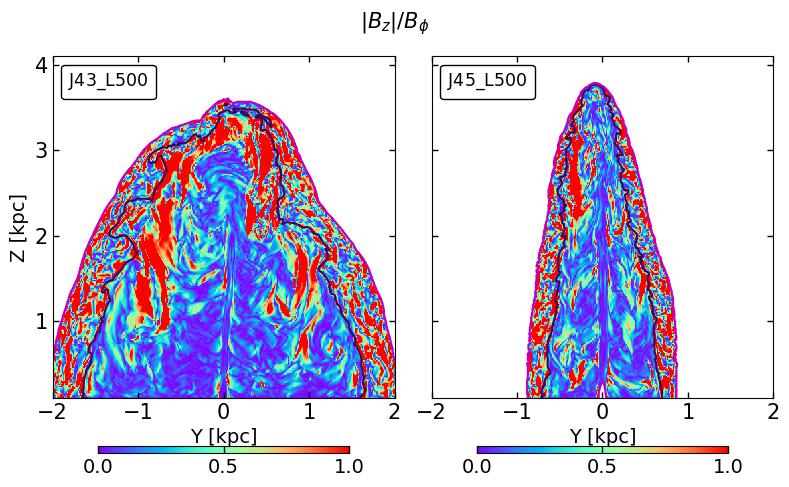} 
    \caption{$Z$ component of the velocity (top) and the ratio of poloidal ($B_z$) and toroidal ($B_{\phi} = \sqrt{{B_x}^2 +{B_y}^2}$) magnetic fields (bottom) in the $Y-Z$ plane for $\mathrm{J43\_L500}$ and $\mathrm{J45\_L500}$. The times for the snapshot and color for the contours are similar to Fig.~\ref{fig:marks}.}
    \label{fig:marks3}
\end{figure}

In terms of the pressure ($P$), initial pressure ($P_0$), jet-tracer (tr1)\footnote{Tracers in \textsc{pluto} are passive scalar quantities that are advected with the fluid. Their values lie between 0 and 1. At $t=0$, the jet-tracer is set to 1 at the injection zone and zero elsewhere.}, and the height of the jet ($L_j$), we define the regions inside the forward shock and shocked ambient medium or SAM (the regions between the contact discontinuity and the forward shock) as follows for different simulations:
    
\begin{gather}
  \textbf{J42}:(P>2 P_0\, \text{or\, tr1}\geq 10^{-20}\,\& \,Z<L_j + 0.5)\, \&\text{\,tr1} \leq 10^{-7} \nonumber \\
  \textbf{J43}:(P>4P_0\,\text{or\, tr1}\geq 10^{-20} \,\&\, Z<L_j + 0.5) \,\& \text{\,tr1} \leq 10^{-7} \nonumber \\
  \textbf{J44\,\&\,J45}:(P>4 P_0 \,\ \&\text{\, tr1}>10^{-30}\,\&\, Z<L_j + 0.5)\, \nonumber \\ \& \text{\,tr1} \leq 10^{-7}, \label{eq:forward}
\end{gather}
where $L_j$ is the maximum height of the jet estimated using a jet-tracer upper cutoff value of $10^{-7}$. In Eq.~\ref{eq:forward}, the regions within the forward shock, namely the cocoon and SAM, are chosen by the expression enclosed in the brackets. The condition for a limit on the $Z$ value and lower limit on the jet-tracer for $\mathrm{J44}$ and $\mathrm{J45}$ is set to exclude most of the high-pressured regions in the ambient medium that is not part of the forward shock \footnote{Such high-pressured regions are caused by the local compressions in the ambient medium (also in `no-jet' runs), which happens as the gas density tries to follow the distribution of the magnetic field.}. The SAM is identified by the last condition on the jet-tracer ($\mathrm{tr1} \leq 10^{-7}$). In Fig.~\ref{fig:marks}, we show the density map for $\mathrm{J43\_L500}$ and $\mathrm{J45\_L500}$, overlaid with the contours showing the cocoon (black) and forward shock (magenta). Using $\mathrm{tr1}>10^{-7}$ encompasses the entire cocoon along with the outer parts of the mixing layer, which may contain high magnetic fields originating from the jets. The comparison of PDF of the fields in the SAM (see Fig.~\ref{fig:marks2}) depicts that the limit $10^{-5}$ shifts the tail of the PDF towards higher values. Such high magnetic fields in the SAM are unlikely unless some cells with jet's high magnetic fields are now included in the SAM, as compressed fields start decaying after they enter the SAM. This can bias the result for the depolarization from the external fields in the SAM, and that is why we use a limit of $10^{-7}$ for the jet-tracer in this study. We also show the $Z$ component of the velocity and ratio of poloidal and toroidal components of the field in the $Y-Z$ slice (same as Fig.~\ref{fig:marks}) in Fig.~\ref{fig:marks3}. Extended poloidal fields comparable in strength to the toroidal components can be seen near the contact surface regions and close to the jet's head. The central lower regions of the cocoon are primarily dominated by the injected toroidal fields.

\section{Emission and polarization maps for different correlation lengths}
\label{app:figures}
\begin{figure*}
         \centerline{
\def\arraystretch{1.0}
\setlength{\tabcolsep}{0.0pt}
\begin{tabular}{lcr}
    \includegraphics[height=4.5cm,keepaspectratio]{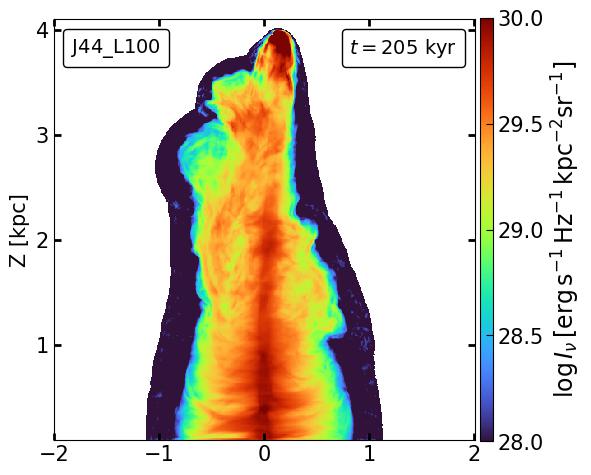} &
    \includegraphics[height=4.5cm,keepaspectratio]{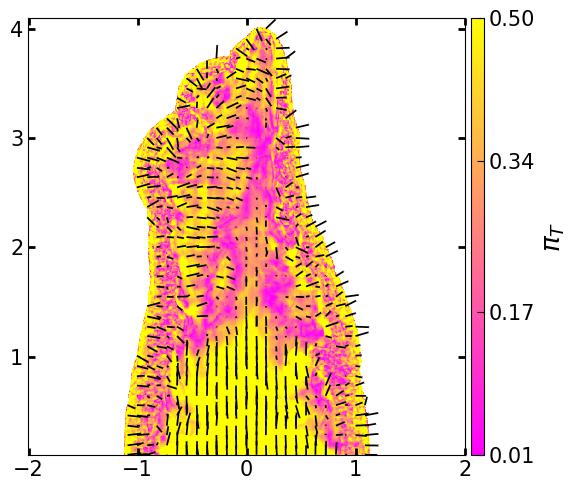} 
    \includegraphics[height=4.5cm,keepaspectratio]{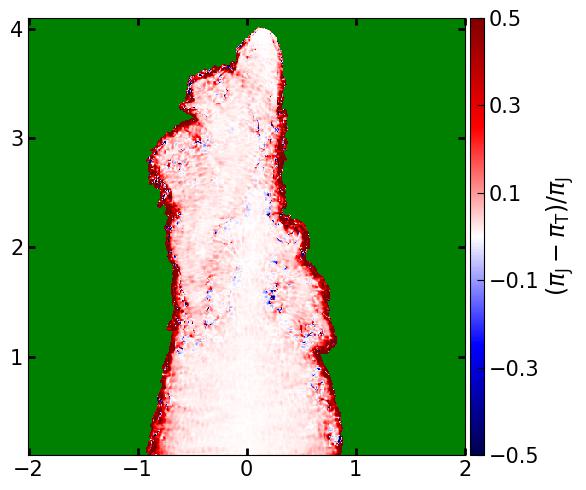} 
     \end{tabular}}
 \centerline{
\def\arraystretch{1.0}
\setlength{\tabcolsep}{0.0pt}
\begin{tabular}{lcr}
      \includegraphics[height=4.65cm,keepaspectratio]{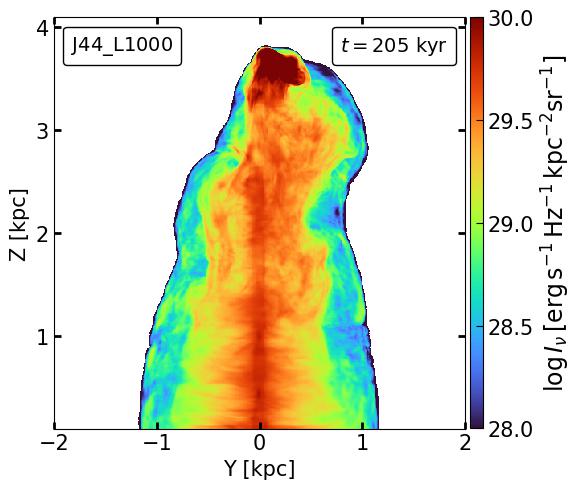} &
    \includegraphics[height=4.65cm,keepaspectratio]{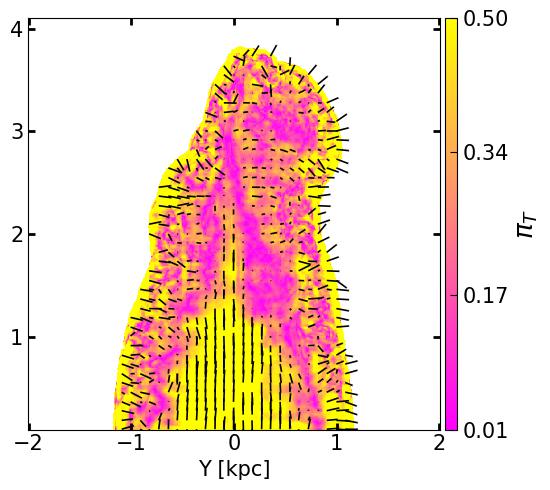} 
    \includegraphics[height=4.65cm,keepaspectratio]{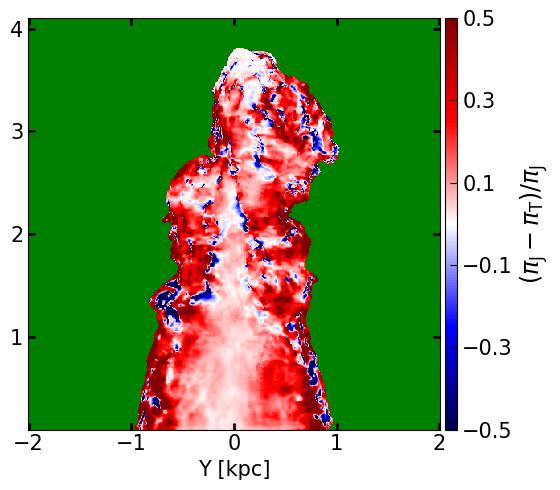} 
     \end{tabular}}
     \caption{\textbf{Top:} Logarithmic total synchrotron emission, polarization fraction ($\pi_T$) and fractional change in polarization ($\Delta\pi$) for $\mathrm{J44\_L100}$ at $\theta_I=90^\circ$ image plane. \textbf{Bottom:} Same for $\mathrm{J44\_L1000}$. }
\label{fig:diff_len_44}
\end{figure*}

\begin{figure*}
         \centerline{
\def\arraystretch{1.0}
\setlength{\tabcolsep}{0.0pt}
\begin{tabular}{lcr}
    \includegraphics[height=4.5cm,keepaspectratio]{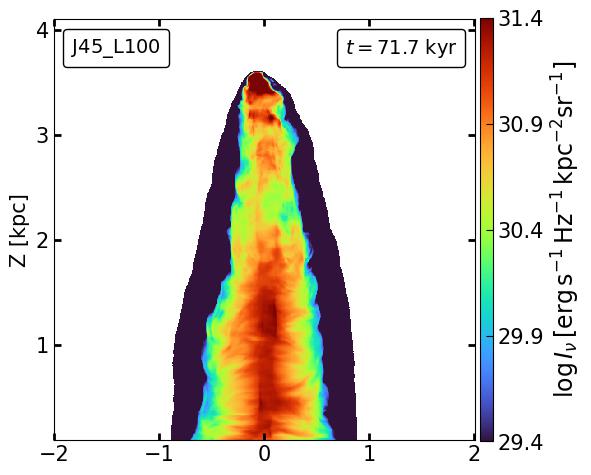} &
    \includegraphics[height=4.5cm,keepaspectratio]{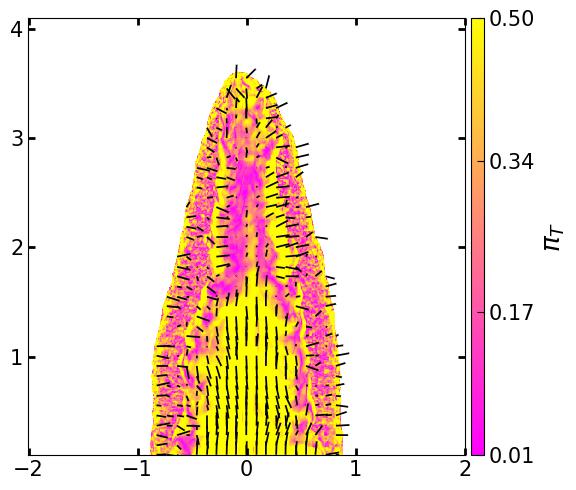} 
    \includegraphics[height=4.5cm,keepaspectratio]{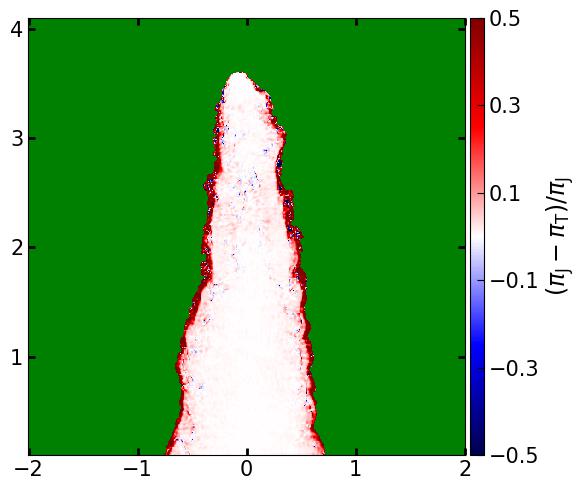} 
     \end{tabular}}
 \centerline{
\def\arraystretch{1.0}
\setlength{\tabcolsep}{0.0pt}
\begin{tabular}{lcr}
      \includegraphics[height=4.65cm,keepaspectratio]{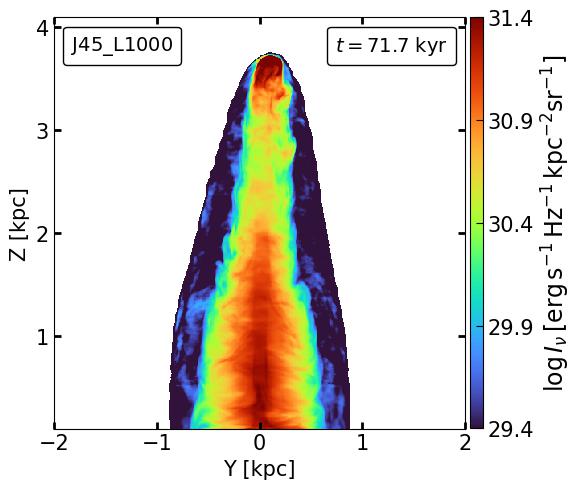} &
    \includegraphics[height=4.65cm,keepaspectratio]{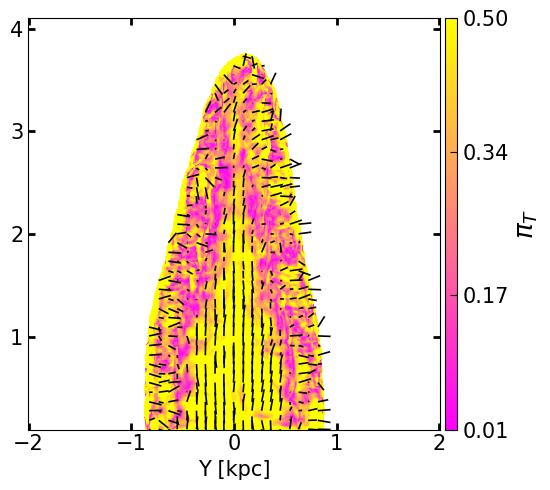} 
    \includegraphics[height=4.65cm,keepaspectratio]{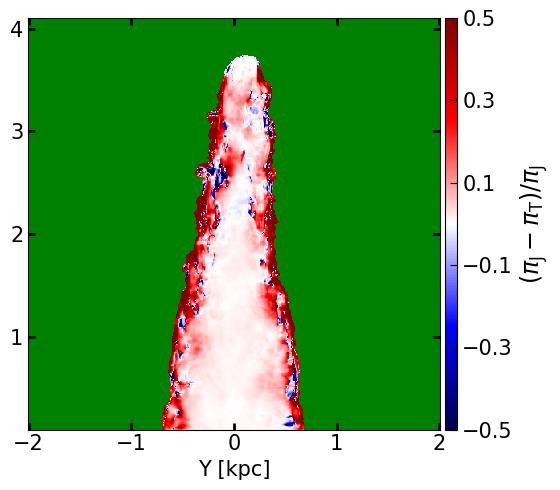} 
     \end{tabular}}
     \caption{\textbf{Top:} Logarithmic total synchrotron emission, polarization fraction ($\pi_T$) and fractional change in polarization ($\Delta\pi$) for $\mathrm{J45\_L100}$ at $\theta_I=90^\circ$ image plane. \textbf{Bottom:} Same for $\mathrm{J45\_L1000}$.}
\label{fig:diff_len_45}
\end{figure*}

In Fig.~\ref{fig:diff_len_44} and~\ref{fig:diff_len_45}, we show the maps for integrated synchrotron flux, polarization fraction ($\pi_T$), and the fractional change in the polarization fractions ($\Delta\pi$) due to the contribution from the SAM for different correlation lengths of the magnetic field, for $\mathrm{J44\_L(100,1000)}$ and $\mathrm{J45\_L(100,1000)}$, respectively.

\bsp	
\label{lastpage}
\end{document}